\documentclass{aa} 
\usepackage{graphicx}
\usepackage{url}
\usepackage[breaklinks=true]{hyperref}
\hypersetup{colorlinks=true,linkcolor=blue,citecolor=blue,filecolor=blue,urlcolor=blue}
\usepackage{twoopt}
\usepackage{natbib}
\bibpunct{(}{)}{;}{a}{}{,} 
\usepackage{amssymb}
\usepackage{float}
\usepackage{color}
\usepackage{epstopdf}
\usepackage{ctable}
\usepackage{multirow}
\usepackage[font=small]{caption}
\usepackage{multirow}
\usepackage{comment}
\usepackage{txfonts}

\usepackage{xcolor}

\newcommand{\cluster}{PSZ2G091}

\begin{document} 


\title{The diffuse radio emission in the high-redshift cluster PSZ2\,G091.83+26.11: total intensity and polarisation analysis with Very Large Array 1-4 GHz observations}

\author{
G.~Di Gennaro\inst{\ref{inst:hamb}}\thanks{Alexander von Humboldt Fellow}
\and
M.~Br\"uggen\inst{\ref{inst:hamb}}
\and
R.J.~van Weeren\inst{\ref{inst:leiden}}
\and
A.~Simionescu\inst{\ref{inst:leiden},\ref{inst:sron},\ref{inst:japan}}
\and
G.~Brunetti\inst{\ref{inst:ira}}
\and
R.~Cassano\inst{\ref{inst:ira}}
\and
W.R.~Forman\inst{\ref{inst:cfa}}
\and
M.~Hoeft\inst{\ref{inst:tauten}}
\and
A.~Ignesti\inst{\ref{inst:padova}}
\and
H.J.A.~R\"ottgering\inst{\ref{inst:leiden}}
\and
T.W.~Shimwell\inst{\ref{inst:leiden},\ref{inst:astron}}
}

\institute{
{Hamburger Sternwarte, Universit\"at Hamburg, Gojenbergsweg 112, 21029 Hamburg, Germany}\label{inst:hamb}
\and
{Leiden Observatory, Leiden University, PO Box 9513, 2300 RA Leiden, The Netherlands}\label{inst:leiden}
\and
{SRON Netherlands Institute for Space Research, Sorbonnelaan 2, 3584 CA Utrecht, The Netherlands}\label{inst:sron}
\and
{Kavli Institute for the Physics and Mathematics of the Universe, The University of Tokyo, Kashiwa, Chiba 277-8583, Japan}\label{inst:japan}
\and
{Istituto Nazionale di Astrofisica-Istituto di Radioastronomia, Bologna Via Gobetti 101, I40129 Bologna, Italy}\label{inst:ira}
\and
{Center for Astrophysics $\mid$ Harvard \& Smithsonian, 60 Garden Street, Cambridge, MA 02138, USA}\label{inst:cfa}
\and
{Th\"uringer Landessternwarte, Sternwarte 5, 07778 Tautenburg, Germany}\label{inst:tauten}
\and
{INAF-Padova Astronomical Observatory, Vicolo dell’Osservatorio 5, I-35122 Padova, Italy}\label{inst:padova}
\and
{ASTRON, The Netherlands Institute for Radio Astronomy, Postbus 2, 7990 AA, Dwingeloo, The Netherlands}\label{inst:astron}
}

\date{Received 13 January 2023; Accepted 11 April 2023}

\abstract
{Diffuse radio emission in galaxy clusters, namely radio halos and radio relics, is usually associated with merger events. Despite the tremendous advance in observations in the last decades, the particle (re-)acceleration and magnetic field amplification mechanisms, and the connection with the stage and geometry of the cluster merger are still uncertain.
}
{In this paper, we present the peculiar case of PSZ2\,G091.83+26.11 at $z=0.822$. This cluster hosts a Mpc-scale radio halo and an elongated radio source, whose morphology resembles that of a radio relic. 
However, the location of this diffuse radio source with the respect to the intracluster medium (ICM) distribution and to the cluster centre is not consistent with a simple merger scenario.}
{We use Karl-Jansky Very Large Array (VLA) data at 1--4 GHz to investigate the spectral and polarisation properties of the diffuse radio emission. We combine these data with previously published data from the Low Frequency Array (LOFAR) in the 120--168 MHz band, and from the upgraded Giant Metrewave Radio Telescope (uGMRT) at 250--500 and 550-900 MHz. Finally, we complement the radio data with {\it Chandra} X-ray observations, to compare the thermal and non-thermal emission of the cluster.}
{The elongated radio emission east of the cluster is visible up to 3.0 GHz and has an integrated spectral index of $\alpha^{\rm 3.0GHz}_{\rm 144MHz}=-1.24\pm0.03$, with a steepening from $-0.89\pm0.03$ to $-1.39\pm0.03$. These values correspond to Mach numbers $\mathcal{M}_{\rm radio,int}=3.0\pm0.19$ and $\mathcal{M}_{\rm radio,inj}=2.48\pm0.15$. {\it Chandra} data reveals a surface brightness discontinuity at the location of the radio source, with a compression factor of $\mathcal{C}=2.22^{+0.39}_{-0.30}$ (i.e. $\mathcal{M}_{\rm Xray}=1.93^{+0.42}_{-0.32}$).  We also find that the source is polarised at GHz frequencies. Using {\it QU}-fitting, we estimate an intrinsic polarisation fraction of $p_0\sim0.2$, a Rotation Measure of ${\rm RM}\sim50~{\rm rad~m^{-2}}$ (including the Galactic contribution) and an external depolarisation of $\sigma_{\rm RM}\sim60~{\rm rad~m^{-2}}$. The polarisation $B$-vectors are aligned with the major axis of the source, suggesting magnetic field compression. Hence, we classify this source as a radio relic. Finally, we find a linear/super-linear correlation between the non-thermal and thermal emission.}
{We propose an off-axis merger and/or multiple merger events to explain the position and orientation of the relic with the respect to the ICM emission. Given the properties of the radio relic, we speculate that PSZ2\,G091.83+26.11 is in a fairly young merger state.}


\titlerunning{VLA total intensity and polarisation analysis of PSZ2\,G091.83+26.11}
\authorrunning{G. Di Gennaro et al.}
\maketitle

%

\section{Introduction}

It is well established that galaxy clusters grow over cosmic time via accretion of infalling matter (e.g. sub-clusters or galaxy groups) along the filaments that constitute the cosmic web \citep[e.g.][]{press+schecter74}. Mergers of sub-clusters and galaxy groups build the most massive virialised structures in the Universe, but also generate shock waves and trigger magneto-hydrodynamical turbulence in the intracluster medium \citep[ICM; see][]{markevitch+vikhlinin07}.
The presence of diffuse radio emission in clusters, which is not associated with black holes in galaxies (i.e. active galactic nuclei, or radio galaxies), is usually associated with these merger events. This radio emission is linked to the presence of (re-)accelerated particles (i.e. cosmic rays, CRs), with a Lorentz factor $\gamma_{\rm L}\gtrsim10^3$, and magnetic fields, on average levels of a few $\rm\mu Gauss$ \citep[for a theoretical review]{brunetti+jones14}. 
Diffuse radio emission in merging galaxy clusters can generally be divided in two categories\footnote{Additional sub-classes of diffuse radio emission are the {\it radio phoenices} and {\it mini radio halos}. We will not discuss them in this manuscript, but we refer to \cite{vanweeren+19} for a detailed description of these sources.}: {\it radio relics} and {\it radio halos} \citep[for an observational review]{vanweeren+19}. 

Radio relics are elongated structures, with sizes that can reach up to $\sim2$ Mpc. Recent high-resolution GHz-frequency observations have shown that they exhibit filamentary morphologies \citep{owen+14,digennaro+18,rajpurohit+20,rajpurohit+21,degasperin+22}. They are usually detected in the cluster outskirts \citep[e.g.][]{roettiger+99,vanweeren+10,pearce+17,digennaro+18,vazza+18}, 
in the proximity of shock discontinuities detected in X-rays \citep[e.g.][]{finoguenov+10,akamatsu+15,urdampilleta+18,digennaro+19}. Therefore, these sources are associated with {\it merger shocks}.
The most favourable formation scenario involves particle acceleration and magnetic field compression and amplification due to merger-induced shock propagation. At the shock location, first-order Fermi acceleration mechanisms provide energy to the thermal electrons in the ICM which then become ultra-relativistic (i.e. CRs). This is also known as diffusive shock acceleration mechanism \citep[DSA,][]{blandford+eichler87}. These CRs then lose energy via synchrotron and inverse Compton (IC) radiation, and fade away in the {\it post-shock} region. In support to this, the spectral index ($\alpha$, being $S_\nu\propto\nu^\alpha$) of these sources is flat (i.e. $\alpha\sim-0.8$) at the shock location and steepens towards the cluster centre (i.e. $\alpha\lesssim-1.2$). However, several issues arise 
in the acceleration of thermal electrons from the ICM. For instance, cluster-merger shocks show in X-ray too low Mach numbers ($\mathcal{M}\sim2-3$) to accelerate particles efficiently from the thermal pool \citep[e.g.][]{botteon+20b}. Moreover, a recent study on a sample of relics at $\sim150$ MHz with the Low Frequency Array (LOFAR) strongly suggested that the radio emission in the post-shock region extends too far downstream \citep{digennaro+18,rajpurohit+18} if we simply consider synchrotron and IC energy losses (Jones et al., subm.). Observational evidence \citep[e.g.][]{markevitch+05,vanweeren+17a} and numerical simulations \citep{kang+17} have suggested the presence of pre-existing electrons that are then re-accelerated by the shock. This idea is supported by observations of radio galaxies that appear to be close to radio relics. These could provide the source of fossil plasma needed in the re-acceleration scenario \citep{vanweeren+17a,digennaro+18}. 
Recently, another class of shocks have been proposed, i.e. the so-called {\it equatorial shocks}. These kind of events are more difficult to be observed, as they are thought to be formed in the very first stages of the cluster merger \citep{ha+18}. 
The only equatorial shock known to date is the one in the cluster pair 1E\,2216.0-0401 and 1E\,2215.7-0404 at redshift $z=0.09$ \citep{gu+19}.

Radio relics are also found to be highly polarised \citep[up to 60\% at GHz frequencies,][]{ensslin+98}. This is consistent with the picture of shock-acceleration formation, as a shock wave can compress and amplify cluster magnetic fields \citep{iapichino+bruggen12,donnert+18,wittor+19,dominguez-fernandez+21,hoeft+22}. Recent GHz-frequency high-resolution observations have shown that the filamentary morphology is also detected in polarised emission \citep{digennaro+21b,rajpurohit+22b,degasperin+22}. 

Radio halos have roundish shapes with Mpc sizes, and emission that follows the ICM distribution. Several recent studies have shown a correlation between the cluster X-ray luminosity and mass and the halo radio emission \citep[e.g.][]{cassano+13,cuciti+21b}, as well as a correlation between the presence of radio halos and the cluster disturbance \citep[e.g.][]{cassano+10,cuciti+21b}. The most favourable scenario involves second-order Fermi acceleration due to merger-induced magneto-hydrodynamical turbulence. Turbulence stochastically re-accelerates electrons, and triggers a small-scale dynamo that causes magnetic field amplification \citep[e.g.][]{brunetti+01,petrosian01,brunetti+lazarian07,brunetti+blasi05,brunetti+lazarian16,donnert+13}. Additionally, 
re-acceleration of secondary electrons emerging from proton-proton collisions might also support the synchrotron radiation from radio halos \citep{brunetti+lazarian11,pinzke+17,brunetti+17}, and still being consistent with $\gamma$-ray limits from Fermi-LAT observations 
\citep[e.g.][]{adam+21}. Radio halos are mostly characterised by a rather uniform spectral index distribution ($\alpha\sim-1.3$), although ultra-steep spectra radio halos have also been found \citep[i.e. $\alpha\lesssim-1.5$, e.g.][]{brunetti+08,dallacasa+09,bonafede+12,wilber+18,bruno+21,duchesne+21,digennaro+21c}. 
The existence of these ultra-steep spectra sources is one of the most stringent predictions of turbulent re-acceleration models \citep{cassano+06,cassano+12,cassano+23}. 

Unlike radio relics, radio halos are yet to be detected in polarisation. Magnetic field profiles and average magnetic field values have been obtained for a handful of systems \citep[e.g.][]{bonafede+13,stuardi+19,rajpurohit+22a,osinga+22} via Rotation Measure or depolarisation analysis of polarised background/embedded radio galaxies. These studies suggest average magnetic fields of a few $\rm\mu Gauss$, with higher values in the cluster centre.
Recently, $\rm\mu Gauss$-level magnetic fields in the cluster volume were also estimated in a sample of high-redshift (i.e. $z=0.6-0.9$) clusters observed with the LOw Frequency Array (LOFAR) at 144 MHz \citep{digennaro+21a}. Although these observations do not provide any constraints on the magnetic seeds, they yield strong limits on the evolution of magnetic fields when the Universe was only half of its age, and only a few Gyr after the first large-scale structure formed. Moreover, these observations provide additional evidence on the re-acceleration mechanisms \citep{cassano+19}, as all the less massive clusters (i.e. $M_{\rm 500,SZ}<6\times10^{14}~{\rm M_\odot}$) exhibit steep spectral index halos \citep[i.e. $\alpha<-1.5$,][]{digennaro+21c}.

\section{PSZ2\,G091.83+26.11}
\begin{figure}
\centering
\includegraphics[width=0.5\textwidth]{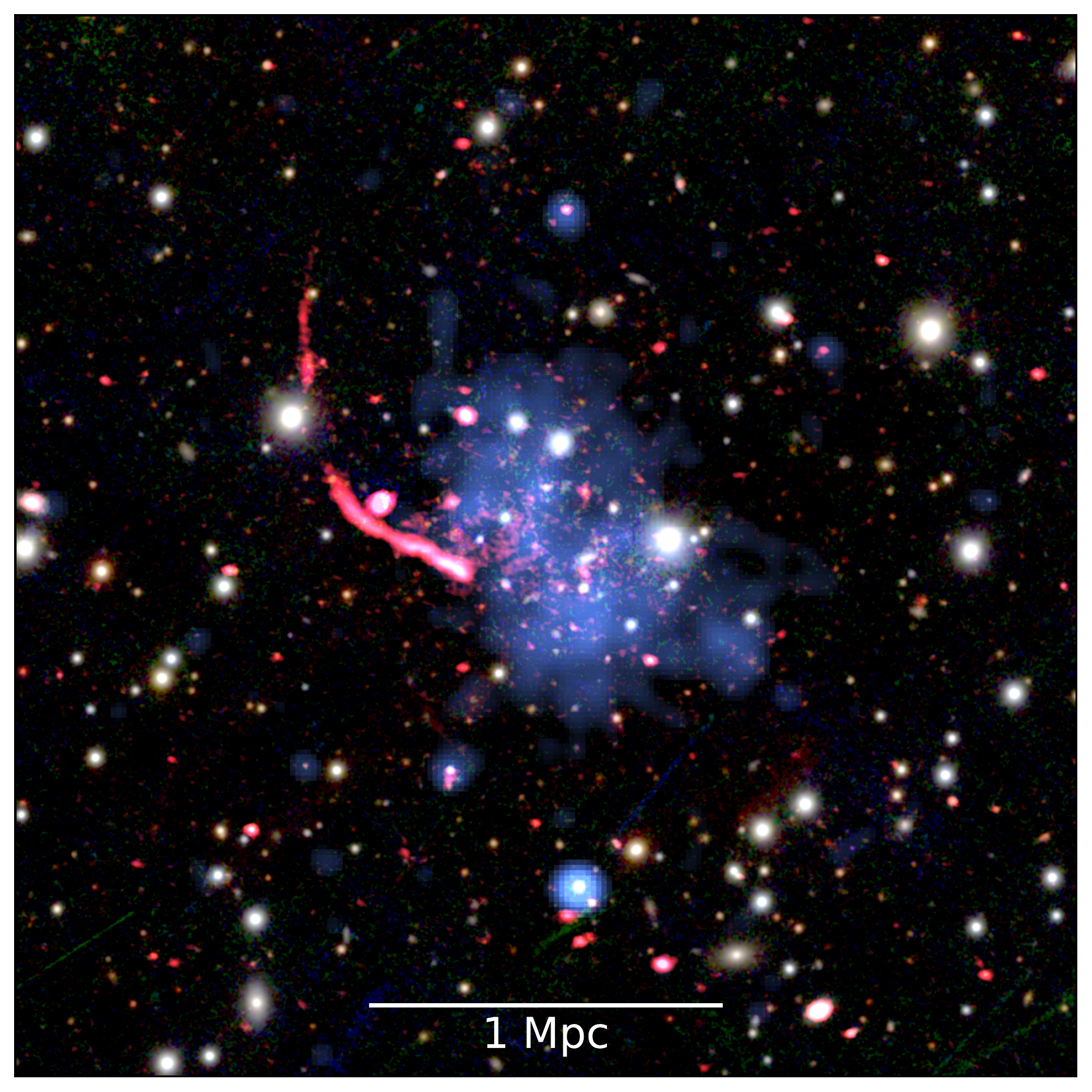}
\caption{Composite image of \cluster. Optical (white): PanSTARSS {\it gri}; Radio (red): 3.0 GHz VLA observations; X-ray (blue): {\it Chandra} 0.8--4.0 keV observations.}
\label{fig:opt_xray_radio}
\end{figure}

PSZ2\,G091.83+26.11 (hereafter, \cluster; Figure \ref{fig:opt_xray_radio} and Table \ref{tab:cluster}) was spectroscopically confirmed to be at $z=0.822$ by \cite{amodeo+18}. With a mass of $M_{\rm SZ, 500}=(7.4\pm0.4)\times10^{14}$ M$_\odot$ estimated from the total Sunyaev-Zel'dovich (SZ) intensity \citep[i.e. $Y_{\rm SZ, 500}=(2.3\pm0.3)\times10^{-4}$ Mpc$^2$; see][]{planckcoll16}, it is the third most massive cluster hosting diffuse radio emission at $z>0.8$ (``el Gordo'' at $z=0.870$ with  $M_{\rm SZ, 500}=(1.17\pm0.17)\times 10^{15}~\rm M_\odot$, \citealt{lindner+14}, and PSZ2\,G160.83+81.66 at $z=0.888$ with $M_{\rm SZ, 500}=5.7^{+0.6}_{-0.7}\times 10^{14}~\rm M_\odot$, \citealt{digennaro+21a,digennaro+21c}, are the other two). In particular, observations with LOFAR at 144 MHz and upgraded Giant Metrewave Radio Telescope (uGMRT) at 400 MHz and 650 MHz showed that the cluster hosts a giant radio halo and an additional elongated source in the east/south-east of the cluster \citep{digennaro+21a,digennaro+21c}. Due to its elongated shape and hints of spectral steepening towards the cluster centre, the latter was associated with a candidate radio relic. Preliminary inspection of the X-ray morphology with {\it Chandra} and XMM-Newton showed that the ICM emission of the cluster is highly disturbed, with two main peaks of emission, and is elongated in the north-east/south-west direction. 
More recently, \cite{artis+22} found that the thermal SZ emission of the cluster with the New IRAM Kids Arrays (NIKA2) at 2 mm has very similar morphology as the X-ray emission. By comparing the X-ray and the SZ surface brightness maps, they suggested that the two sub-clusters, identified by the two peaks of emission, could be in the first stages of a major merger. 

\begin{table}
\caption{Cluster information.}
\vspace{-5mm}
\begin{center}
\begin{tabular}{lr}
\hline
\hline
Cluster name & PSZ2\,G091.83+26.11 \\
Redshift ($z$) & 0.822 \\
Right Ascension (RA) & $\rm 18^h31^m11.136^s$\\
Declination (DEC) & $+62^\circ14^\prime56.04''$\\
Galactic Longitude ($l$ [deg]) & 91.829853 \\
Galactic Latitude ($b$ [deg]) & 26.116268\\
SZ Intensity ($Y_{\rm SZ,500}~[\times10^{-4}~{\rm Mpc^2}]$) & $2.3\pm0.3$ \\
Mass ($M_{\rm SZ,500}~[\times10^{14}~{\rm M_\odot}$]) & $7.4\pm0.4$ \\
Cosmological scale ($\rm kpc/''$) & 7.576 \\
\hline
\end{tabular}
\end{center}
\vspace{-5mm}
\label{tab:cluster}
\end{table}

\begin{table*}[h!]
\caption{Radio observation details.}
\vspace{-5mm}
\begin{center}
\begin{tabular}{lcccccccccc}
\hline
\hline
Project & \multicolumn{3}{c}{19B-080, 21A-025} & \multicolumn{3}{c}{15A-270, 21A-025}   \\
Band & \multicolumn{3}{c}{S-band} & \multicolumn{3}{c}{L-band}  \\
Freq. range [GHz] & \multicolumn{3}{c}{2--4} & \multicolumn{3}{c}{1--2}  \\
Central Freq. [GHz] & \multicolumn{3}{c}{3.0} & \multicolumn{3}{c}{1.5}  \\
No. Channels & \multicolumn{3}{c}{1024} & \multicolumn{3}{c}{1024} \\
Channel width [MHz] & \multicolumn{3}{c}{2} & \multicolumn{3}{c}{1}  \\
Configurations  & B & C & D & B & C & D  \\ 
Obs. Length [h] & 5 & 5 & 5 & 0.5, 4.5 & 5 & 5 \\ 
Obs. Date [dd-mm-yyyy] & 17-09-2021 & 13-06-2021 & 16-11-2019$^\dagger$ & 22-03-2015 & 20-06-2021 & 21-11-2021 \\ 
                     & & & 17-11-2019$^\dagger$ & 30-11-2021 & & 23-11-2021  \\ 
                     & & & 19-11-2019 & & & 24-11-2021 \\ 
Minimum {\it uv} coverage [$\lambda$] & 900 & 170 & 190 & 480 & 170 & 120  \\
Largest angular scale [$''$] & 120 & 970 & 970 & 58 & 490 & 490 \\
\hline
\end{tabular}
\end{center}
\vspace{-5mm}
\tablefoot{The observing length includes also the time on the calibrators. Number of channels and channel width are stated before 
the averaging in frequency. 
$^\dagger$During these days, the observation was split in two rounds.
}\label{tab:obs}
\end{table*}

Interestingly, while the central radio emission follows well the ICM emission resulting in the radio halo classification, the position of the candidate radio relic is parallel to the ICM elongation, and not perpendicular as is usually observed. 
In this paper, we therefore focus on the total intensity and polarisation emission of the candidate relic in \cluster, to shed lights on the origin of this radio source and on the cluster merger state. We used Karl Jansky Very Large Array (VLA) observations in L- and S-band (i.e. covering the 1--4 GHz frequency range). We complement the radio data with {\it Chandra}  X-ray observations. The paper is organised as follows: in Sect. \ref{sec:obs} we describe our observations; in Sects. \ref{sec:results} and \ref{sec:disc} we describe and discuss the results of our analysis; we conclude with the paper summary in Sect. \ref{sec:concl}.

Throughout the paper, we assume a standard $\Lambda$CDM cosmology, with $H_0 = 70$ km s$^{-1}$ Mpc$^{-1}$, $\Omega_m = 0.3$ and $\Omega_\Lambda = 0.7$. This translates to a luminosity distance of $D_{\rm L}=5187.7$ Mpc, and a scale of 7.576 kpc/$^{\prime\prime}$ at the cluster redshift, $z = 0.822$.

\section{Observations and data reduction}\label{sec:obs}
In this section, we describe the VLA and {\it Chandra} observation and calibration. We refer to \cite{digennaro+21a,digennaro+21c} for the LOFAR (120--168 MHz) and uGMRT (250--500 MHz, band 3; 550--900 MHz, band 4) data, for which we repeat the imaging to match them with the present radio images.

\subsection{VLA data}
\cluster\ was observed with the Karl Jansky Very Large Array (VLA) in the L- (1--2 GHz) and S-bands (2--4 GHz) in D-, C- and B-configurations (Table \ref{tab:obs}). The observations were mostly carried out during 2019 and 2021 (project codes: 19B-080 and 21A-025, PI: Di Gennaro), for a total observing time of 13 and 15 hours, in L- and S-band respectively. L-band B-configuration observations were complemented with archival snapshot observations (30 minutes on target, project code: 15A-270, PI: van Weeren).

The data reduction was carried out with \texttt{CASA v5.4} \citep{mcmullin+07}, using 3C286 as bandpass, flux and polarisation angle calibrator and J1927+6117 as phase calibrator. In the absence of a standard leakage calibrator (e.g. 3C147), we used instead J1407+2827\footnote{For the project code  15A-270, 3C147 was observed, therefore it was used as leakage calibrator.}. For both project codes, we followed the same data reduction strategy as described in \cite{digennaro+18}. Briefly, we split the wide-band observations in single spectral window ({\it spw}) datasets and, after a first removal of radio frequency interference (RFI) using the {\it tfcrop} mode, we calibrated the antenna delays, bandpass, cross-hand delays, and polarisation leakage and angle. Finally, we merged the {\it spw} together and applied all the solutions to the target, averaging in time and frequency (factor of two and four, respectively). A final round of \texttt{AOFlagger} \citep{offringa+10} on the cross-hand polarisation (LR and RL) was run to remove additional RFI. Bad {\it spw} were discarded from the final imaging. Due to bad quality, one L-band D-configuration observation was also discarded. We performed self-calibration on the single configurations to refine the amplitude and phase solutions, using \texttt{WSClean v2.10} \citep{offringa+14,offringa+17} to produce the model images. We finally combined the {\it uv}-data for all the configurations together, to produce the final images of the cluster, at different resolutions (see Table \ref{tab:images}). For the L-band dataset, an additional self-calibration on a bright source located at the edge of the primary beam ($\rm RA=18^h31^m24.6^s~DEC=+62^\circ30^\prime34.32''$) was needed to improve the quality of the image (i.e. {\it peeling}; see Appendix \ref{apx:Lband_extracal}). Additionally, we performed a bandpass calibration on the source, using the model derived from the self-calibration. 

In order to perform a polarisation analysis, we created images with a spectral resolution of $\Delta\nu=8$ MHz 
for all the Stokes parameters (i.e., $I$, $Q$ and $U$).
Following \cite{digennaro+21b}, Stokes-$Q$ and -$U$ images for each channel were produced with \texttt{WSClean} with the options \texttt{-join-channels} and \texttt{-join-polarizations}. We note that the option \texttt{-squared-channel-joining}, which is used to prevent the polarisation to average to zero, was not used as the Stokes-$Q$ and -$U$ were imaged separately\footnote{\url{https://wsclean.readthedocs.io/en/latest/rm_synthesis.html}}, using the Stokes-$I$ image to search for clean components (i.e. \texttt{-pol IQ} and \texttt{-pol IU}). 
We removed all the noisy or low-quality images, as well as those covering flagged {\it spw}, and 
finally regridded them to the same pixel grid and convolved to the same angular resolutions. 

All the VLA images presented in the manuscript are made with  \texttt{WSClean v2.10} with \texttt{Briggs} weighting and \texttt{robust=0}, unless stated otherwise, and using \texttt{multiscale} deconvolution.

\begin{table}
\caption{X-ray observation details.}
\vspace{-5mm}
\begin{center}
\begin{tabular}{lr}
\hline
\hline
ObsID & 18285 \\
Instrument & ACIS-I \\
Mode & \texttt{VFAINT} \\
Obs. Date [dd-mm-yyy] & 19-06-2016 \\
Right Ascension (RA) & $\rm 18^h30^m52.1^s$\\
Declination (DEC) & $+62^\circ17^\prime06.9''$\\
Obs. Cycle & 17 \\
Exposure time [ks] & 23.0 \\
Clean time [ks] &  23.0 \\
\hline
\end{tabular}
\end{center}
\vspace{-5mm}
\tablefoot{RA and DEC refer to the pointing coordinates.}
\label{tab:xray_info}
\end{table}
\subsection{Chandra}
We complemented our VLA radio data with {\it Chandra} observations. \cluster\ was observed on 19 June 2016 with the Advanced CCD Imaging Spectrometer (ACIS), using the ACIS-I CCD configuration, for a total time of 23 ks (see Table \ref{tab:xray_info}; PI: Rossetti). 
We processed the data with 
\texttt{CIAO v4.14} \citep{fruscione+06}, which applies the \texttt{CALDB 4.9.8} calibration files. We used \texttt{chandra\_repro} to generate the \texttt{level=2} event file, including the option \texttt{check\_vf\_pha=yes} that screens the particle background in very faint mode observations for `bad' events, most likely associated with cosmic rays.
We screened the light curve of the observation using the \texttt{CIAO} tasks \texttt{deflare} and \texttt{lc\_clean}. No time bins were identified where the count rate deviated from the mean by more than 2$\sigma$; the observation is therefore not affected by flares, and the full 23 ks exposure can be used for further science analysis.
Standard blank-sky files, scaled to match the hard-band count rate of the observation, were used for background subtraction.
The final exposure-corrected image was made in the 0.8--4.0 keV energy band, where the signal to detector background ratio is optimal. Where appropriate, compact sources unrelated to the diffuse ICM were removed from the spatial and spectral analysis.


\begin{figure*}[h!]
\centering
\includegraphics[width=\textwidth]{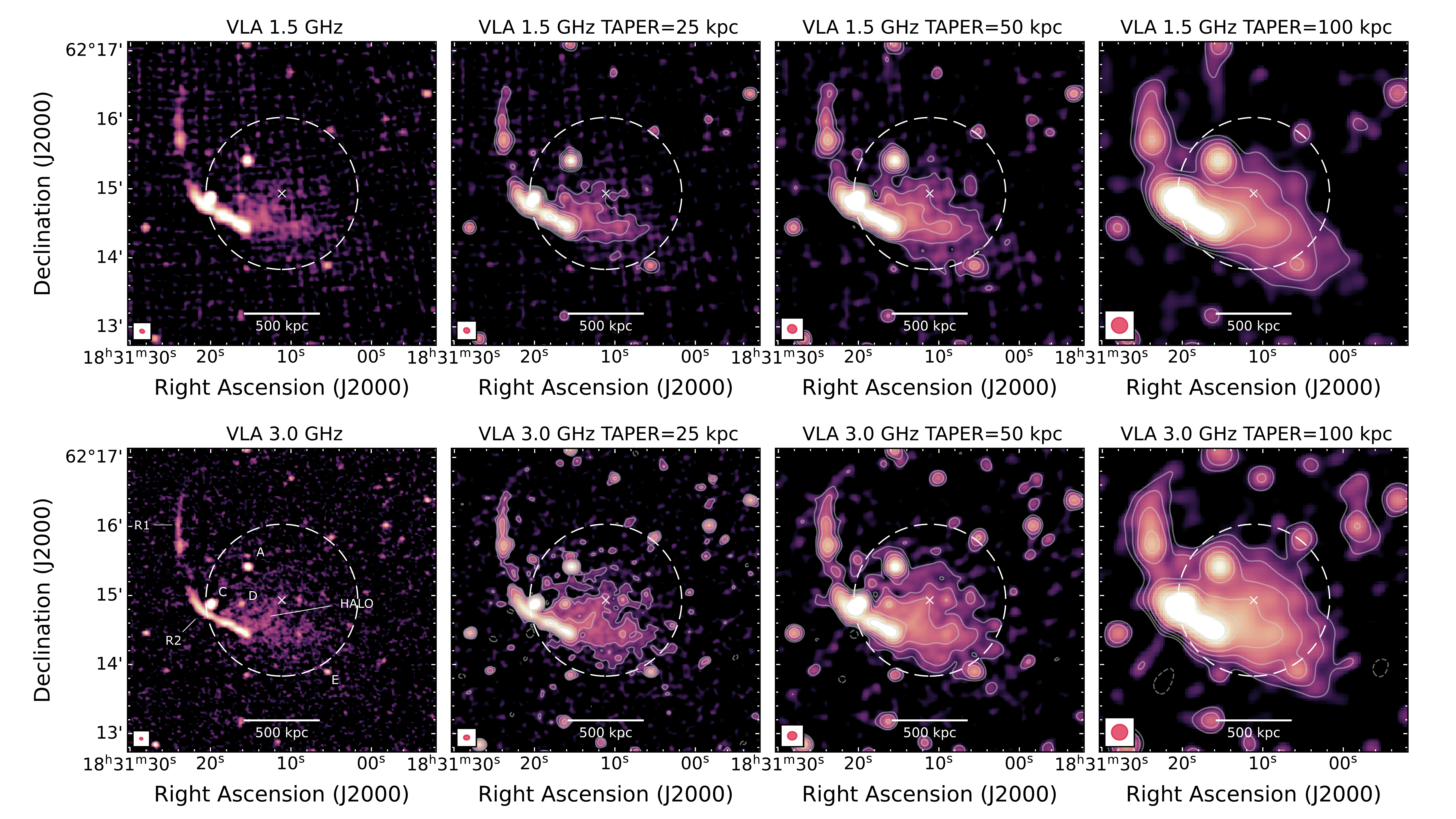}
\caption{VLA images of \cluster\ at 1.5 GHz (top row) and 3.0 GHz (bottom row). From left to right: full-resolution (i.e. no taper applied), taper of 25 kpc, 50 kpc and 100 kpc (see Table \ref{tab:images} for the final resolutions and map noise). The beam size is shown in the bottom left corner of each panel. Labels in the leftmost panel follow the one shown in \cite{digennaro+21c}. 
The white dashed circle shows the $R=0.5R_{\rm SZ,500}$ region, and the cluster centre is marked with a white cross. Radio contours, when showed, are drawn with white solid lines at the $2.5\sigma_{\rm rms}\times[1,2,4,8,16,32]$ levels; a $-2.5\sigma_{\rm rms}$ level is also drawn with a dashed line.}
\label{fig:vla_full_low_res}
\end{figure*}

\begin{figure*}
\centering
\includegraphics[width=0.19\textwidth]{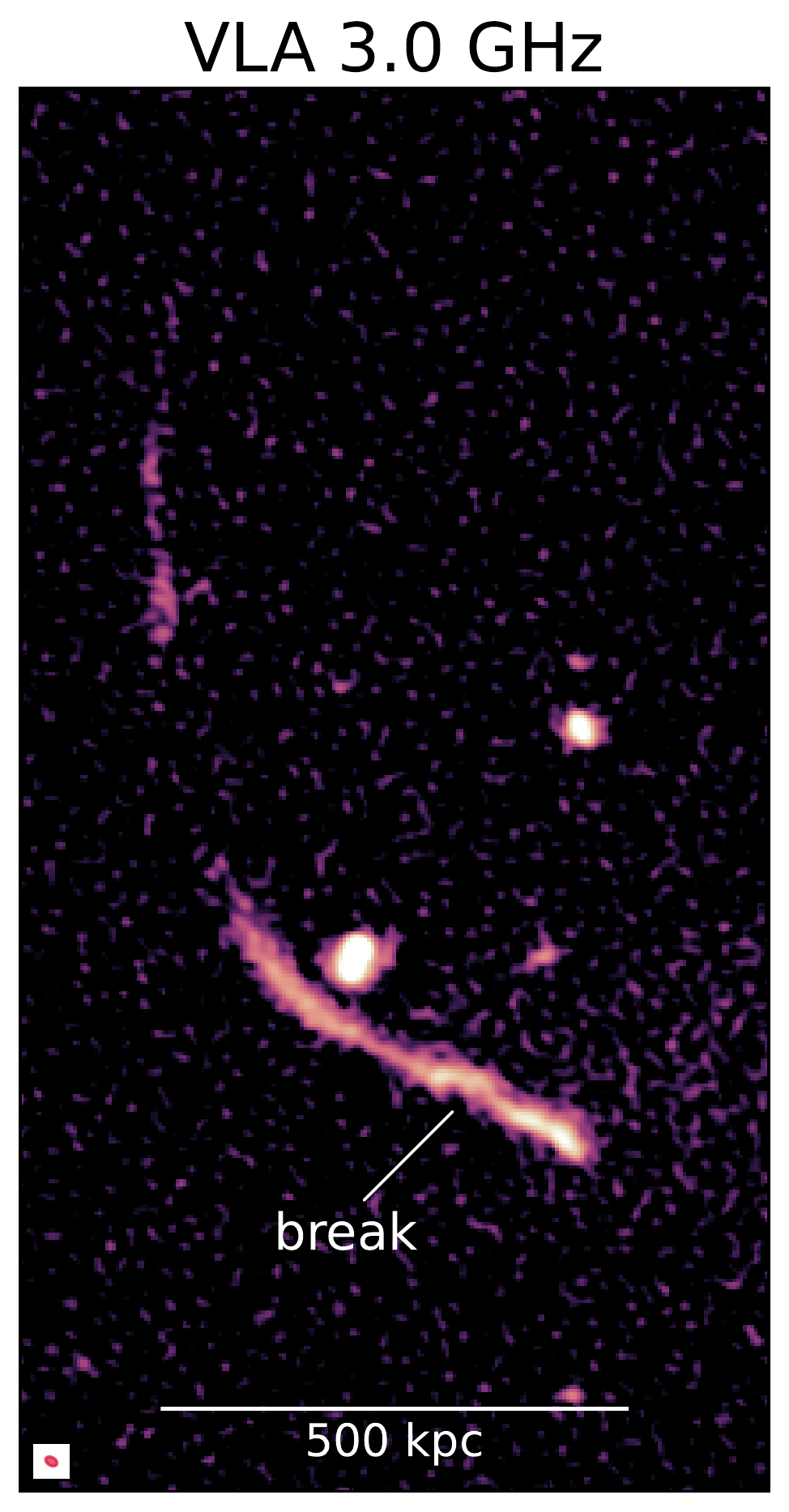}
\includegraphics[width=0.19\textwidth]{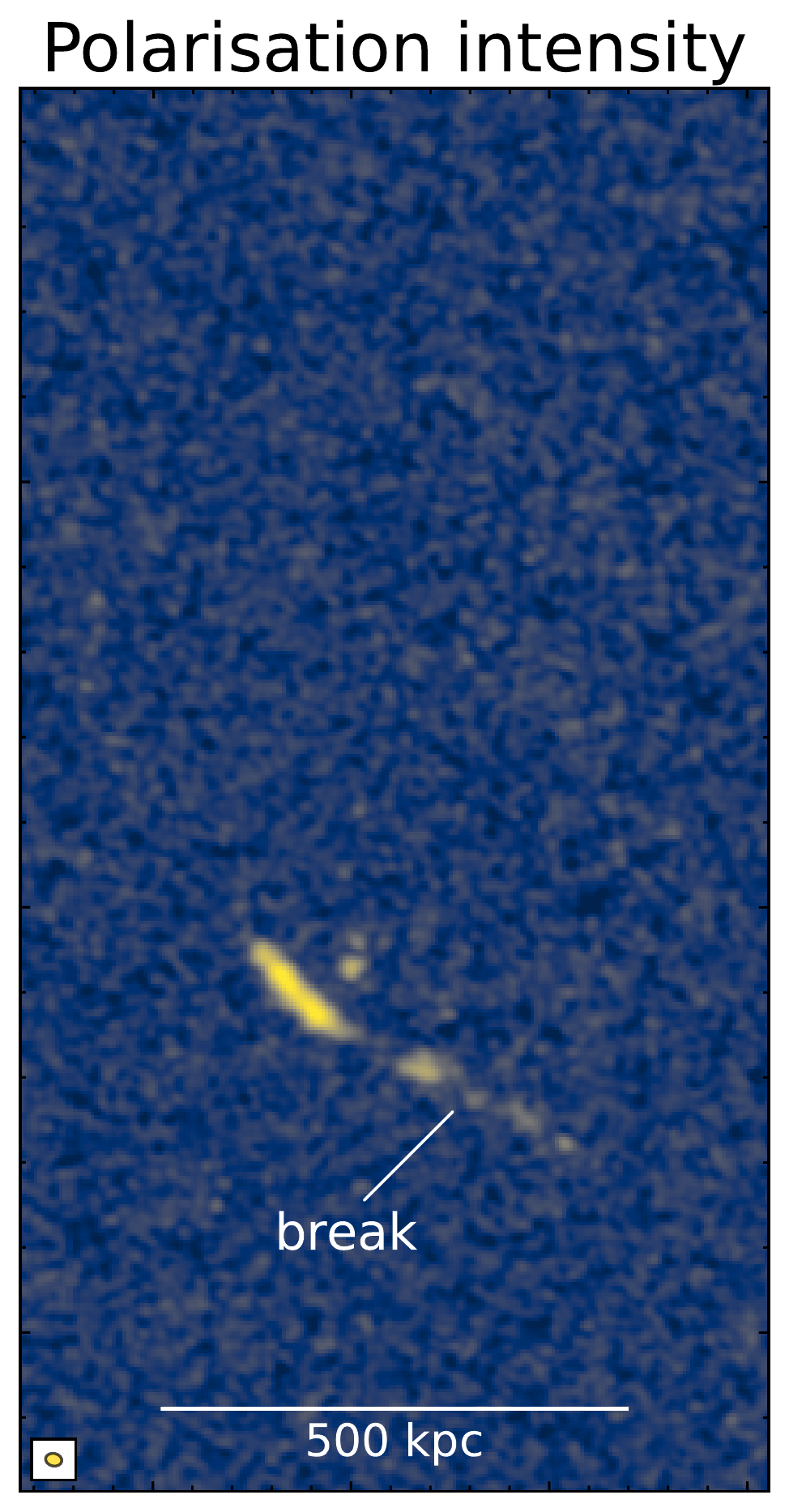}
\includegraphics[width=0.19\textwidth]{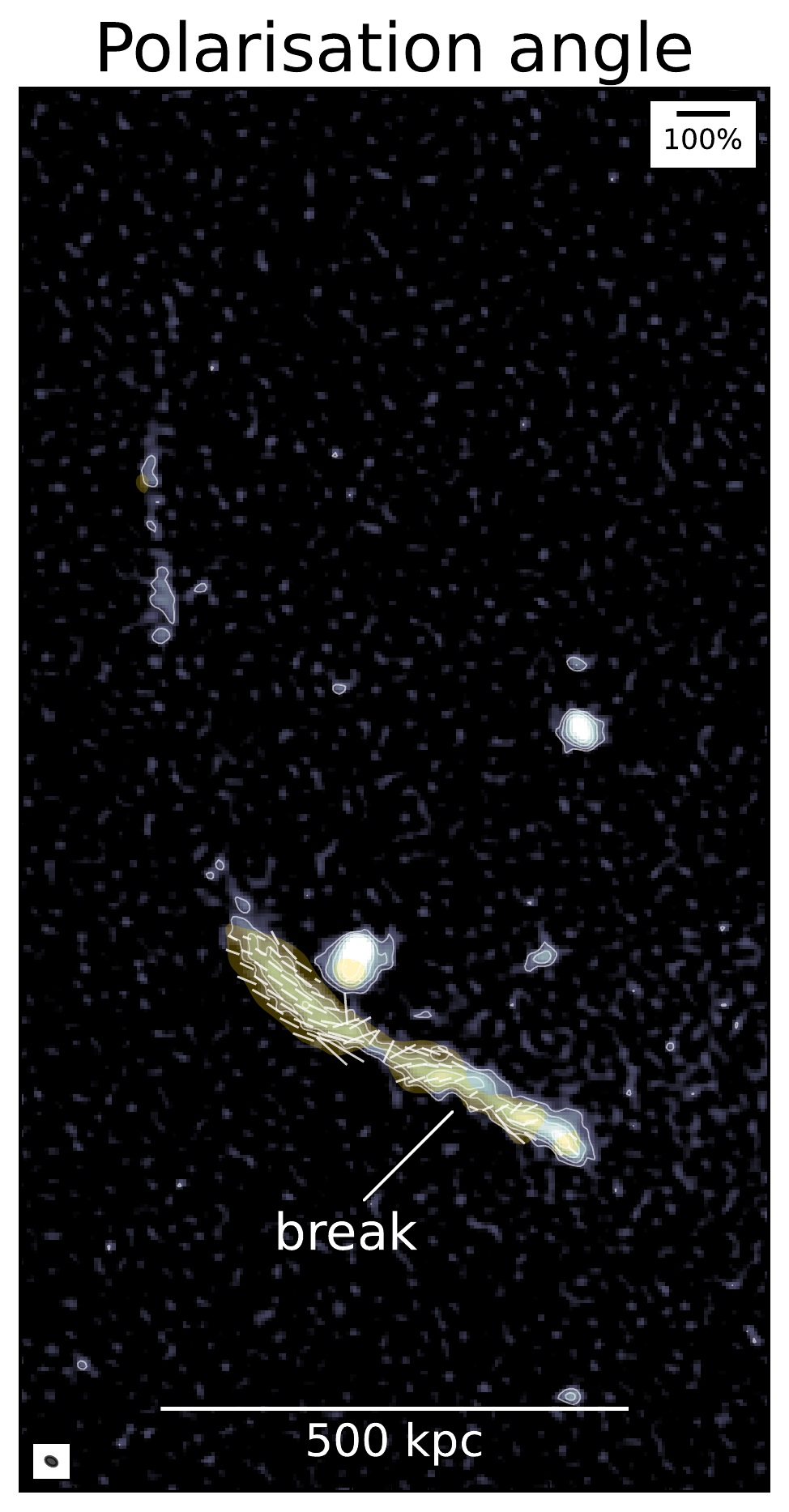}
\includegraphics[width=0.19\textwidth]{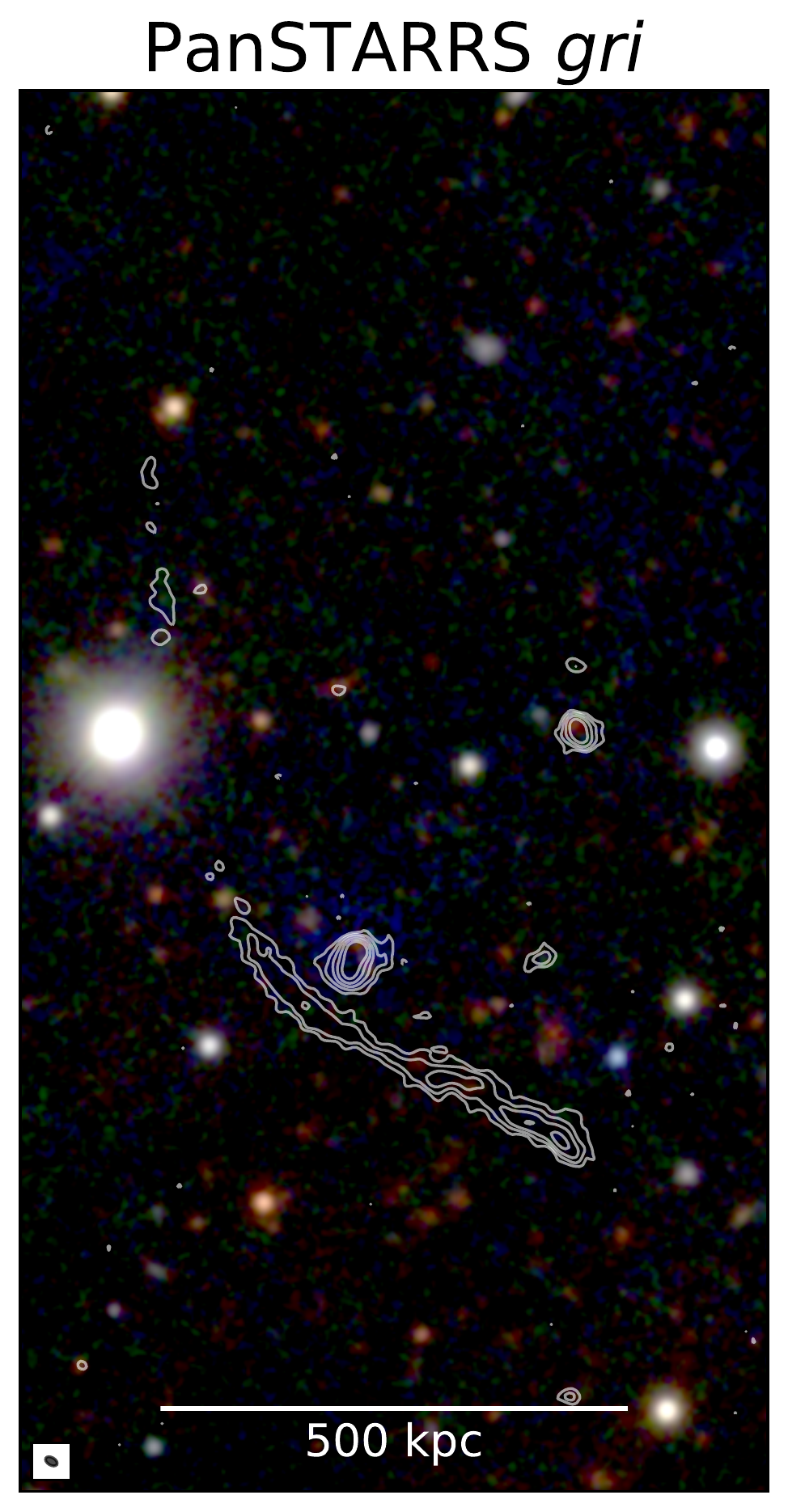}
\includegraphics[width=0.19\textwidth]{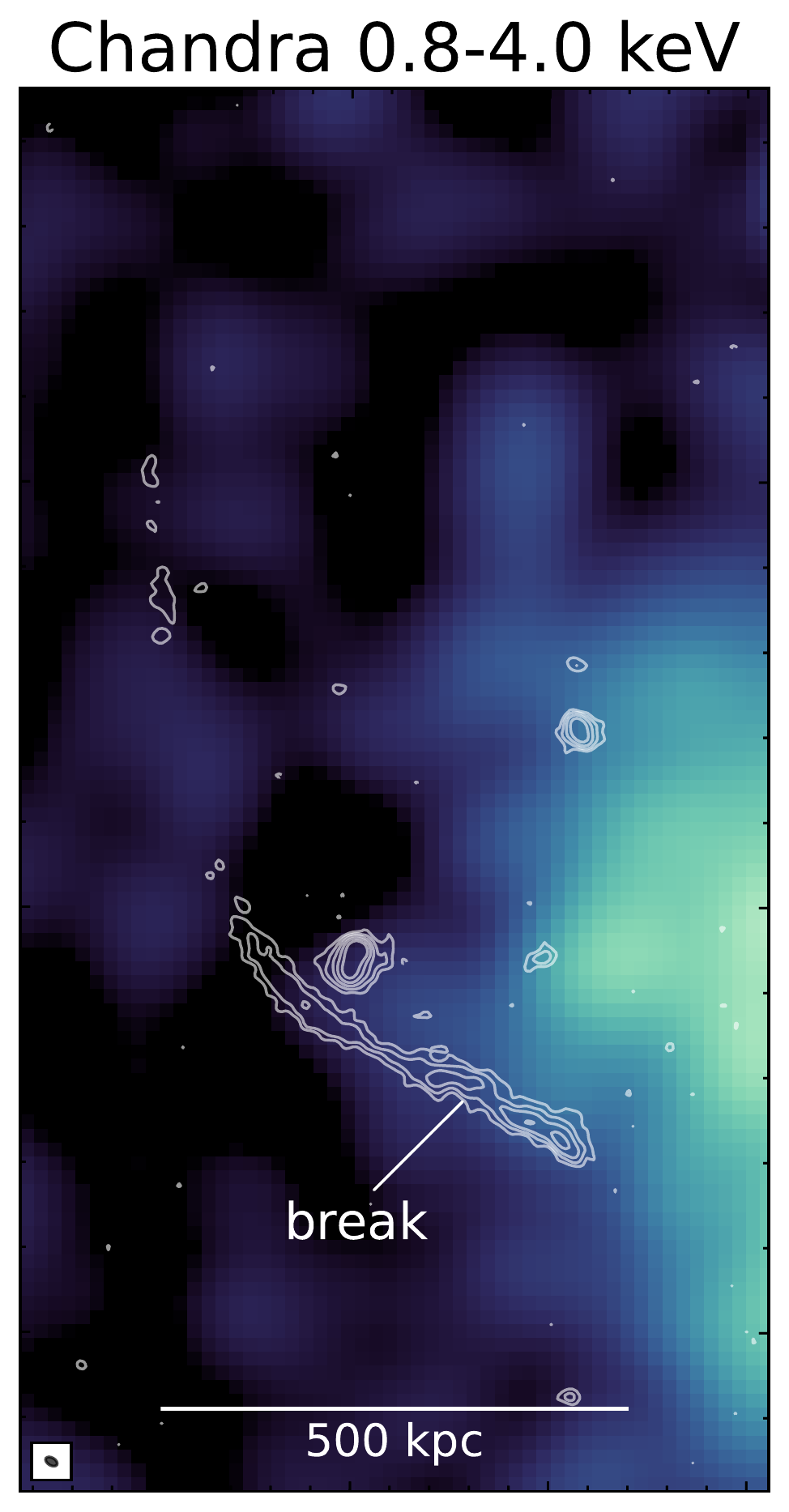}
\caption{Zoom on the radio relic. From left to right: highest-resolution total intensity VLA image (i.e. 3.0 GHz at $1.8''\times1.1''$; see Tab. \ref{tab:images}); total averaged polarisation intensity \citep[without correction for Ricean bias, see][]{digennaro+21b} in the 2--4 GHz band (effective frequency 3.1 GHz) at $5''$ resolution; high-resolution total intensity image at 3.0 GHz with the intrinsic polarisation magnetic field vectors at $5''$ resolution, corrected for Faraday rotation; optical PanSTARRS {\it gri} image with radio contours from the leftmost panel, starting from $2\sigma_{\rm rms}$; Chandra 0.8--4.0 keV image with radio contours from the leftmost panel, starting from $2\sigma_{\rm rms}$.}
\label{fig:relic_zoom}
\end{figure*}

\begin{table*}
\centering
\caption{Details on the polarisation images. Column 1: observing resolution; Column 2: observing bandwidth (i.e. first and last frequency channels); Column 3: single-channel width; Column 4: Number of channels for the L- and S-band; Column 5: map noise for the Stokes {\it I}, {\it Q} and {\it U}, calculated as the standard deviation of the datacube, and for the total averaged polarised intensity {\it P} (not corrected for the Ricean bias); Column 6: maximum observable Faraday depth; Column 7: resolution in the Faraday space; Column 8: resolution in the Faraday space; Column 9: largest observable Faraday scale.}
\resizebox{\textwidth}{!}{
\begin{tabular}{cccccccccccc}
\hline\hline
Resolution & Bandwidth & Channel width & \multicolumn{2}{c}{Number of channels} & \multicolumn{4}{c}{Map noise}  & max Faraday depth & Faraday resolution & max recovered scale \\
$\Theta~['']$ & $\Delta\nu$ [GHz] & $\delta\nu$ [MHz] & & & \multicolumn{4}{c}{$\rm \sigma_{rms}~[\mu Jy~beam^{-1}]$} & $||\rm \phi_{max}||~[rad~m^{-2}]$ & $\rm\delta\phi ~[rad~m^{-2}]$ & $\rm \Delta\phi_{max}~[rad~m^{-2}]$ \\
 & & & L-band & S-band & $I$ & $Q$ & $U$ & $P$ \\
\hline
5 & 2.0--4.0 & 8 & -- & 168 & 6.4 & 4.7 & 4.7 & 9.9 & 1400 & 205 & 558 \\
12.5 & 1.34--4.0 & 8 & 41 & 155 & 21.5 & 11.5 & 11.0 & 23.4 & 1400 & 78 & 558 \\
\hline
\end{tabular}}
\label{tab:poldetails}
\end{table*}

\section{Results}\label{sec:results}
\subsection{Total intensity and polarised images}
The VLA 1.5 GHz and 3.0 GHz images of \cluster\ (Fig.~\ref{fig:vla_full_low_res}) are remarkably similar to those at lower frequencies \citep[i.e. at 144 MHz, 400 MHz and 650 MHz,][]{digennaro+21c}. At both frequencies, the radio halo is visible also in the full-resolution images (i.e. $\Theta_{1.5}=4.3''\times3.2''$ and $\Theta_{3.0}=2.9''\times2.2''$). It is elongated in the north-east/south-west direction, following the emission from the X-ray {\it Chandra} image and filling almost completely the $0.5R_{\rm SZ,500}$ area (see Sect. \ref{sect:xray}). 
The brightest radio feature, however, is the candidate radio relic located about 500 kpc to the east of the cluster. It is a narrow filamentary structure that is sharply broken into two separate pieces, i.e. R1 and R2, forming an angle of about $130^\circ$. 
Among the two pieces of emission, R1 is much fainter than R2, and it is not completely detected in the full-resolution images at both frequencies. The length of the two sub-filaments is about $80''$ and $40''$ at low resolution, for R1 and R2 respectively, corresponding to about 640 kpc and 300 kpc at the cluster redshift. This remains unchanged in both the 1.5 GHz and 3.0 GHz images, and also compared to the LOFAR and uGMRT images \citep{digennaro+21c}. 
The radio source remains unresolved in its width, even at the highest resolution, i.e. $1.8''\times1.1''$ (3.0 GHz image, with \texttt{robust=-1.25}), meaning that the filamentary structure is about 15 kpc (deconvolved, i.e. full width half maximum).
At this  resolution we notice that the south-west side of R2, i.e. the brightest part of the candidate relic, appears to be broken (Fig.~\ref{fig:relic_zoom}). No particular feature in the X-ray image is visible at the same location, although we stress that the X-ray observation is quite shallow (i.e. 23 ks), and the radio source is located where the X-ray emission is  faint due to the low density of the ICM (see last panel in Fig. \ref{fig:relic_zoom}). Moreover, no optical galaxy appears to be associated with the radio source (see fourth panel in Fig. \ref{fig:relic_zoom}).

The location of this radio feature is quite interesting. Commonly, radio relics are located perpendicularly to the merger axis, as shock waves are generated after the collision of two or more clusters. The X-ray extension of the ICM emission of \cluster\ suggests that the merger event happened - or is happening - in the north-east/south-west direction, and the candidate radio relic (at least R2) appears to be parallel to it. 


In order to investigate the degree of polarisation of the cluster, we 
run the \texttt{pyrmsynth} tool\footnote{\url{https://github.com/mrbell/pyrmsynth}}, which combines the single-channel Stokes-{\it Q} and -{\it U} images to generate Faraday cubes through Rotation Measure (RM)-Synthesis technique \citep[][]{brentjens+debruyn05}. We assume a Faraday depth in the range of $\rm\pm4000~rad~m^{-2}$, with a sampling bin of $\rm 2~rad~m^{-2}$. We produced Faraday spectra both at high and low resolutions (i.e. $5''$ and $12.5''$), the former only using observations in the 2--4 GHz band. 
These frequency bandwidth and resolution correspond to a maximum Faraday depth ($||\phi_{\rm max}||=\sqrt{(3)}/\delta\lambda^2$, being $\delta\lambda^2$ the channel width in wavelength squared) of $\rm\sim1400~rad~m^{-2}$, a resolution in Faraday space ($\delta\phi=2\sqrt{(3)}/\Delta\lambda^2$, being $\Delta\lambda^2=\lambda_{\rm max}^2-\lambda_{\rm min}^2$ the total bandwidth in wavelength squared) of $\rm\sim78~rad~m^{-2}$ and a maximum recovered scale in Faraday space ($\rm \Delta\phi_{max}=\pi/\lambda_{min}^2$) of $\rm\sim558~rad~m^{-2}$ \citep[see Tab. \ref{tab:poldetails}]{brentjens+debruyn05}.
The only part of the cluster where we detect linear polarisation is R2, with the polarisation magnetic field vectors well aligned with the direction of the radio surface brightness edge (see second and third panels in Fig.~\ref{fig:relic_zoom}; see also Sect. \ref{sec:polariz}). However, we note that the southern part of R2 is fainter in polarisation than the northern part. This part of the candidate relic coincident with the ICM emission (see last panel in Fig.~\ref{fig:relic_zoom}), which may suggest a larger depolarisation effect on this piece of diffuse radio emission. No polarised emission is detected from compact sources in the cluster area. 

\begin{table}
\caption{Radio imaging details.}
\vspace{-5mm}
\begin{center}
\resizebox{0.5\textwidth}{!}{
\begin{tabular}{cccccccccc}
\hline
\hline
Central frequency & Resolution & $uv$-taper & Map noise \\
$\nu$ [GHz] & $\Theta$ [$''\times'',^\circ$] & [kpc] & $\rm \sigma_{rms}~[\mu Jy~beam^{-1}]$ \\
\hline
1.5 & $2.9\times2.2$, 60 &  None$^\dagger$ & 13.6 \\
    & $4.3\times3.2$, 67 & None & 8.3 \\
    & $5.4\times4.7$, 65 & 25 & 8.4 \\
    & $8.3\times7.6$, 67 & 50 & 11.2 \\
    & $14.0\times13.7$, 71 & 100 & 18.6 \\
3.0 & $1.8\times1.1$, 60 & None$^\dagger$ & 3.8 \\
    & $2.9\times2.2$, 75 & None & 2.4 \\
    & $5.4\times4.3$, 92 & 25 & 3.0 \\
    & $7.9\times7.2$, 92 & 50 & 4.1 \\
    & $14.0\times13.7$, 103 & 100 & 6.4 \\
\hline
\end{tabular}
}
\end{center}
\vspace{-5mm}
\tablefoot{$^\dagger$These images are obtained with \texttt{robust=-1.25}.}
\label{tab:images}
\end{table}

\subsection{X-ray morphology}\label{sect:xray}
\begin{figure}
\centering
\includegraphics[width=0.45\textwidth]{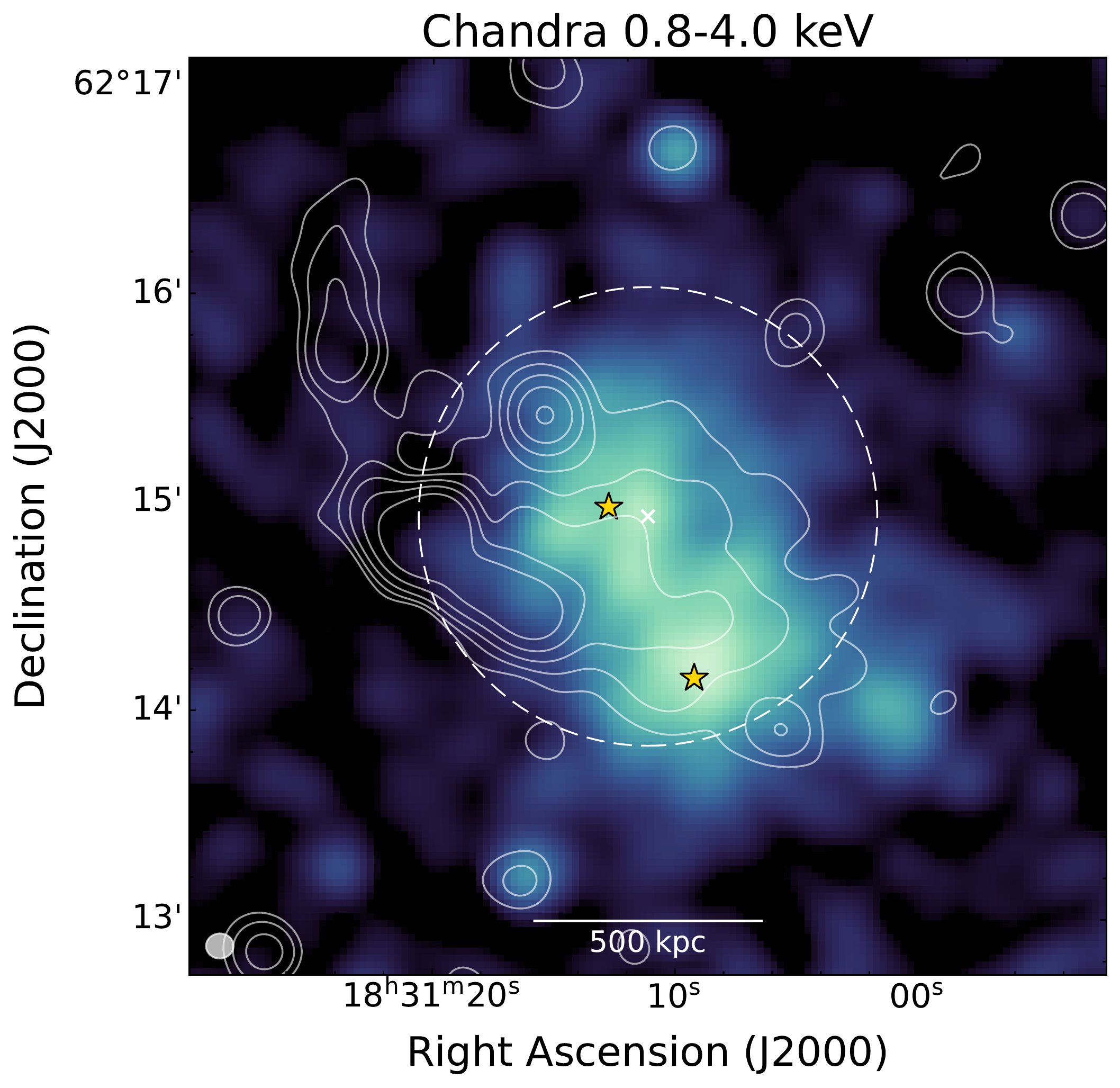}
\caption{Thermal/non-thermal comparison. X-ray 0.8--4.0 keV {\it Chandra} image with 3.0 GHz radio contours tapered at 50 kpc (contour levels as for Fig.~\ref{fig:vla_full_low_res}), whose beam size is shown in the bottom left corner. The white dashed circle shows the $R=0.5R_{\rm SZ,500}$ region, and the white cross denotes the cluster centre. The two yellow stars show the cluster X-ray peaks, which are associated with the two candidate sub-clusters \citep{artis+22}.}
\label{fig:chandra_vla_contours}
\end{figure}

As also shown in \cite{digennaro+21a} (see Fig. 2 in that manuscript) and in \cite{artis+22}, the ICM morphology of \cluster\ is strongly disturbed, with the X-ray emission elongating in the north-east/south-west direction. We identify two main peaks of emission, one in the north and the other in the south, at coordinates $\rm RA_S = 18^h31^m09.2^s ~ DEC_S = +62^\circ 14^\prime09.5''$ and $\rm RA_N = 18^h31^m12.8^s ~ DEC_N = +62^\circ 14^\prime58.7''$ (Fig.~\ref{fig:chandra_vla_contours}). The two peaks dist $\sim420$ kpc from each other, if we assume the cluster is perfectly on the plane of the sky.

In order to determine the global X-ray cluster properties (i.e. temperature and luminosity), we extracted the spectrum within a circle of $R=1.01$ Mpc (i.e. $\sim2.2^\prime$), which corresponds to $R_{\rm SZ,500}$ given the cluster mass $M_{\rm SZ,500}$, centred on the cluster coordinates (see Tab. \ref{tab:cluster}). We modelled the spectrum with a single-temperature (\texttt{phabs$\ast$APEC}) model, including the absorption from the hydrogen column density ($N_{\rm H}$) of our Galaxy. We used the mean total value\footnote{\url{https://www.swift.ac.uk/analysis/nhtot/}}, which includes both the molecular (H$_2$) and atomic (HI) hydrogen \citep{willingale+13}, i.e. $N_{\rm H}=4.86\times10^{20}$ atoms cm$^{-2}$.


We found a global cluster temperature of $kT_{500}=12.7^{+2.1}_{-1.5}$ keV, and an unabsorbed luminosity in the 0.1--2.4 keV range of $L_{[0.1-2.4 ~{\rm keV}], 500}= (3.21 \pm 0.14) \times 10^{45}$ erg s$^{-1}$. Emission from highly ionised Fe is marginally detected even in this shallow data, constraining the average gas metallicity at 0.77$\pm0.29$ Solar using the reference units of \citet{asplund+09}.

\begin{figure*}
\centering
\includegraphics[width=0.8\textwidth]{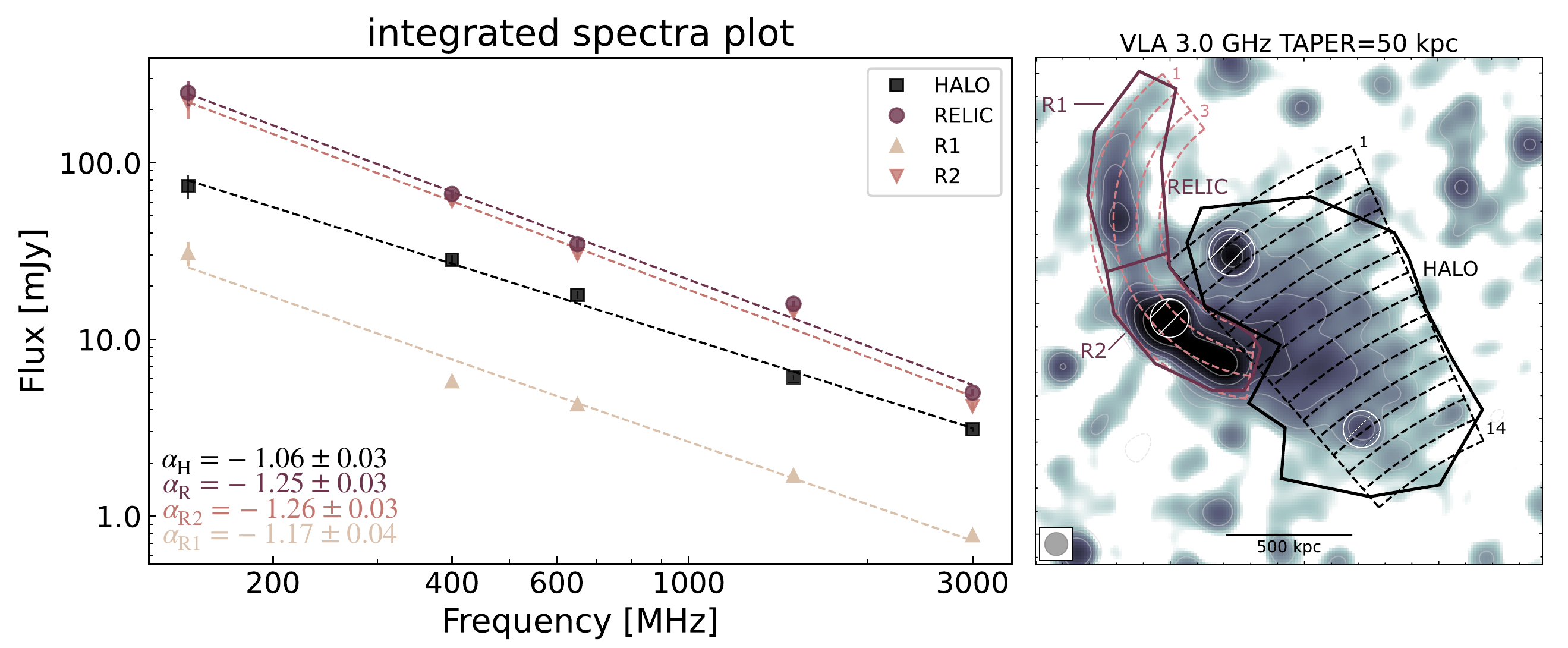}
\caption{Integrated spectra of the diffuse radio emission in \cluster\ (left panel) . The flux densities are measured from the solid-line regions in the right panel, and the spectral indices are reported in the bottom left corner in the panel (black for the halo and shades of purple for the relic). Radio contours are from the 3.0 GHz VLA image, at $12''$ resolution. The separation of R1 and R2 is given by the purple solid line in the right panel. The dashed dashed-line sectors in the right panel refer to the profiles displayed in Fig.~\ref{fig:profiles}. White circles mark the position of the compact sources that have been masked for the flux density, spectral and curvature profiles.}
\label{fig:intspix}
\end{figure*}

\subsection{Flux densities and integrated spectral indices}
We obtained the relic flux density, both in total intensity and total average polarisation, from the 50 kpc tapered image. In this way, we are able to retrieve all the diffuse radio emission while minimising the blending with the radio halo.
For the total intensity, in order to have a wide frequency coverage, we also used the 144 MHz LOFAR and the 400 MHz and 650 MHz uGMRT observations, using the same tapering value (see Appendix \ref{apx:other_freqs}). All these images were regridded at the same pixel size and convolved at the same common resolution (i.e. $12''$, see Tab. \ref{tab:flux})\footnote{We used the \texttt{CASA} tasks \texttt{imsmooth} and \texttt{imregrid}.}. The flux densities were obtained from the same region at all frequencies, 
which also includes source C (see Fig. \ref{fig:vla_full_low_res} and right panel in Fig. \ref{fig:intspix}). The flux density of this source was then mathematically subtracted\footnote{We decided to follow this method as the proximity of source C to the relic makes it difficult to exclude it in the flux density measurement.}.
The uncertainty on the radio flux densities for the radio relic were calculated as

\begin{equation}
\Delta S_{\nu, \rm relic} = \sqrt{\Delta S_{\nu, \rm total} + \Delta S_{\nu, \rm source~C}} \, ,
\end{equation}
where
\begin{equation}
\Delta S_\nu = \sqrt{ (fS_\nu)^2 + N_{\rm beam}\sigma_{\rm rms}^2 } \, .
\end{equation}
Here, $f$ is the systematic uncertainty due to residual amplitude errors \citep[we assume 15\% for LOFAR, 5\% for uGMRT, and 3\% for VLA, see][respectively]{shimwell+22,chandra+04,perley+butler13}, $\sigma_{\rm rms}$ is the map noise level and $N_{\rm beam}$ is the number of beams covering the relic region and source C.
We measure $248.8\pm40.9$ mJy, $66.5\pm3.9$ mJy, $34.6\pm2.0$ mJy, $15.9\pm0.6$ mJy and $5.0\pm0.2$ mJy, for the full relic (i.e. R1+R2) at 144 MHz, 400 MHz, 650 MHz, 1.5 GHz and 3.0 GHz respectively (see Table \ref{tab:flux}). We also calculated the flux density for R1 and R2 separately, at the same frequencies and resolution (Table \ref{tab:flux}). As clear from the images, R2 represents the main source of emission, while R1 is barely detected at the highest frequency. 
We fit the flux densities with a power law of the type $y=ax+b$, where $y=\log S$, $x=\log\nu$ and the slope $a=\alpha^{\rm 3.0GHz}_{\rm 144MHz}$ the spectral index between 144 MHz and 3.0 GHz. We obtain $\alpha^{\rm 3.0GHz}_{\rm 144MHz}=-1.24\pm0.03$, $\alpha^{\rm 3.0GHz}_{\rm 144MHz}=-1.16\pm0.03$ and $\alpha^{\rm 3.0GHz}_{\rm 144MHz}=-1.25\pm0.03$, for the full relic, R1 and R2 respectively (see Fig.~\ref{fig:intspix}). Given these spectral indices and flux densities, we calculate the radio luminosity at each frequency, $L_\nu$, according to:
\begin{equation}
L_\nu = \frac{4 \pi D_L^2}{(1+z)^{1+\alpha}} S_\nu \, ,
\end{equation}
where $D_L$ is the luminosity distance at the cluster redshift (i.e. 5188 Mpc). These are reported in the last column in Tab. \ref{tab:flux}. Uncertainties on the radio luminosity take into account both the flux density and the spectral index uncertainties through Monte Carlo simulations.

\begin{table}
\caption{Integrated radio fluxes and luminosities.}
\vspace{-5mm}
\begin{center}
\resizebox{0.5\textwidth}{!}{
\begin{tabular}{cccccccccc}
\hline
\hline
Source  & Resolution & Frequency & Flux Density & Luminosity \\
        & $\Theta$ [$''$] & $\nu$ [GHz] & $S_\nu$ [mJy] & $\log (L_\nu$ [W Hz$^{-1}$]) \\
\hline
R1  & 12 & 3.0 & $0.78\pm0.04$ & $24.44 \pm 0.02$ \\
    & 12 & 1.5 & $1.7\pm0.1$ & $24.78 \pm 0.03$ \\
    & 12 & 0.650 & $4.3\pm0.3$ & $25.18 \pm 0.03$ \\
    & 12 & 0.400 & $5.8\pm0.5$ & $25.31 \pm 0.04$ \\
    & 12 & 0.144 & $30.8\pm4.7$ & $26.04 \pm 0.07$\\
R2  & 12 & 3.0 & $4.2\pm0.2$ & $25.2 \pm 0.02$ \\
    & 12 & 1.5 & $14.2\pm0.6$ & $25.73 \pm 0.02$ \\
    & 12 & 0.650  & $30.3\pm2.0$ & $26.06 \pm 0.03 $ \\
    & 12 & 0.400  & $60.7\pm3.9$ & $26.36 \pm 0.02$ \\
    & 12 & 0.144  & $218.0\pm41.2$ & $26.91 \pm 0.09$ \\
RELIC$^\dagger$   & 12 & 3.0 & $5.0\pm0.2$ & $25.27 \pm 0.02$ \\
        & 12 & 1.5 & $15.9\pm0.6$ & $25.77 \pm 0.02$\\
        & 12 & 0.650  & $34.6\pm2.0$ & $26.11 \pm 0.02$ \\
        & 12 & 0.400  & $66.5\pm3.9$ & $26.39 \pm 0.03$ \\    
        & 12 & 0.144  & $248.8\pm40.9$ & $26.97 \pm 0.07$ \\
HALO& 19 & 3.0 & $3.1\pm0.1$ &  $25.0 \pm 0.02$ \\
    & 19 & 1.5 & $6.1\pm0.2$ & $25.31 \pm 0.02$ \\
    & 19 & 0.650 & $17.9\pm1.0$ & $25.77 \pm 0.02$ \\
    & 19 & 0.400 & $28.3\pm1.7$ & $25.97 \pm 0.03$ \\
    & 19 & 0.144 & $73.8\pm11.1$ & $26.39 \pm 0.06$ \\
\hline
\end{tabular}
}
\end{center}
\vspace{-5mm}
\tablefoot{$^\dagger$ With ``relic'' we refer to the sum of R1 and R2. The LOFAR and uGMRT images (i.e. $\nu=0.144,~0.400,~0.650$ GHz) are obtained with a \texttt{robust=-0.5}. The map noises at $12''$ are 6.2, 17.3, 41.6, 98.7 and 137.0 $\rm \mu Jy~beam^{-1}$ for the 3.0 GHz, 1.5 GHz, 650 MHz, 400 MHz and 144 MHz observations respectively. The map noises at $19''$ are 9.6, 26.6, 70.3, 156.2, 178.5 $\rm \mu Jy~beam^{-1}$ for the 3.0 GHz, 1.5 GHz, 650 MHz, 400 MHz and 144 MHz observations respectively.}
\label{tab:flux}
\end{table}

In order to estimate reliable values for the averaged polarised flux, we need to correct for the Ricean bias \citep{rice45}. Indeed, the polarised flux is given by:
\begin{equation}
P = \sqrt{Q^2 + U^2} \, .
\end{equation}
This means that we always measure a positive polarised emission, even where no ``real'' polarised signal is present (i.e. the Stokes-$Q$ and -$U$ noise dominates). Following \cite{stuardi+21}, we correct for the Ricean bias as $P=\sqrt{|F(\phi_{\rm peak})|^2 - 2.3\sigma_{QU}^2}$ \citep{george+12}. Here, $|F(\phi_{\rm peak})|$ is the peak of the Faraday dispersion function $F(\phi)$ obtained from the RM-Synthesis technique, and $\sigma_{QU}$ is the average noise in the Stokes-$Q$ and -$U$ maps ($\sigma_{QU}=3.3~{\rm \mu Jy~beam^{-1}}$ at $12.5''$). 
After this correction, we obtain $P_{\rm 3.0GHz}=0.5$ mJy\footnote{We calculate the total average polarised intensity at the effective frequency of our observations.} for R2 at $12.5''$ resolution, while we set an upper limit for R1 of $P_{\rm 3.0GHz}<0.06$ mJy.

\begin{figure*}[h!]
\centering
\includegraphics[height=0.3\textwidth]{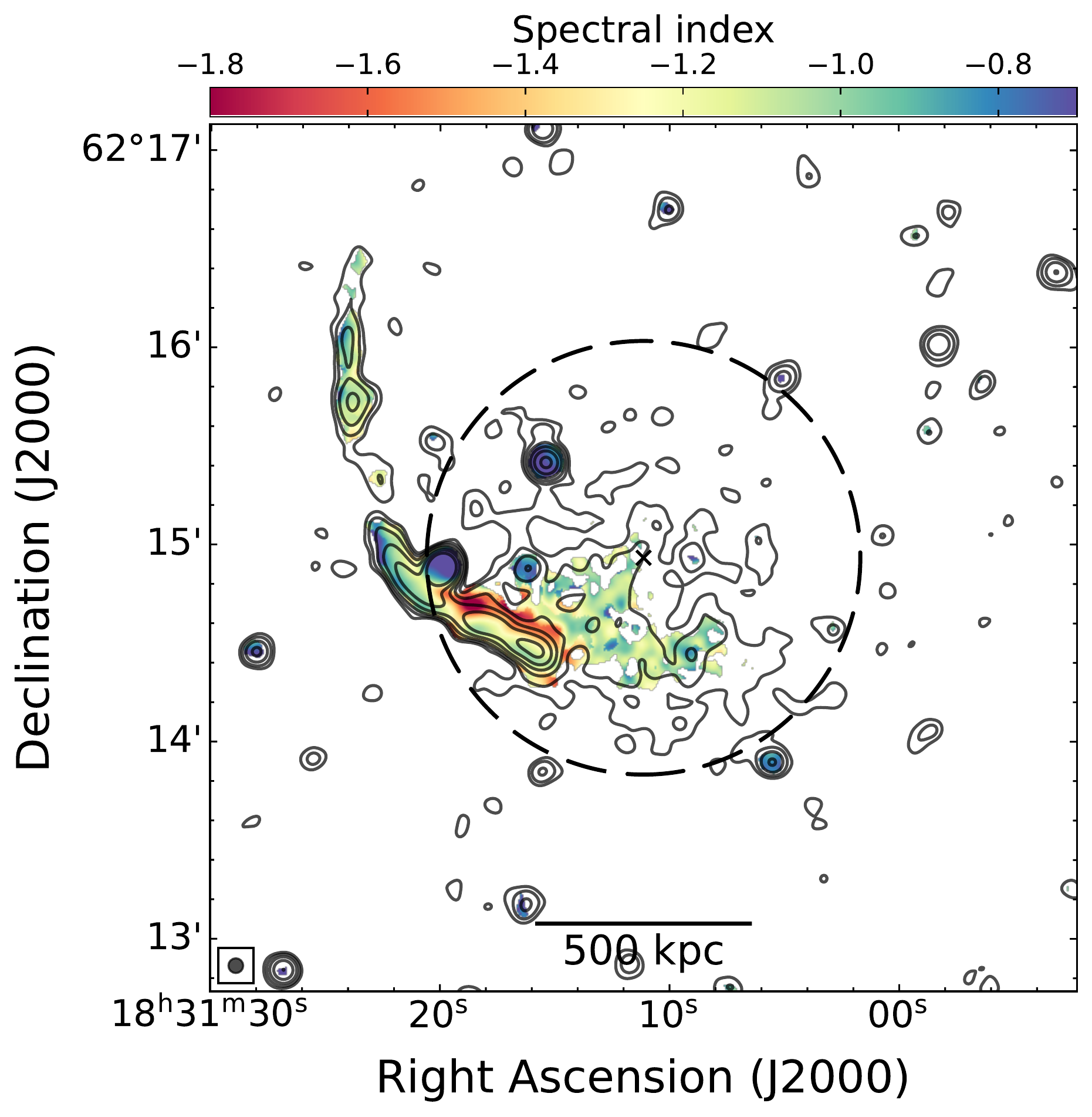}
\includegraphics[height=0.3\textwidth]{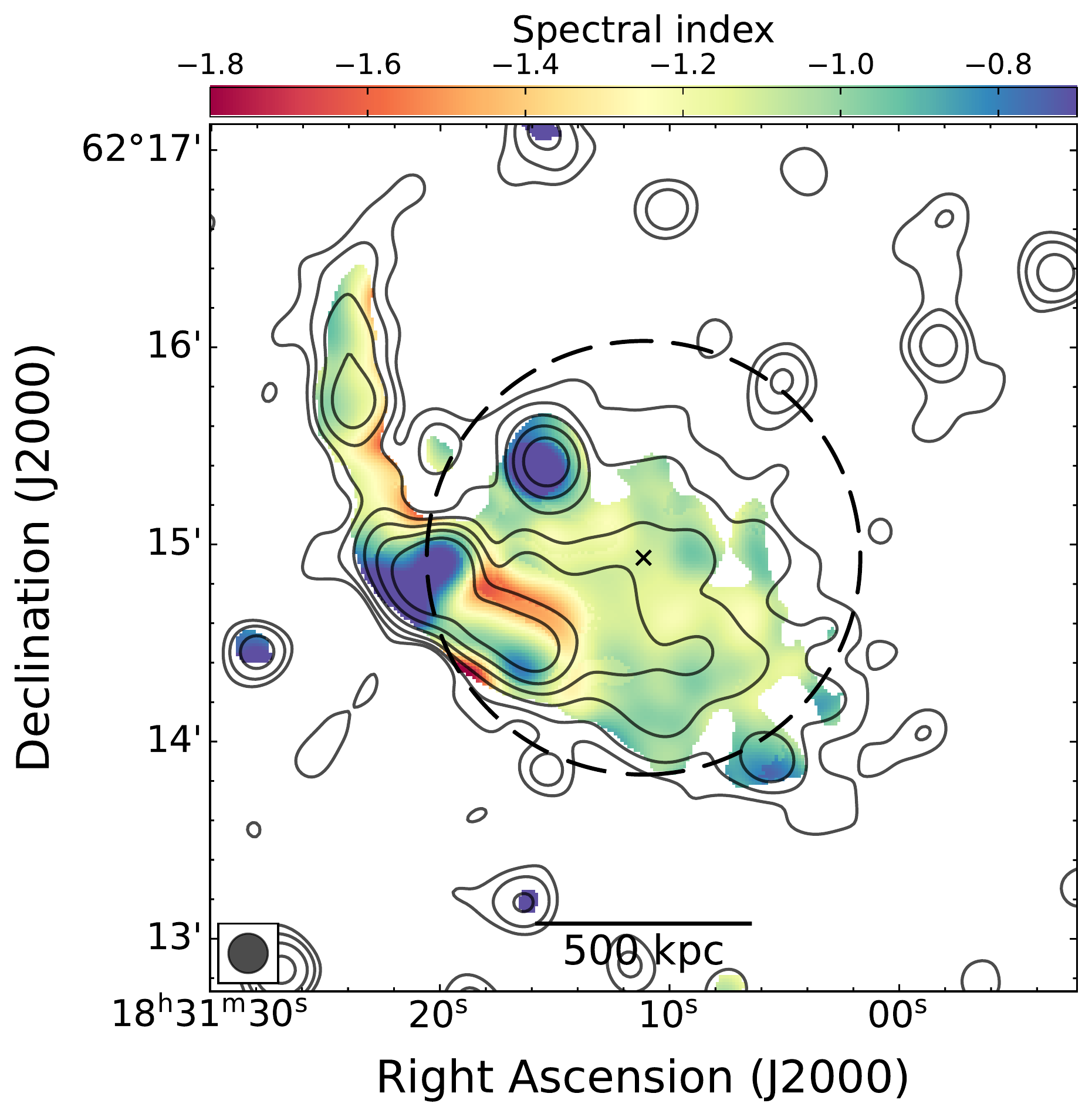}
\includegraphics[height=0.3\textwidth]{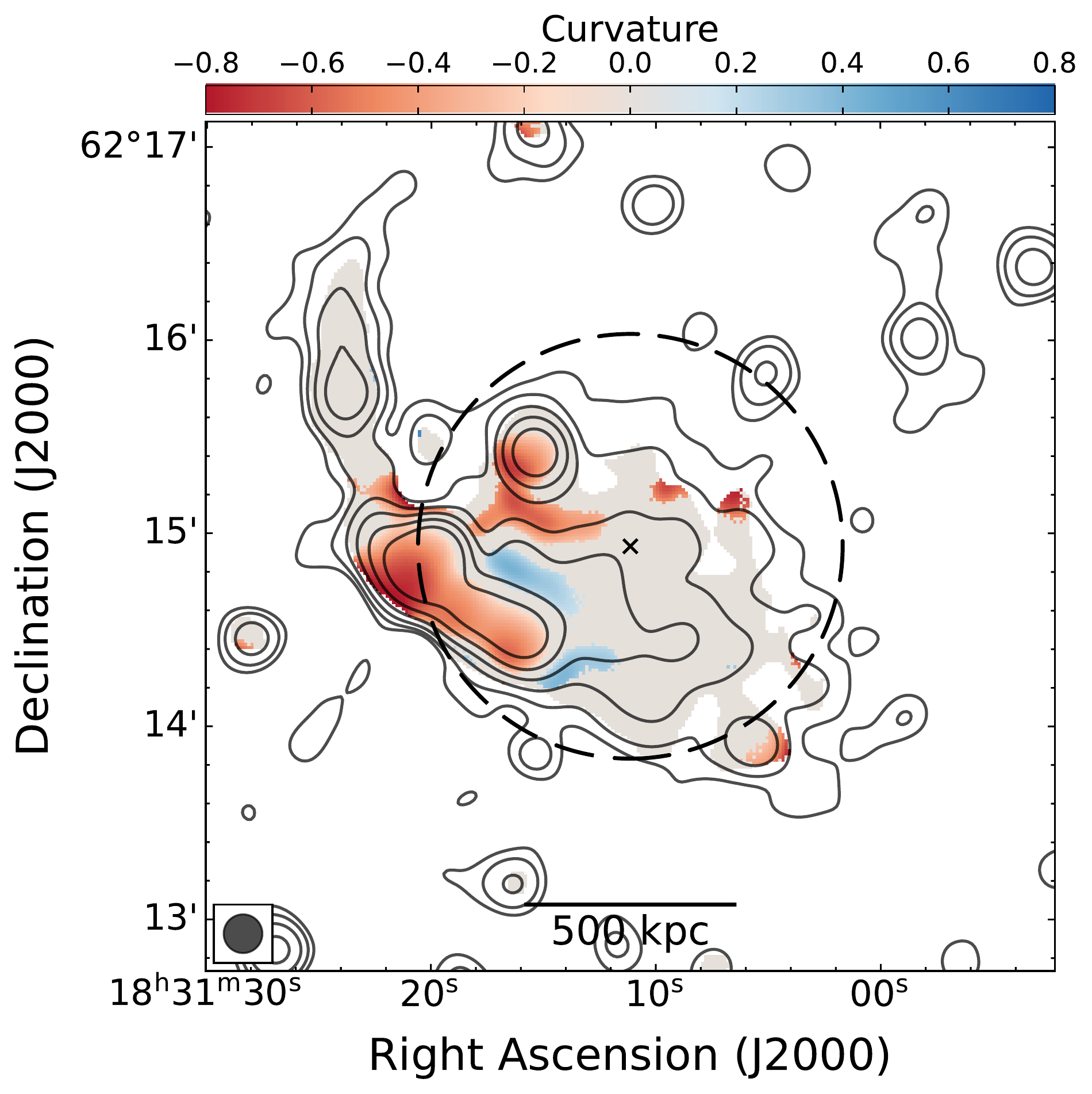} \\
\includegraphics[height=0.3\textwidth]{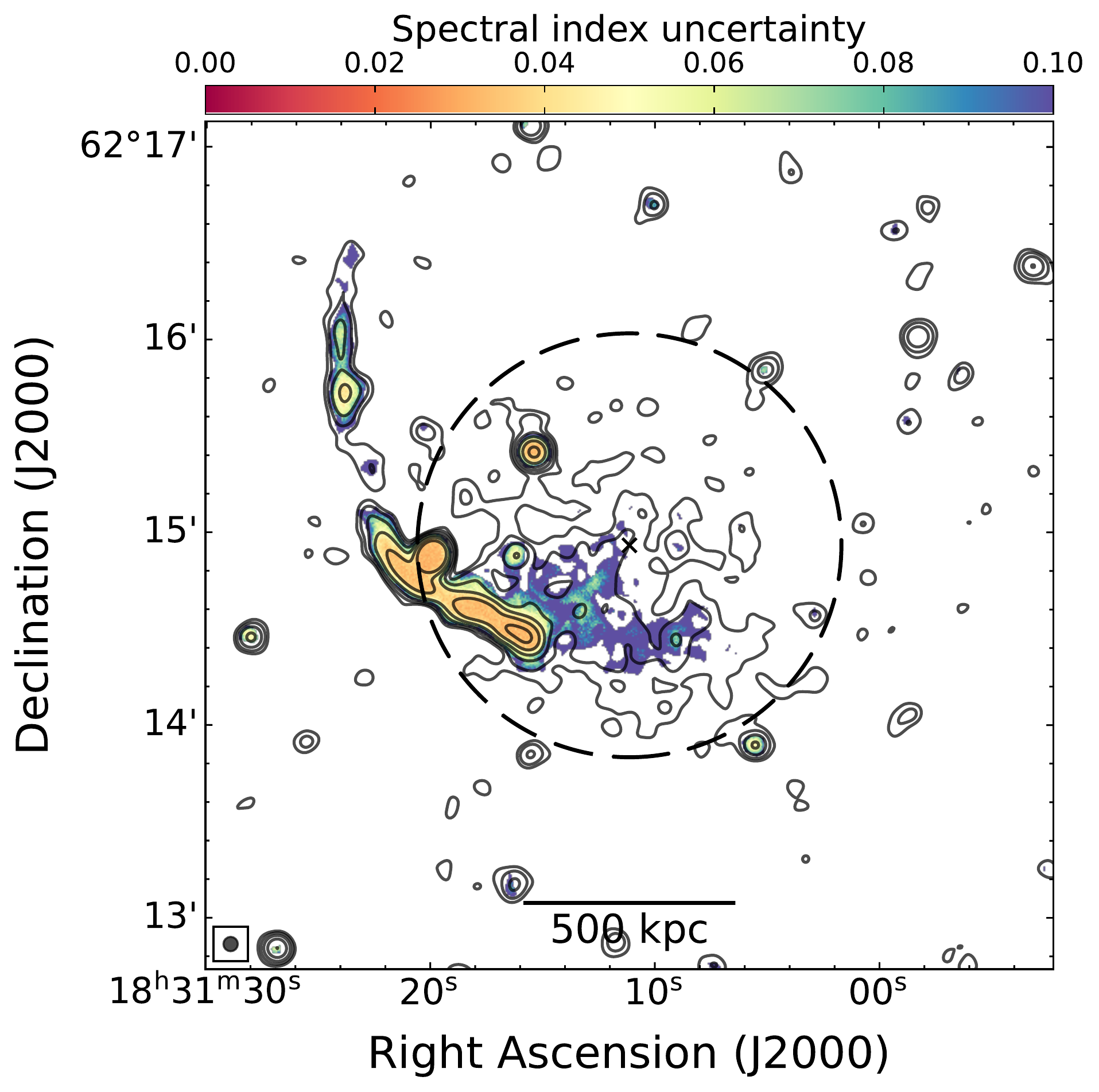}
\includegraphics[height=0.3\textwidth]{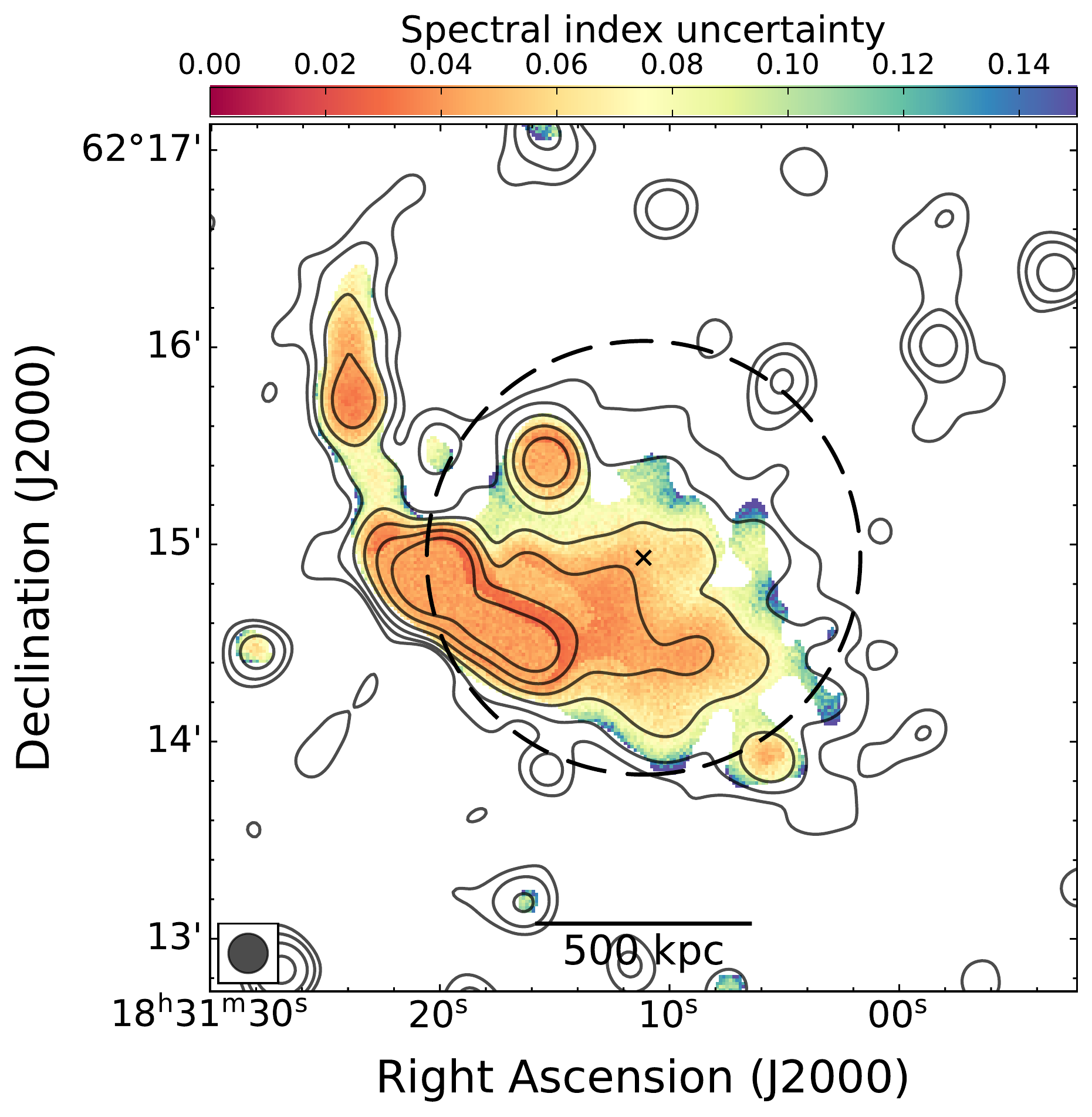}
\includegraphics[height=0.3\textwidth]{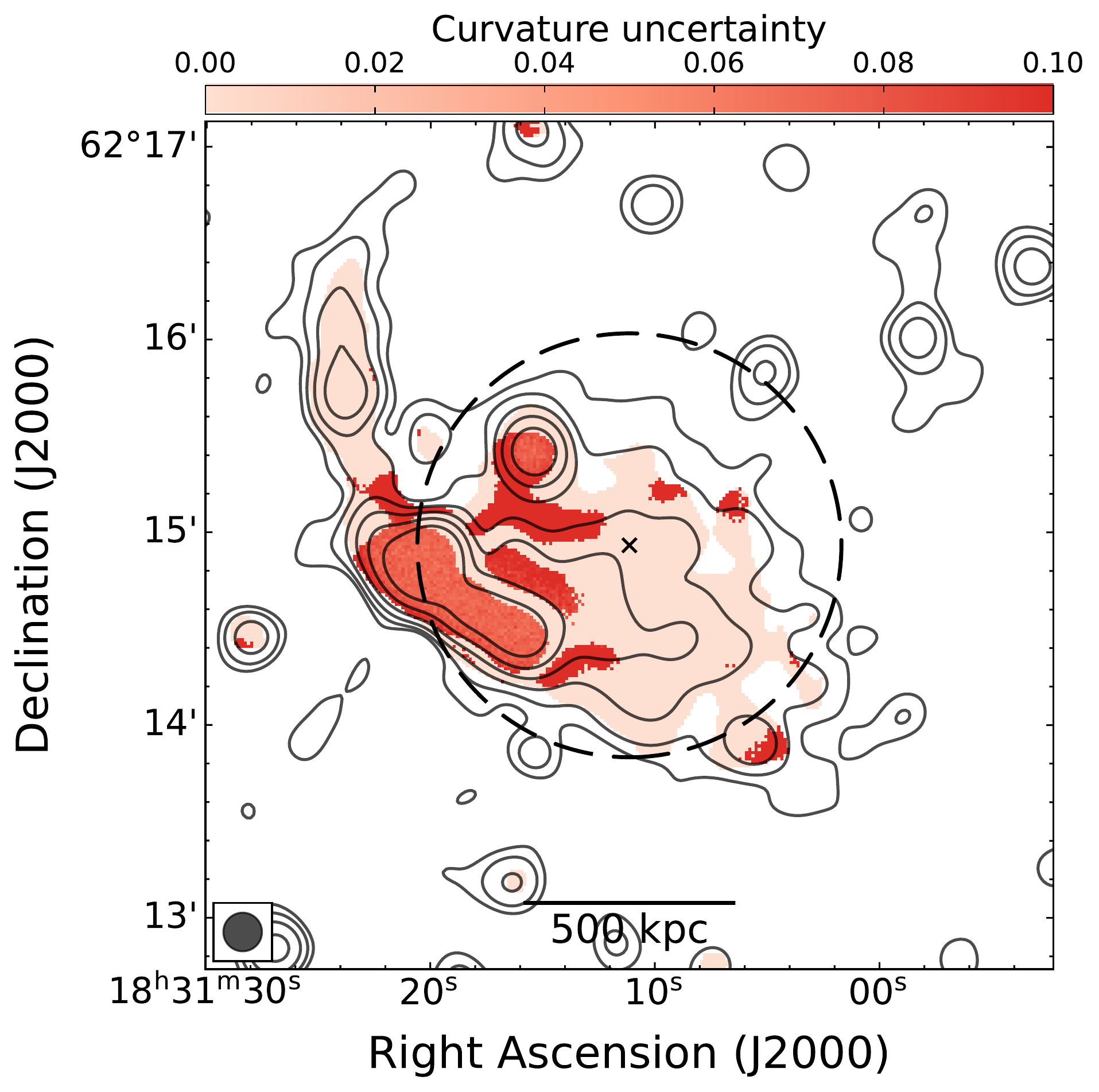}
\caption{Spectral index and curvature maps (top) with corresponding uncertainty maps (bottom) of \cluster. Left column: $4.5''$ using the 144 MHz, 650 MHz, 1.5 GHz and 3.0 GHz observations; Central and Right columns: $12''$ resolution using the 144 MHz, 400 MHz, 650 MHz, 1.5 GHz and 3.0 GHz observations. Radio contours are from the 3.0 GHz observations, drawn at the $2.5\sigma_{\rm rms}\times[1,2,4,8,16,32]$ levels (with $\rm \sigma_{rms, 4.5''}=3.1~\mu Jy~beam^{-1}$) and $\rm \sigma_{rms, 12''}=6.2~\mu Jy~beam^{-1}$).}
\label{fig:spix_curv}
\end{figure*}

\begin{figure}
\centering
\includegraphics[height=0.43\textwidth]{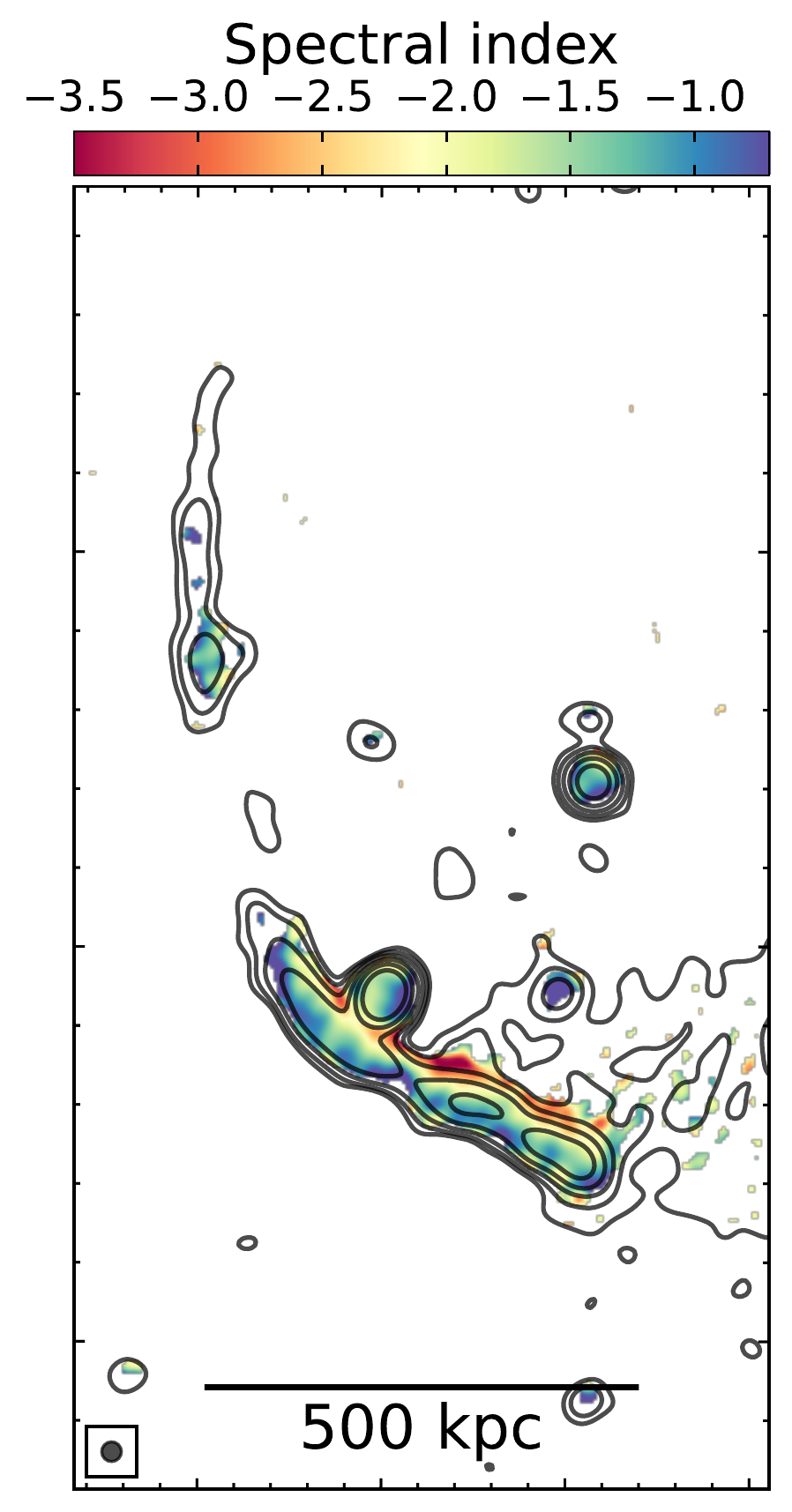}
\includegraphics[height=0.43\textwidth]{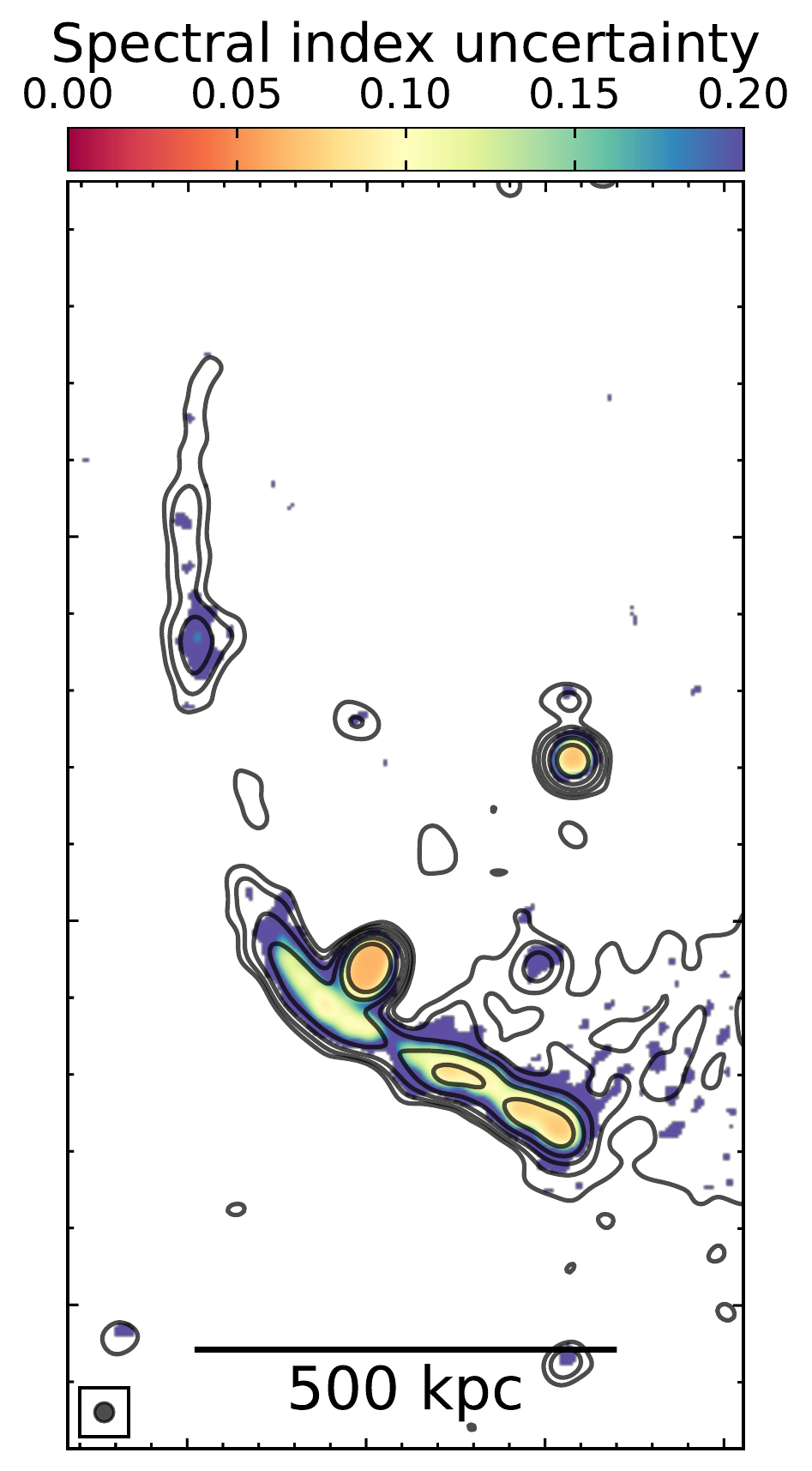}
\caption{Spectral index and uncertainty maps on the radio relic at $3''$ resolution using the 1.5 GHz and 3.0 GHz observations. Radio contours are from the 3.0 GHz observations, drawn at the $2.5\sigma_{\rm rms}\times[1,2,4,8,16,32]$ levels (with $\rm \sigma_{rms}=2.7~\mu Jy~beam^{-1}$).}
\label{fig:spix_relic}
\end{figure}

We also measure the flux density from the radio halo. In order to do so, we produced source-subtracted images, i.e. we cut all the visibilities corresponding to scales larger than 500 kpc to create a model of ``compact'' sources which then were subtracted from the visibilities. We then tapered the $uv$-data at lower resolutions. This approach was also performed on LOFAR and uGMRT data \citep{digennaro+21c}, and it also included \texttt{multiscale} deconvolution to take into account the diffuse emission associated with radio galaxies. Similarly to the LOFAR and uGMRT cases, this procedure leaves, however, part of the emission of the candidate radio relic, that needs to be manually masked. We chose the 100 kpc tapered images to retrieve all the diffuse emission from the radio halo (see rightmost panel in Fig.~\ref{fig:vla_full_low_res}), at the VLA, uGMRT and LOFAR frequencies. Similarly to the radio relic, to retrieve the integrated spectral index we also added the information from the 144 MHz LOFAR and 400 MHz and 650 MHz uGMRT observations. In this case, the uncertainties on the flux densities are given by $\Delta S_\nu = \sqrt{(f S_\nu)^2 + N_{\rm beam}\sigma_{\rm rms}^2 + \sigma_{\rm sub}^2}$, where $\sigma_{\rm sub}$ are the residuals on the subtraction \citep[i.e. a few percent of the residual flux from compact sources; see][]{cassano+13}. We measure $73.8\pm11.1$ mJy, $28.3\pm1.7$ mJy, $17.9\pm1.0$ mJy, $6.1\pm0.2$ mJy, and $3.1\pm0.1$ mJy at 144 MHz, 400 MHz, 650 MHz, 1.5 GHz and 3.0 GHz, respectively. We note that for the LOFAR and uGMRT flux densities we get values that are slightly higher than those reported in \cite{digennaro+21c}, although within the errorbars. Assuming a power law, the total integrated spectrum is $\alpha^{\rm 3.0GHz}_{\rm 144MHz}=-1.06\pm0.03$ \citep[in agreement with that reported in ][]{digennaro+21c}. 
The radio luminosities at each frequency are reported in Table \ref{tab:flux}.

\begin{figure*}
\centering
\includegraphics[width=0.45\textwidth]{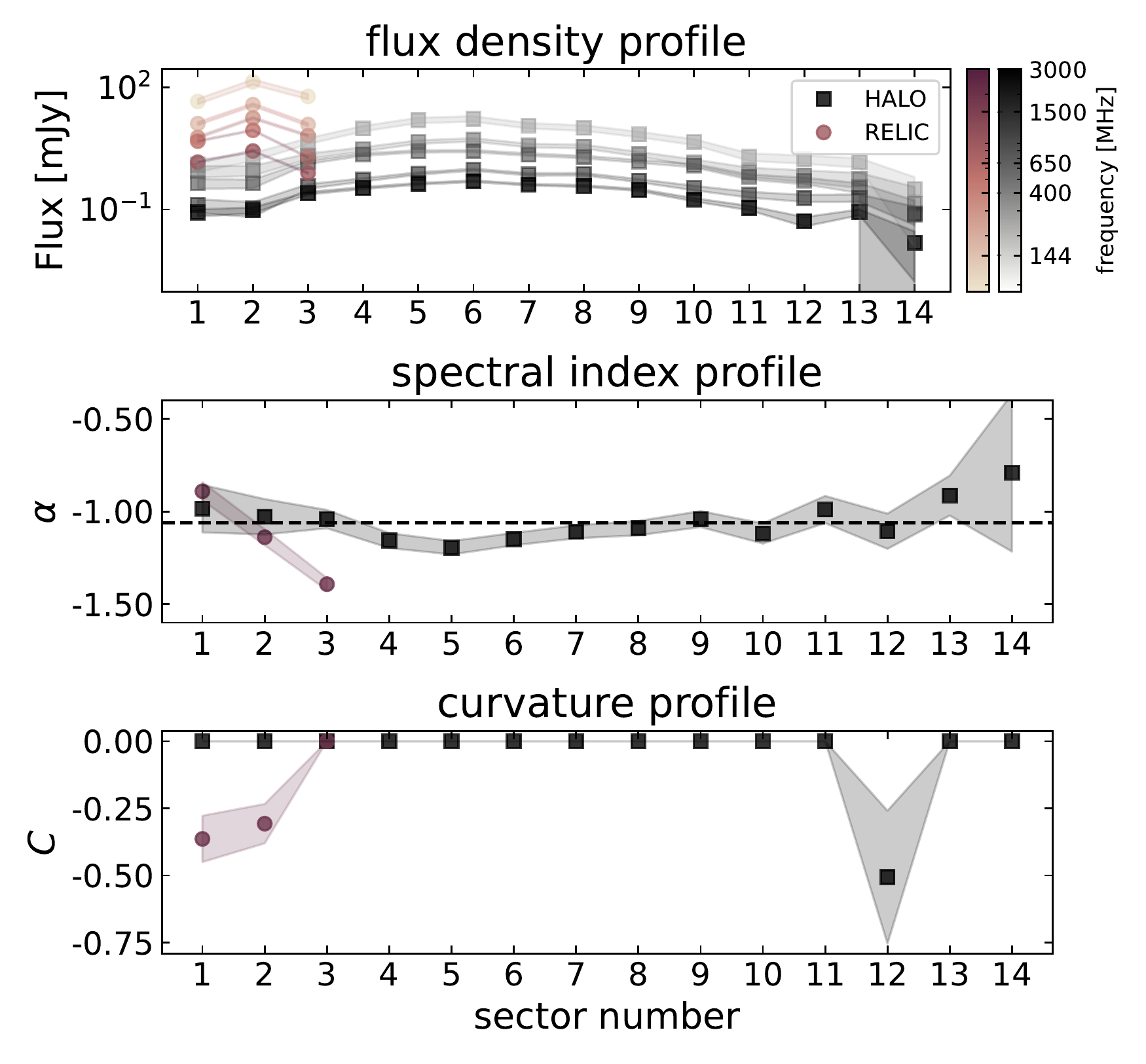}
\includegraphics[width=0.47\textwidth]{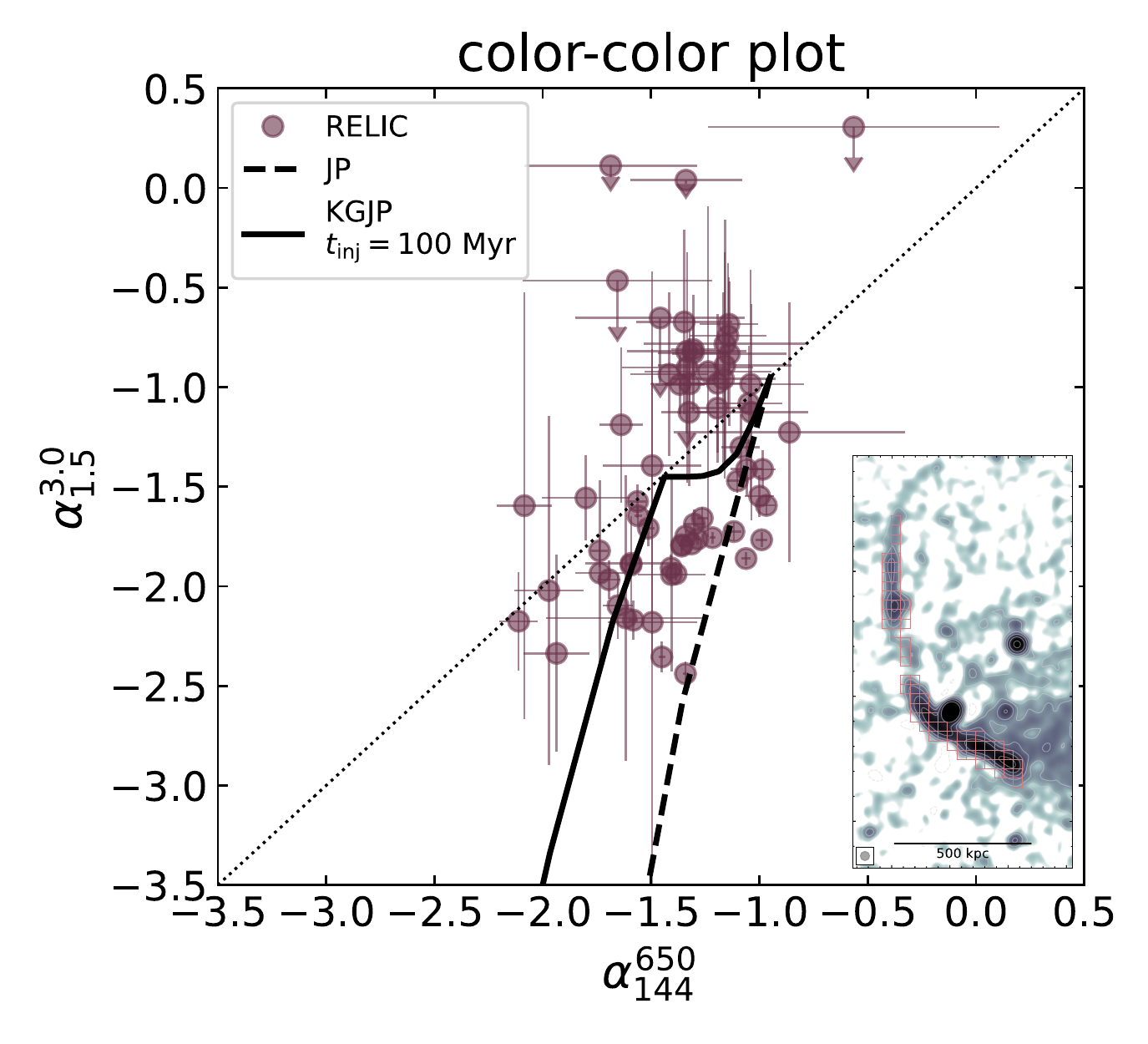}
\caption{Spectral and curvature analysis of the relic in \cluster. Left panel: From top to the bottom: flux density, spectral index and curvature profiles across the relic (purple circles) and the halo (black squares) at $12''$ resolution; the box and sector numbers are shown in the right panel in Fig.~\ref{fig:intspix}; the black dashed line in the spectral index profile plot shows the integrated spectral index for the radio halo (i.e. $\alpha^{\rm 3.0GHz}_{\rm 144MHz}=-1.06\pm0.03$). Right panel: Radio colour-colour (i.e. $\alpha^{\rm 650MHz}_{\rm 144MHz}-\alpha^{\rm 3.0GHz}_{\rm 1.5GHz}$) plot for the radio relic at $4.5''$ resolution (see inset for the location of the boxes); arrows refer to the upper limit in the spectral index; the dotted line shows the power law case across the spectrum; the solid and dashed lines display the KGJP (with a particle injection time $t_{\rm inj}=100$ Myr) and JP ageing models.}
\label{fig:profiles}
\end{figure*}

\subsection{Spectral index and curvature maps}\label{sec:spixmaps}
We produce spectral index ($\alpha$) and curvature ($C$) maps for the full cluster at different resolutions, using the approach already described in \cite{digennaro+21c}. The images were re-gridded at the same pixel scale and convolved and at the same resolution. Moreover, we checked and fixed for possible off-set in the astrometry at the different frequencies, selecting point-like sources and cross-matching their flux peak. Using then the VLA 3.0 GHz image as the reference due to the most accurate astrometry\footnote{\url{https://science.nrao.edu/facilities/vla/docs/manuals/oss/performance/positional-accuracy}}, we found a separation of $\Delta_{\rm RA}=1''$ and $\Delta_{\rm Dec}=1.5''$ for the LOFAR and the uGMRT observations, and about $\Delta_{\rm RA}=\Delta_{\rm Dec}=-0.5''$ for the VLA 1.5 GHz.

To produce the low-resolutions ($12''$) spectral index and curvature maps, we used all the frequencies available (i.e. the LOFAR, uGMRT and VLA) with a 50 kpc taper. 
For each pixel, we fit a second-order polynomial (i.e. $y=ax^2 + bx + c$) if the curvature parameter is larger than $2\sigma$, where $\sigma$ is the uncertainty associated with the second-order term. In this case, the spectral index and the curvature were calculated at the median of the total band, i.e. $\nu_{\rm ref}=650$ MHz, as $2a\log \nu_{\rm ref} + b$ and $C_{\rm 650MHz}=a$, respectively; otherwise, we fit a first-order polynomial and the spectral index is simply given by the slope of the fit, while the curvature parameter is set to zero. According to this convention, in the second-order fit, negative values of curvature correspond to steeper spectral indices at high frequencies, i.e. the spectum is convex \citep[see also, e.g.][]{digennaro+18,stuardi+19,rajpurohit+20}.
We blanked all the pixels below the $2.5\sigma_{\rm rms}$ threshold for each frequency. The spectral index uncertainty maps are obtained via 150 Monte Carlo simulations of the first-/second-order polynomial fit. We assumed that the uncertainty of each flux given by the sum in quadrature of the noise map and the systematic flux uncertainties, i.e. $\Delta S_\nu=\sqrt{(f S_\nu)^2 + \sigma_{\rm rms}^2}$. 
We find hints of steepening from the outer edge of the relic towards the cluster centre, i.e. from $\alpha_{\rm 650MHz}\sim-1$ to $\alpha_{\rm 650MHz}\sim-1.5$. For R2, we also find a significant curvature $C_{\rm 650MHz}\sim-0.4$ (see Fig.~\ref{fig:spix_curv}), meaning that the spectral index steepens at higher frequencies. The flattening nearby source C is probably due to the blending of the radio galaxy spectral index and the outermost edge of the relic, as this is not visible in the spectral index maps at higher resolutions (i.e. at $4.5''$ using the 144 MHz, 650 MHz, 1.5 GHz and 3.0 GHz maps, see left panel in Fig.~\ref{fig:spix_curv}, and at $3''$ using the 1.5 GHz and 3.0 GHz maps, see Fig.~\ref{fig:spix_relic}). The highest-resolution spectral index map (i.e. $3''$) also shows steepening towards the cluster centre as for the lower resolutions (Fig.~\ref{fig:spix_relic}).

The radio halo shows hints of spectral index variation around the integrated value (i.e. $\alpha^{\rm 3.0GHz}_{\rm 144MHz}=-1.06\pm0.03$; see also dashed line in the right-middle panel in Fig.~\ref{fig:profiles}).  No significant curvature has been found (see Fig.~\ref{fig:spix_curv} and last bottom panel in Fig.~\ref{fig:profiles}).


\section{Discussion}\label{sec:disc}

\cluster\ is a remarkably peculiar cluster at $z=0.822$. Together with a central radio halo, this cluster also hosts an elongated diffuse radio source oriented parallel to the merger axis, which in turn is inferred from the X-ray morphology. The radio morphology in total and polarised intensity, the spectral index and the lack of a clear optical counterpart suggest that this source can be classified as a radio relic. \cluster\ is then the second most-massive galaxy cluster hosting a radio relic at $z>0.6$\footnote{``el Gordo'' at $z=0.870$ \citep{lindner+14} is the most distant and the most massive cluster hosting a radio relic observed to date; we also mention the recent observation of a third cluster hosting  a radio relic, i.e. PSZ2\,G069.39+68.05 at $0.762$ with a mass of $M_{\rm SZ,500}=5.7\times10^{14}~{\rm M}_\odot$ \citep{jones+23}.}. In the following sections we discuss the possible scenarios for the origin of such emission.



\subsection{Radio spectral analysis}

In the framework of the Diffusive Shock Acceleration (DSA) scenario, an initial population of electrons with a momentum ($p$) distribution of the type $N(p) \propto p^{-\delta}$, with $\delta=1-2\alpha$, 
is (re-)energised by the passage of the merger-shock. 
As a consequence, 
the freshly (re-)accelerated particles will still follow the similar energy distribution\footnote{For re-acceleration of fossil plasma, this depends on the spectra of the underlying electrons \citep{markevitch+05}.} $dN(E)/dE \propto E^{-\delta_{\rm inj}}$ at the shock location, with a ``flat'' injected spectral index ($\alpha_{\rm inj}$), while in the post-shock region the spectral index becomes steeper due to 
loss of energy through synchrotron and inverse Compton (IC) radiation. 

Across the width of the relic, we observe a similar trend. We extract the spectral index and curvature profiles, using the $12''$-resolution LOFAR, uGMRT (band 3 and 4) and VLA (L- and S-band) images, from beam-size sectors covering the full relic (see light purple regions on right panel in Fig.~\ref{fig:intspix}). Each sector corresponds to a physical size of $\sim 90$ kpc, at the cluster redshift. We find an increasing spectral steepening towards the cluster centre, as expected from the DSA prediction, and a negative curvature at least in the two outermost sectors (see left panel in Fig.~\ref{fig:profiles}). 
However, the curvature is decreasing downstream, contrary to what is expected in the post-shock region. This
may be due to 
mixing of different populations of electrons: those in the post-shock region that after the (re-)energisation lose energy due to synchrotron and IC radiation, and those in the halo region that are experiencing turbulent re-acceleration \citep{digennaro+21a}.
The presence of steepening, and hints of curvature, in the post-shock region is also observed in the higher-resolution\footnote{In this case, we do not use the uGMRT band 3 image, as it is limited by poorer resolution.} (i.e. $4.5''$) colour-colour plots \citep{katz-stone+93}, i.e. the comparison of the spectral index in the ``low'' and ``high'' frequency bands ($\alpha^{\rm 650MHz}_{\rm 144MHz}$ and $\alpha^{\rm 3.0GHz}_{\rm 1.5GHz}$, respectively, see right panel in Fig.~\ref{fig:profiles}). We obtain the ``low''- and ``high''-frequency spectral indices from beam-size boxes covering the full extension of the peripheral diffuse radio emission (see inset in the right panel in Fig.~\ref{fig:profiles}) and calculate the spectral indices as $\alpha=\log(S_1/S_2)/\log(\nu_1/\nu_2)$. In these kind of plots, points laying on the line $\alpha^{\rm 650MHz}_{\rm 144MHz}=\alpha^{\rm 3.0GHz}_{\rm 1.5GHz}$ follow a power-law spectrum, while points below it have convex spectra, i.e. are characterised by steeper spectral index at high frequencies. We also overlay two theoretical ageing models, i.e. the Jaffe-Perola \citep[JP;][]{jaffe+perola73} and \citep[KGJP;][]{komissarov+gubanov94}
spectral ageing models, which take into account the injection spectral index, the magnetic fields and, for the latter, the time of injection, to describe their radiative losses. Here, we assume $\alpha_{\rm inj}=-0.9$, $B=5~\mu$G and, for the KGJP, $t_{\rm inj}=100$ Myr (which mimics projection effects). As for other relics \citep[e.g.][]{digennaro+18,rajpurohit+21}, the KGJP ageing model better represents the data.

If we assume that the radio relic traces a shock wave, we can relate the injected spectral index and the shock Mach number as
\begin{equation}\label{eq:machinj}
\mathcal{M_{\rm radio, inj}} = \sqrt{\frac{2\alpha_{\rm inj} - 3}{2\alpha_{\rm inj} + 1}} \, .
\end{equation}
From the spectral index profile, we measure a value in the outermost sector of $\alpha_{\rm inj}=-0.89\pm0.03$, which results in a Mach number of $\mathcal{M_{\rm radio,inj}}=2.48\pm0.15$.
In the case that the particle cooling time is much shorter than the shock lifetime, we can relate the injected spectral index to the integrated spectral index ($\alpha_{\rm int}$) through the relation $\alpha_{\rm inj} = \alpha_{\rm int} + 0.5$ \citep{kardashev62}. 
In this case, the shock Mach number is
\begin{equation}\label{eq:machint}
\mathcal{M_{\rm radio, int}} = \sqrt{\frac{\alpha_{\rm int} - 1}{\alpha_{\rm int} + 1}} \, .
\end{equation}
For the relic in \cluster, the injected spectral index obtained directly from the map is steeper than the one obtained by the integrated spectrum ($\sim-0.89$ compared to $-0.75$), as also observed for other radio relics \citep[e.g. ``Sausage'' relic, ``Toothbrush'' relic;][respectively]{digennaro+18,rajpurohit+18}. The integrated spectral index of the relic, i.e. $\alpha_{\rm int}=-1.25\pm0.3$ corresponds to a Mach number of $\mathcal{M_{\rm radio, int}}=3.0\pm0.19$.

\subsection{X-ray/radio shock analysis}\label{sec:xraySB}
In order to reliably classify a patch of diffuse radio emission as a radio relic, it is necessary to detect a jump in the pressure ($P$) profile across the source \citep{markevitch+vikhlinin07}. Particularly, to be defined as a shock discontinuity we would need to measure $p_{\rm post}/p_{\rm pre}>1$, where $p=n_ek_BT$ (here, $n_e$ is the electron density, $k_B$ is the Boltzmann constant and $T$ the ICM temperature). However, our X-ray observations are too shallow to provide clear temperature measurements in the pre- and post-shock regions, and only hints on the surface brightness profile could be derived.

\begin{figure}
\centering
\includegraphics[width=0.45\textwidth]{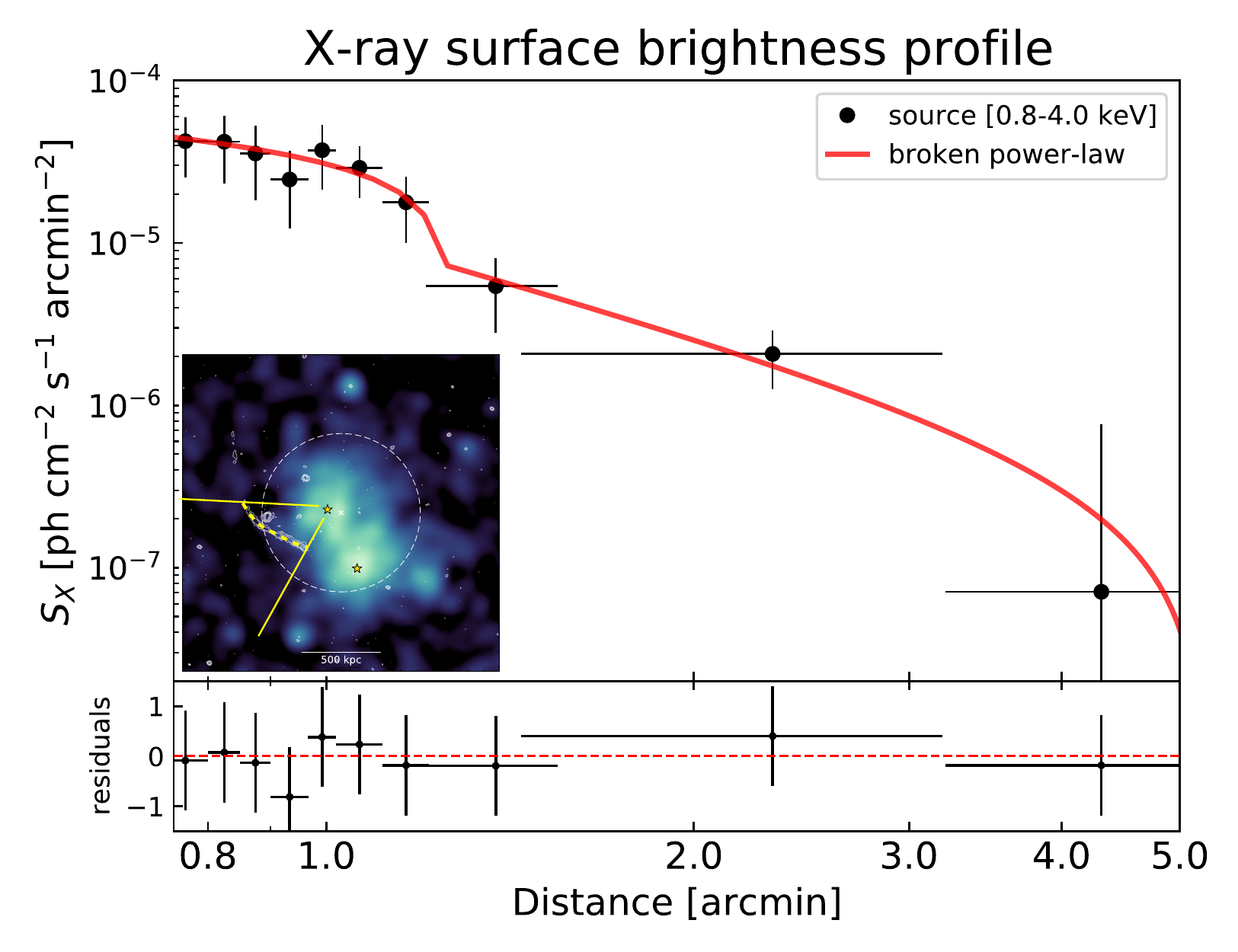}
\caption{Surface brightness profiles across R2 (see inset; the dashed line shows the best-fit position of the X-ray discontinuity) with the best-fit model overlaid. The model corresponds to the emission from a projected broken power-law density distribution, plus a constant surface brightness that accounts for the residual sky background. The data has been binned to a minimum signal to noise of 2. 
On the bottom, the residuals of the best-fit (i.e. $(S_{X{\rm ,obs}}-S_{X{\rm ,mod}})/\Delta S_{X{\rm ,obs}}$) are displayed.}
\label{fig:xraySB}
\end{figure}

We made use of the {\it Chandra} observations to investigate whether a surface brightness discontinuity can be found at the location of the relic. We modelled the underlying density distribution with a broken power-law \citep{markevitch+vikhlinin07}, assuming spherical symmetry,  
where the compression factor is defined as the ratio of the electron densities in the post- and pre-shock regions, i.e. $\mathcal{C}=n_{\rm post}/n_{\rm pre}$. We show in Fig.~\ref{fig:xraySB} the background-subtracted surface brightness profile across an elliptical sector centred at the coordinates of the northern sub-cluster, and with an ellipticity matching the shape of the candidate radio relic (see inset in the Figure). 
The observed profile was binned to a minimum signal to noise ratio of 2. 
We estimated the residual background level due to any imperfections in the blank sky approximation by using a ``local background'' region, namely a $2^\prime$-radius circle located on the same CCD (the ACIS-I3 chip) but as far as possible (i.e. $4.7^\prime$) from the cluster. The local background count rate is statistically consistent with that obtained from the blank sky background; their difference is $\rm -3.9\pm2.9\times10^{-7}~ ph~cm^{-2}s^{-1}arcmin^{-2}$. This value was included as a fixed \texttt{constant} model in the surface brightness profile fitting. 
The fit was performed with \texttt{proffit} \citep{eckert16} and the model was evaluated using Cash statistics \citep{cash79}. 
We detect a density jump $\mathcal{C}=2.22^{+0.37}_{-0.30}$ at the location of the candidate radio relic, i.e. at $r\sim1.2^\prime$ (i.e. $r\sim550$ kpc; $\rm stat/dof=32.3/6$).
Assuming that the measured discontinuity can be associated with a shock wave, the X-ray Mach number can be calculated as:
\begin{equation}\label{eq:machxray}
\mathcal{M_{\rm Xray}} = \sqrt{\frac{2\mathcal{C}}{\gamma +1 - \mathcal{C}(\gamma -1) } } \, ,
\end{equation}
where $\gamma=5/3$ is the adiabatic index of a monoatomic gas. 
Given the measured compression factor, we obtain an X-ray Mach number of $\mathcal{M}_{\rm Xray}=1.93^{+0.42}_{-0.32}$.
We notice these values is marginally smaller than those measured from the radio band. 
\cite{dominguez-fernandez+21} showed that an initially uniform Mach number can result in a distribution when it encounters a turbulent medium such as the ICM \citep[e.g.][]{zhuravleva+19} and that relics trace the high end of such a distribution. Moreover, \cite{wittor+21} pointed out that also viewing angles affect such a distribution, and that projection effects tend to impact more X-ray observations.

We have checked that our results are robust to uncertainties in the background subtraction, by fixing the residual background constant included in the model to (conservatively) $\pm20\%$ of the total blank sky count rate. 
In these cases, the density jump varies in the range  $\mathcal{C}=2.04-2.27$, if we over- and under-subtract the background ($\rm stat/dof=41.4/6$ and $\rm stat/dof=48.1/6$, respectively). For a single power-law model we obtain 
$\rm stat/dof=88.9/8$, therefore this model is disfavoured by the data.

\subsection{Polarisation properties}\label{sec:polariz}
In general, three polarisation models are used to describe the polarised emission \citep{burn66,tribble91,sokoloff+98}: the ``simple'' case of a single Faraday screen which only rotates the polarisation vectors (i.e. no depolarisation, ND); the case where an external Faraday screen (e.g. ICM) varies the magnetic fields on scales smaller than the restoring beam (i.e. external Faraday depolarisation, EFD); the case where the Faraday screen is internal to the emitting source (i.e. internal Faraday depolarisation, IFD). These cases can be expressed as:

\begin{equation}\label{eq:polmodels}
P(\lambda^2) = 
\begin{cases}
\smallskip
p_0 I \exp[2i(\chi_0 + {\rm RM}\lambda^2)]  \hfill \qquad (\rm ND)\\
\smallskip
p_0 I \exp(-2\sigma^2_{\rm RM}\lambda^4) \exp[2i(\chi_0 + {\rm RM}\lambda^2)]
 \hfill \qquad (\rm EFD) \\ 
\smallskip
p_0 I \left [ \frac{1-\exp(-2\varsigma^2_{\rm RM}\lambda^4)}{2\varsigma^2_{\rm RM}\lambda^4} \right ] \exp[2i(\chi_0 + {\rm RM}\lambda^2)]  \hfill \qquad (\rm IFD)
\end{cases}
,
\end{equation}
with $\lambda$ being the wavelength, $p_0$ the intrinsic polarisation fraction, $\chi_0$ the intrinsic polarisation angle, RM the Rotation Measure, and $\sigma_{\rm RM}$ and $\varsigma_{\rm RM}$ the external and the internal depolarisation respectively.

In order to assess the cluster polarisation properties, we run the {\it QU}-fitting code\footnote{\url{https://github.com/gdigennaro/QUfitting}} presented in \cite{digennaro+21b} (see also Appendix \ref{apx:qufitting_plots}) to obtain the intrinsic polarisation fraction ($p_0$), intrinsic polarisation angle ($\chi_0$), Rotation Measure (RM) and external depolarisation ($\sigma_{\rm RM}$).
We created Stokes-{\it Q}, -{\it U} and -{\it I} cubes, at $12.5''$ resolution, with 41 and 155 channels in the L- and S-band respectively and with $\delta\nu=8$ MHz. The final frequency coverage is $\Delta\nu=1.34-4.0$ GHz (i.e. between 0.075 m and 0.22 m), with an effective frequency of 3.0 GHz ($\lambda=0.1$ m). These same single-channel images were also used to generate the total averaged polarised intensity and the Faraday spectrum through the \texttt{pyrmsynth} tool (see Tab. \ref{tab:poldetails}). 
We run the fit for all the pixels above a given threshold, defined starting from the $\tilde{\sigma}_{{\rm rms,}P}$ value, i.e. the root mean square noise of the total averaged polarised emission obtained from the RM-Synthesis technique, including the Ricean bias. We used $1.5\tilde{\sigma}_{{\rm rms,}P}$ (with $\tilde{\sigma}_{{\rm rms,}P}=23.4~\mu{\rm Jy~beam^{-1}}$) and $1.8\tilde{\sigma}_{{\rm rms,}P}$ (with $\tilde{\sigma}_{{\rm rms,}P}=9.9~\mu{\rm Jy~beam^{-1}}$) for the 1--4 GHz low-resolution ($12.5''$) and for the 2--4 GHz high-resolution ($5''$) images. 

\begin{figure*}
\centering
\includegraphics[width=\textwidth]{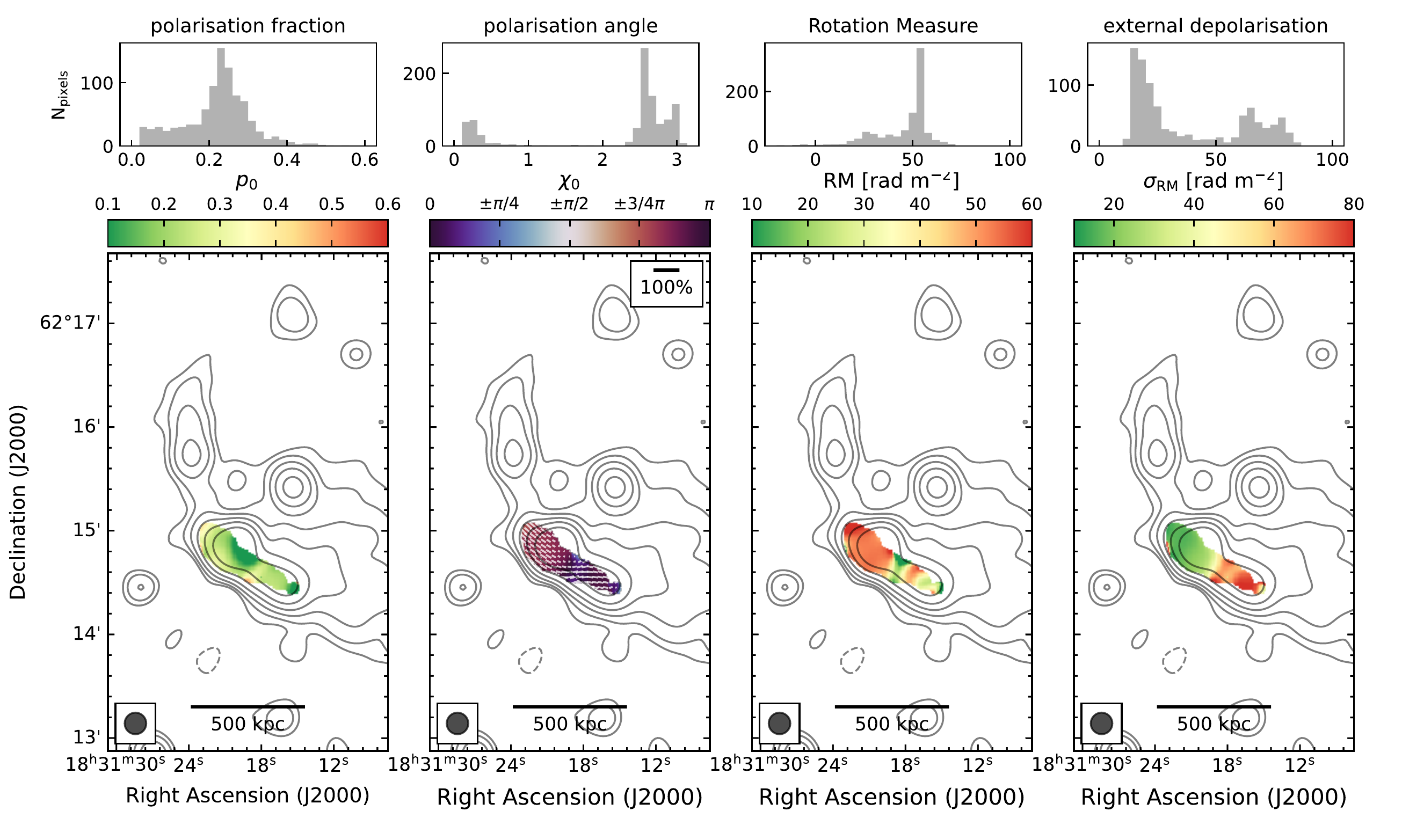}
\caption{Polarisation parameter maps for \cluster\ at $12.5''$ resolution. From left to right: intrinsic polarisation fraction ($p_0$); intrinsic polarisation angle ($\chi_0$), with the polarisation vectors overlaid in white; Rotation Measure (RM), including the Galactic contribution; external depolarisation ($\sigma_{\rm RM}$). On top of each map, the grey histogram shows the distribution per pixel of the polarisation parameters.}
\label{fig:polparam}
\end{figure*}

The maps of the polarisation parameters for the external depolarisation model at low resolution are displayed in Fig.~\ref{fig:polparam}.
The intrinsic polarisation fraction is overall constant, around a value of 0.2, and the polarisation vectors appear to follow well the elongation of the relic. 
In the presence of both ordered and random magnetic field, the theoretical maximum intrinsic polarisation fraction is set by the spectral index \citep{sokoloff+98,govoni+04}, i.e. 
\begin{equation}
p_0 = \dfrac{3\delta+3}{3\delta+7}\dfrac{1}{ 1+ \left ( \dfrac{B_{\rm rand}}{B_{\rm ord}} \right )^2}\, ,
\end{equation}
with $\delta = 1-2\alpha$, and $B_{\rm ord}$ and $B_{\rm rand}$ the ``ordered'' (i.e. aligned with the shock surface) and the ``random'' (i.e. isotropic) components of the magnetic fields respectively. From our spectral index analysis, we obtain a maximum intrinsic polarisation fraction of 0.74. This is much higher than our measured value implying that either the ratio $B_{\rm rand}/B_{\rm ord}$ should be greater than 1 \citep{sokoloff+98,govoni+04} or that the shock viewing angle should be smaller than $40^\circ$ \citep{ensslin+98}.

The Rotation Measure (RM) shows a sharp transition from almost constant values around $\sim50-60$ rad m$^{-2}$ in $\rm R2_{N}$ to more mixed values in $\rm R2_{S}$ (${\rm RM}\sim20-60$ rad m$^{-2}$; see Fig. \ref{fig:polparam}). At the same location where we observe this RM change, the external depolarisation shifts from ``low'' ($\sigma_{\rm RM}\sim10-20$ rad m$^{-2}$) to ``high'' ($\sigma_{\rm RM}\sim60-70$ rad m$^{-2}$) values. Interestingly, we note that this transition happens where, in projection, the relic locates in a denser region of the ICM (see last panel in Fig.~\ref{fig:relic_zoom}, and Fig.~\ref{fig:chandra_vla_contours}). This indicates that part of the denser ICM is in front of the relic and acting as a Faraday screen.

\begin{table}
\centering
\caption{{\it QU}-fitting results on the relic (R2) in \cluster\ assuming an External Depolarisation Model (EDF in Eq. \ref{eq:polmodels}). The uncertainties are from the {\it QU}-fitting procedure.}
\begin{tabular}{lrr}
\hline\hline
Polarisation parameter & ${\rm R2_N}$ & ${\rm R2_S}$\\
\hline
$p_0$ & $0.142\pm0.05$ & $0.190^{+0.023}_{-0.021}$ \\
$\chi_0$  ~[rad] & $2.66\pm0.02$ & $0.12^{+0.08}_{-2.96}$\\ 
RM ~[rad m$^{-2}$] & $52.4\pm1.1$ & $32.3^{+10.1}_{-10.4}$\\
$\sigma_{\rm RM}$ ~[rad m$^{-2}$] & $21.0^{+0.7}_{-0.7}$ & $66.0^{+5.8}_{-5.3}$\\
\hline
\end{tabular}
\label{tab:polfit}
\end{table}

The Rotation Measure of an extragalactic source at redshift $z$ is given as:
\begin{equation}
{\rm RM} = 
0.81 \int_{\rm source}^{\rm observer} \frac{n_e B_\parallel}{(1+z)^2} \, dl\quad {\rm [rad~ m^{-2}]} \, .
\end{equation}
Here, $n_e$ is the electron density (in cm$^{-3}$), $B_\parallel$ the magnetic field (in $\mu$Gauss) along the line of sight, $l$ the path length through the magneto-ionic medium (in pc). Since the observed polarisation signal is a combination of all that comes along the line of sight, the measured RM is actually the combination of different RM layers:
\begin{equation}
{\rm RM} = {\rm RM_{Gal}} + {\rm RM_{extragal}} +{\rm RM_{cluster}} \, .
\end{equation}
We used the latest Galactic Faraday Rotation catalogue \citep{hutschenreuter+22} and estimated the mean value of ${\rm RM_{Gal}}$ at the cluster location ($l=91.830$, $b=+26.116$; see Tab. \ref{tab:cluster}) within a radius of $0.1^\circ$\footnote{This value represents the element resolution provided by \cite{hutschenreuter+22}} (corresponding to $\sim2.7$ Mpc) to be ${\rm RM_{Gal}}=60\pm12~{\rm rad~m^{-2}}$. 
This Galactic RM value is consistent with what we measure in ${\rm R2_N}$, while the larger spread in ${\rm R2_S}$ suggests that there is also some contribution from the cluster. 
If we assume that ${\rm RM_{extragal}}=0$, i.e. the ICM is the only Faraday screen, the mean cluster RM value\footnote{The large scatter in RM does not allow a precise measurement.} is of a few ${\rm rad~m^{-2}}$. 
For a density column of $n_e=2\times10^{-5}~{\rm cm^{-3}}$ (i.e. the value we obtain from the X-ray surface brightness profile, see Sect. \ref{sec:xraySB}) and for a path length of the magnetised plasma of $L=260$ kpc \citep[i.e. $L\approx2\sqrt{2d_s\,r_s}$, where $d_s=15$ kpc and $r_s\sim500$ kpc are the intrinsic width of the shock and its distance from the cluster centre, respectively; see][]{kierdorf+17}, we obtain $B_\parallel\sim5-10~\mu$G. These values are similar to low-$z$ clusters \citep[i.e. CIZA\,J2242.8+5301, see][]{vanweeren+10,digennaro+21b}.


\subsection{A more complex merger event?}
The position of the radio relic in \cluster\ is unusual compared to other relics in literature. 
An idealised binary merger in the plane of the sky would produce double radio relics perpendicular to the merger axis. The double radio relic in CIZA\,J2242.8+5301 is a textbook example of this \citep{vanweeren+10,hoang+17,digennaro+18}. Instead, the radio relic in \cluster\ is located eastward the ICM distribution, parallel to the merger direction. 

\begin{figure*}
\centering
\includegraphics[width=0.3\textwidth]{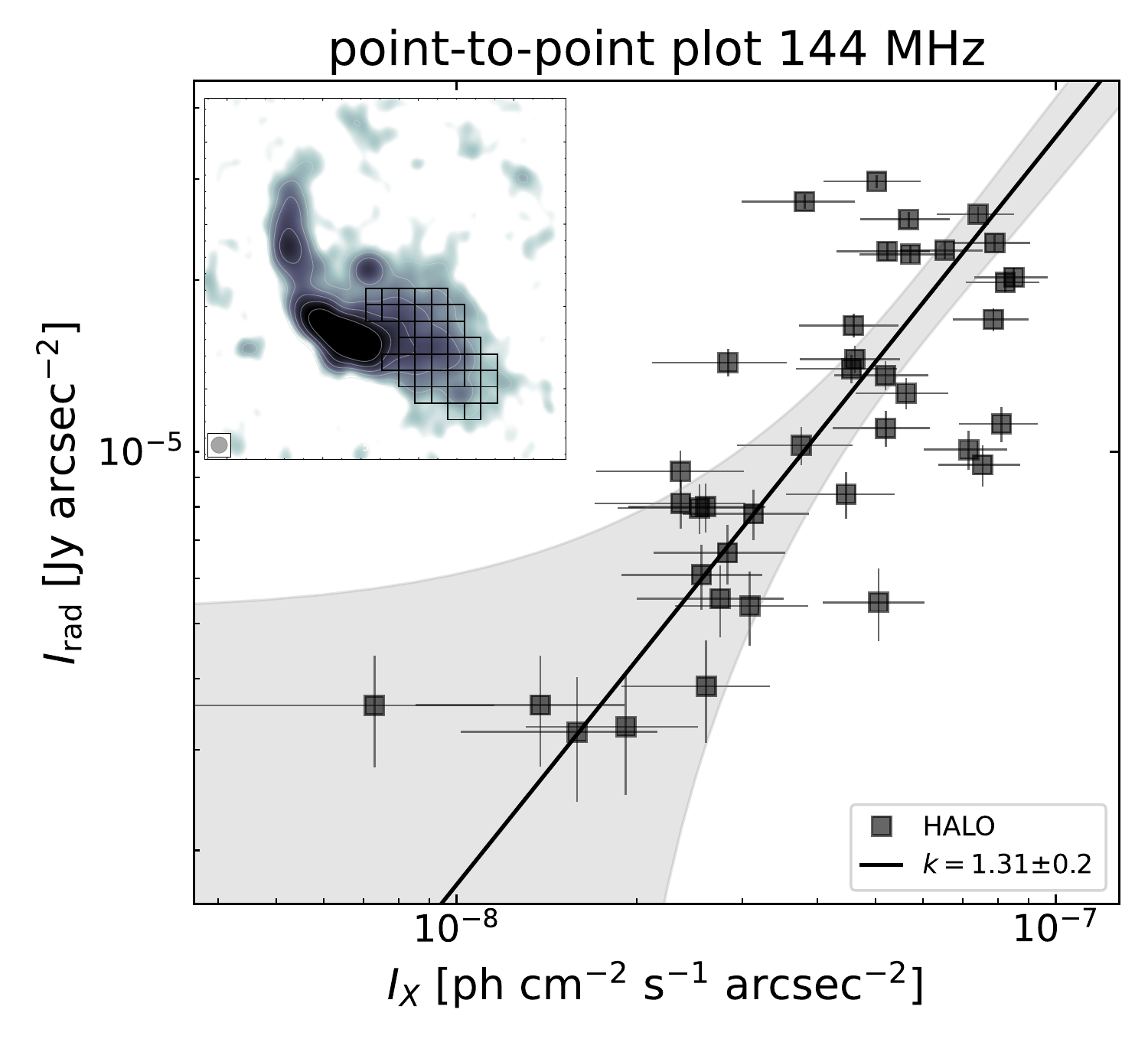}
\includegraphics[width=0.3\textwidth]{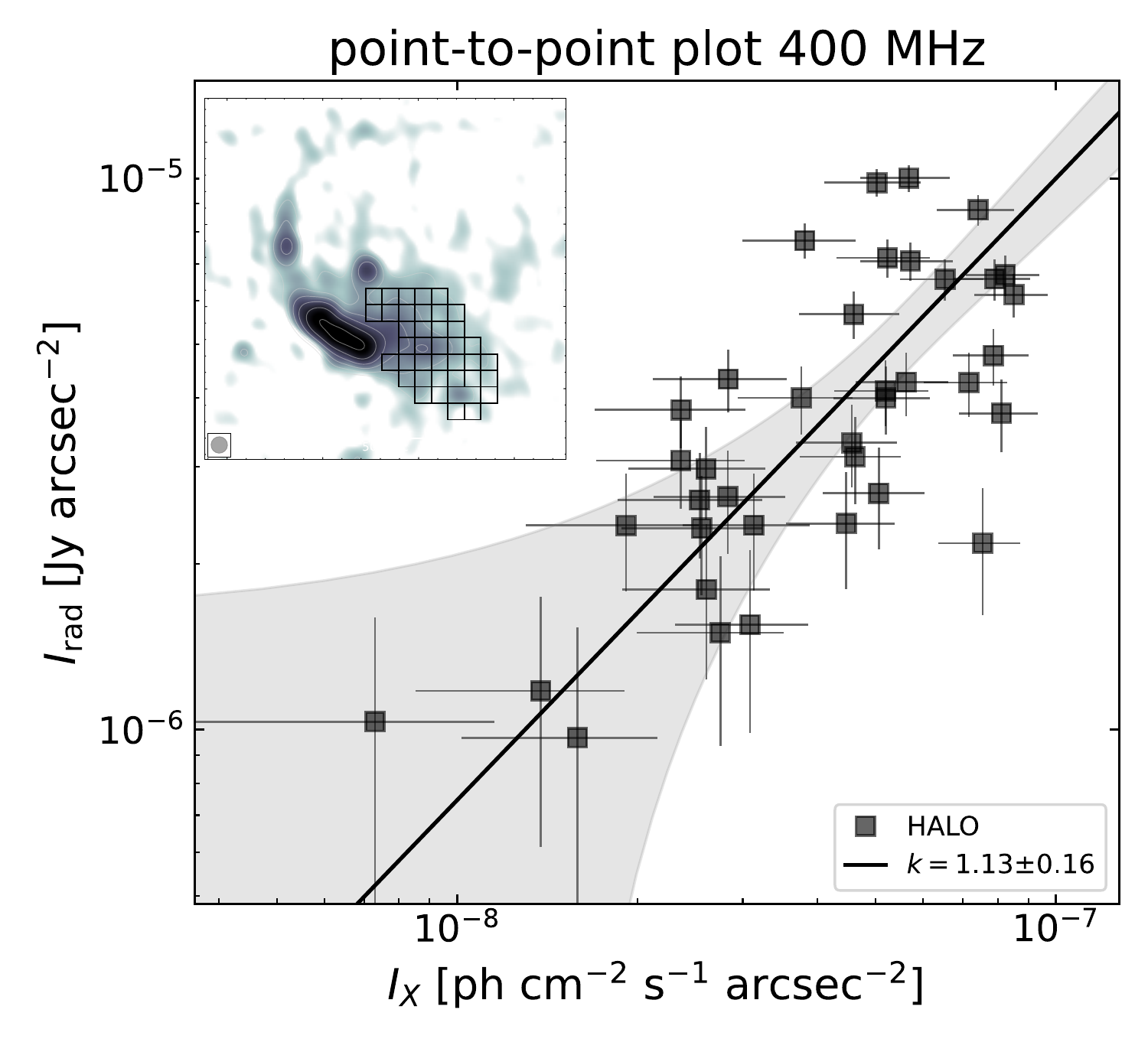}
\includegraphics[width=0.3\textwidth]{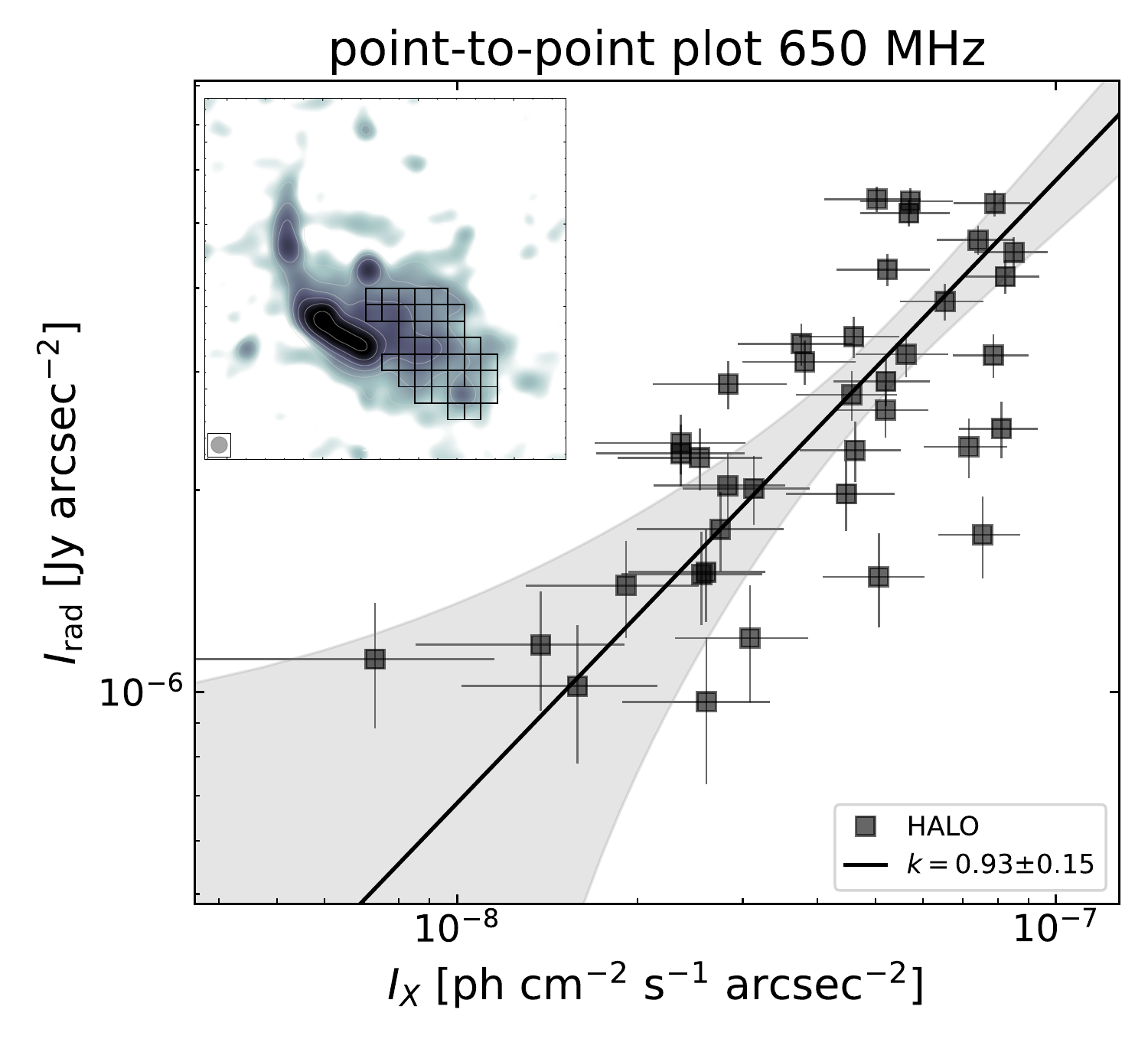}
\includegraphics[width=0.3\textwidth]{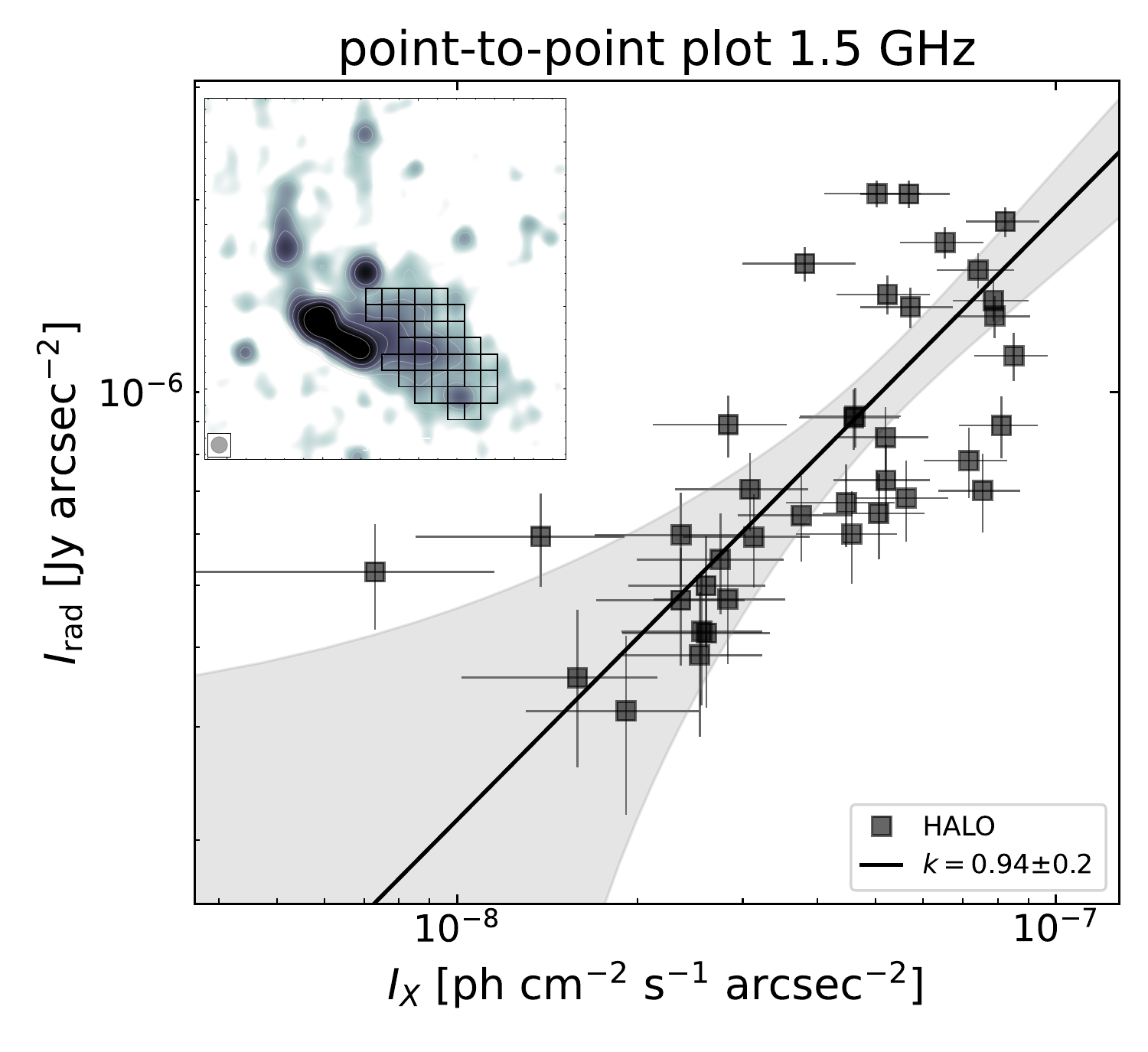}
\includegraphics[width=0.3\textwidth]{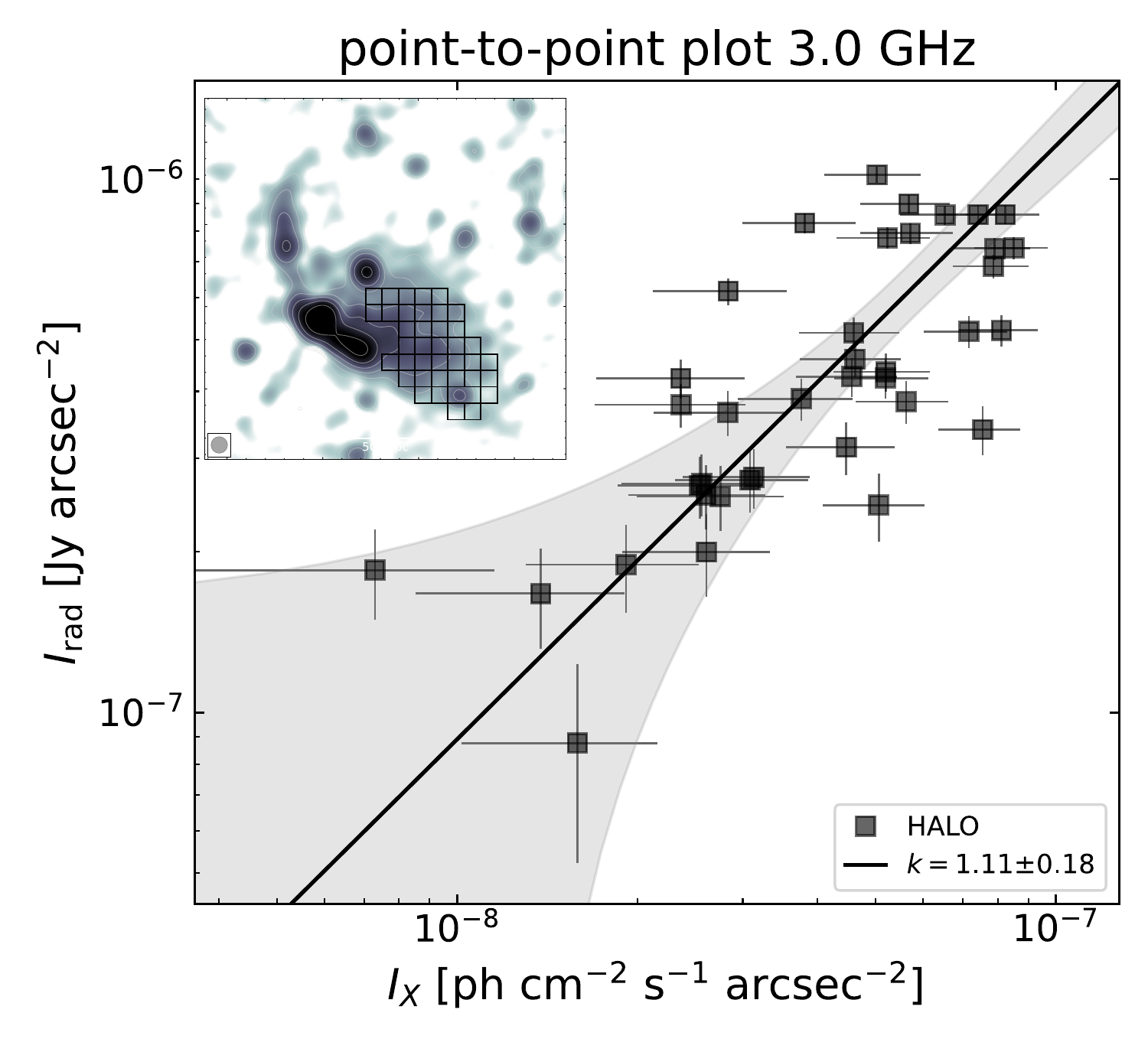}
\caption{Point-to-point analysis of the radio halo in \cluster\ at 144 MHz, 400 MHz, 650 MHz, 1.5 GHz and 3.0 GHz at $12''$ resolution. The inset of each panel shows the corresponding radio image with the beam-sized boxes used to compare the thermal and non-thermal emission.}
\label{fig:ptp_allfreq}
\end{figure*}

One possible explanation is that the two sub-clusters in \cluster\ are merging with a large impact parameter (i.e. $b>0$). We compared our observations with simulations presented in \cite{vanweeren+11b}, which used the FLASH code \citep{fryxell+00} based on adaptive mesh refinement (AMR) to follow the merger with high resolution only in the places of interest, such as shock discontinuities \citep{zuhone10}. According to their simulations, asymmetrical spiral-like shocks would develop in case of an impact factor of $b=4r_c$, with $r_c$ the core radius, after about 1 Gyr from the collision (see their Fig. 2). This could represent the case of the radio relic in \cluster, and the lack of the shock counterpart could be explained by a large mass ratio, or by the presence of pre-existing plasma only at the R2 location being re-accelerated by the mild shock wave \citep{vanweeren+17a}.
This early-merger scenario would also explain the flat spectral index of the radio halo and the small spectral fluctuations, as only the most turbulent regions had time to re-accelerate particles \citep[$t_{\rm acc}<t_{\rm merger}$, see][]{donnert+13}.

Another possibility is that the northern sub-cluster is itself undergoing a merger event, i.e. \cluster\ is a triple merger. In this scenario, the northern sub-cluster merged in the north-west/south-east direction, generating the radio relic, while the southern sub-cluster is moving in the north/north-east direction towards the northern sub-cluster. Deep optical observations, aiming to determine the spectroscopic redshifts of the cluster galaxy members, could provide insights on this possible scenario. 

Finally, we cannot completely exclude the possibility that the X-ray surface brightness discontinuity is associated with an equatorial shock and that the radio relic in \cluster\ is a consequence of that. However, it is no clear the role of these type of shocks in accelerating particles. Moreover, the presence of the radio halo, which would require $\rm\sim few~Gyr$ to form after the merger event \citep[e.g.][]{beresnyak+miniati16}, would disfavour this scenario.



\subsection{Thermal/non-thermal correlation for the radio halo}
Correlations between emission at different wavelengths from the same spatial regions can provide hints on common physical processes. Particularly, the emission from radio halos is expected to be related with that of the ICM \citep{brunetti+jones14} according to the following relation: 
\begin{equation}
I_{\rm rad} \propto I_X^k \, ,
\end{equation}
with $I_{\rm rad}$ and $I_X$ the radio and the X-ray surface brightness, respectively.
In order to investigate the existence of such a correlation at all the frequencies available, we performed point-to-point analyses \citep{govoni+01} using \texttt{PT-REX}\footnote{\url{https://github.com/AIgnesti/PT-REX}} \citep[Point-to-point TRend EXtractor;][]{ignesti22}. At each frequency, we calculated $I_{\rm rad}$ and $I_X$ from boxes of $12''$ resolution with a radio surface brightness above $2.5\sigma_{\rm rms,\nu}$ at each frequency (see Tab. \ref{tab:flux} for the map noises). This threshold also provides enough X-ray counts in each single box, which is the limiting factor due to the short exposure, and enough statistics to investigate the correlation. Other radio sources, such as the radio relic and the radio galaxies, were masked. We then fit the data using the BCES orthogonal method \citep{arkitas+96}.

The point-to-point analysis on radio halos commonly shows sub-linear trends \citep[i.e. $k<1$;][]{hoang+19,xie+20,rajpurohit+21a,rajpurohit+21b,bonafede+22,rajpurohit+22c}. In \cluster\ we find a linear/super-linear correlation between the radio and the ICM distribution (i.e. $k\gtrsim1$, see Fig.~\ref{fig:ptp_allfreq} and Tab. \ref{tab:ptp_allfreq}). 
Although the super-linear correlation would be in line with the early merger scenario, as the turbulence is not uniformly spread in the cluster volume, we point out that the combination of shallow X-ray observations, possible mixing of radio emission from electrons in the radio halo and in the radio relic, the large intrinsic scatter of the best-fit (i.e. $\sigma_k=0.14-0.20$), and the large physical scale of the cells (i.e. $\sim90$ kpc) which hides possible small-scale fluctuations does not allow a clear investigation of the $I_{\rm rad}-I_X$ correlation.
Finally, hints of a trend of $k$ with the frequency is noted, similarly to what has been found by \cite{rajpurohit+21} and \cite{hoang+21}, in the radio halos in MACS\,J0717.5+3745 and CLG\,0217+70 respectively. 


\begin{table}[]
\centering
\resizebox{0.45\textwidth}{!}{
\begin{tabular}{cccc}
\hline\hline
Frequency & Slope & Pearson ($p$-value) & Spearman ($p$-value) \\
$\nu$ & $k\pm\sigma_k$ \\
\hline
144 MHz & $1.31\pm0.20$ & 0.76 ($4.3\times10^{-8}$) & 0.72 ($2.6\times10^{-7}$) \\
400 MHz & $1.13\pm0.16$ & 0.73 ($2.0\times10^{-7}$) & 0.68 ($3.0\times10^{-6}$)\\
650 MHz & $0.93\pm0.14$ & 0.71 ($5.0\times10^{-7}$) & 0.70 ($1.0\times10^{-6}$)\\
1.5 GHz & $0.94\pm0.20$ & 0.69 ($1.6\times10^{-6}$) & 0.78 ($7.8\times10^{-9}$)\\
3.0 GHz & $1.11\pm0.18$ & 0.75 ($6.8\times10^{-8}$) & 0.74 ($8.7\times10^{-8}$)\\
\hline
\end{tabular}}
\caption{Results of the point-to-point (ptp) $I_{\rm rad} \propto I_X^k$ analysis at all the frequencies available for the radio halo in \cluster\ (see Fig.~\ref{fig:ptp_allfreq}). Column 1: frequency for $I_{\rm rad}$.  Column 2: slope of the ptp correlation. Column 3 and 4: Pearson and Spearman coefficients, respectively, and corresponding $p$-value in brackets.}
\label{tab:ptp_allfreq}
\end{table}

\section{Conclusions}\label{sec:concl}
In this paper, we present a multi-frequency analysis of the high-redshift galaxy cluster PSZ2\,G091.83+26.11 ($z=0.822$). We made use of LOFAR (120--168 MHz), uGMRT (250--500 MHz and 550--900 MHz) and VLA (1--2 GHz and 2--4 GHz) radio observations and data from the {\it Chandra} X-ray satellite to investigate the properties of the diffuse radio emission in the cluster, with a particular focus on the candidate radio relic. Below, we summarise the main results of our work:

\begin{itemize}
\item[$\bullet$] The diffuse radio emission in PSZ2\,G091.83+26.11 is visible up to 3.0 GHz. Particularly, the radio halo extends in the north-east/south-west direction for about 1.2 Mpc, while at high resolution ($\sim2''$) the bright extended relic-like source eastward of the cluster looks to be broken into two pieces, with length of about 640 kpc and 300 kpc each and width of about 15 kpc;

\smallskip
\item[$\bullet$] The diffuse radio source eastward of the cluster (i.e. the candidate radio relic) has a flux density of $15.9\pm0.6$ mJy and $5.0\pm0.2$ mJy at 1.5 GHz and 3.0 GHz, respectively, at $12''$ resolution. The resulting integrated spectral index, including also the LOFAR and uGMRT data, is $\alpha=-1.25\pm0.03$;

\smallskip
\item[$\bullet$] The spectral index map and profile across the candidate radio relic steepens towards the cluster centre, from $-0.89\pm0.03$ to $-1.39\pm0.03$, with hints of non-negligible spectral curvature. Moreover, part of the source is also visible in polarisation, with polarisation vectors that follow the source surface. All these pieces of evidence suggest that this source can be classified as a radio relic. Using {\it QU}-fitting, and assuming the only screens to rotate the polarisation vectors are the Galactic foreground and the ICM, we obtain an intrinsic polarisation fraction of about 20\%. From the Rotation Measure we estimate a relic magnetic field of $\rm 5-10~\mu Gauss$;

\smallskip
\item[$\bullet$] X-ray {\it Chandra} analysis reveals a surface brightness jump at the location of the cluster elongated radio source, i.e. $\mathcal{C}=2.22^{+0.37}_{-0.30}$. Assuming this is due to shock compression, this corresponds to a Mach number $\mathcal{M}=1.93^{+0.42}_{-0.32}$;

\smallskip
\item[$\bullet$] We compared our observational results with simulations based on adaptive mesh refinement (AMR), and we find that the location of the radio relic can be explained by an off-set merger, with impact factor $b=4r_c$. However, we cannot entirely exclude also a scenario involving multiple merger events, or an equatorial shocks;


\smallskip
\item[$\bullet$] The emission from the radio halo in PSZ2\,G091.83+26.11 is in line with the previously published studies \citep{digennaro+21a,digennaro+21c}. Its flux density is $6.1\pm0.2$ mJy and $3.1\pm0.1$ mJy at 1.5 GHz and 3.0 GHz, respectively, at $19''$ resolution. Combining these with the flux densities at LOFAR and uGMRT frequencies, we obtain an integrated spectral index $\alpha=-1.06\pm0.03$. The $I_{\rm rad}-I_X$ point-to-point analysis suggests a linear/super-linear trend.

\end{itemize}

Given the above results -- the off-set position of the radio relic, its relatively short distance from the cluster centre (i.e. about 500 kpc) and low polarisation (i.e. $\sim20\%$), the flat spectral index of the radio halo with no sign of curvature nor frequency break, and the cluster redshift -- we speculate that \cluster\ could be in the early phase of its merger event.

\begin{acknowledgements}
We thank the referee for the suggestions which improved the quality of the manuscript.
GDG acknowledges support from the Alexander von Humboldt Foundation. 
MB acknowledges support from the Deutsche Forschungsgemeinschaft under Germany's Excellence Strategy - EXC 2121 “Quantum Universe” - 390833306. 
RJvW acknowledges support from the ERC Starting Grant ClusterWeb 804208. 
AS is supported by the Women In Science Excel (WISE) programme of the Netherlands Organisation for Scientific Research (NWO), and acknowledges the Kavli IPMU for the continued hospitality. SRON Netherlands Institute for Space Research is supported financially by NWO. 
GB and RC acknowledge support from INAF through the mainstream project ``Cluster science with LOFAR''. 
WRF acknowledges support from the Smithsonian Institution, the Chandra High Resolution Camera Project through NASA contract NAS8-03060, and NASA Grants 80NSSC19K0116, GO1-22132X, and GO9-20109X. 
AI acknowledges funding from the European Research Council (ERC) under the European Union's Horizon 2020 research and innovation programme (grant agreement No. 833824).
HJAR acknowledge support from the ERC Advanced Investigator programme NewClusters 321271.
The National Radio Astronomy Observatory is a facility of the National Science Foundation operated under cooperative agreement by Associated Universities, Inc.
This paper is based on data obtained with the LOw Frequency Array (LOFAR). LOFAR \citep{vanhaarlem+13} is the Low Frequency Array designed and constructed by ASTRON. It has observing, data processing, and data storage facilities in several countries, which are owned by various parties (each with their own funding sources), and which are collectively operated by the ILT foundation under a joint scientific policy. The ILT resources have benefited from the following recent major funding sources: CNRS-INSU, Observatoire de Paris and Universit\'e d'Orl\'eans, France; BMBF, MIWF-NRW, MPG, Germany; Science Foundation Ireland (SFI), Department of Business, Enterprise and Innovation (DBEI), Ireland; NWO, The Netherlands; The Science and Technology Facilities Council, UK; Ministry of Science and Higher Education, Poland; The Istituto Nazionale di Astrofisica (INAF), Italy.  This research made use of the Dutch national e-infrastructure with support of the SURF Cooperative (e-infra 180169) and the LOFAR e-infra group. The J\"ulich LOFAR Long Term Archive and the German LOFAR network are both coordinated and operated by the J\"ulich Supercomputing Centre (JSC), and computing resources on the supercomputer JUWELS at JSC were provided by the Gauss Centre for Supercomputing e.V. (grant CHTB00) through the John von Neumann Institute for Computing (NIC). This research made use of the University of Hertfordshire high-performance computing facility and the LOFAR-UK computing facility located at the University of Hertfordshire and supported by STFC [ST/P000096/1], and of the Italian LOFAR IT computing infrastructure supported and operated by INAF, and by the Physics Department of Turin university (under an agreement with Consorzio Interuniversitario per la Fisica Spaziale) at the C3S Supercomputing Centre, Italy.
This paper is based on data obtained with the Giant Metrewave Radio Telescope (GMRT). We thank the staff of the GMRT that made these observations possible. GMRT is run by the National Centre for Radio Astrophysics of the Tata Institute of Fundamental Research.
This research has made use of data obtained from the Chandra Data Archive and the Chandra Source Catalog, and software provided by the Chandra X-ray Center (CXC) in the application packages CIAO and Sherpa.
This research made use of APLpy, an open-source plotting package for Python \citep{aplpy}. 
GDG acknowledges Luca Di Mascolo for the help and support while dealing with APLpy.
\end{acknowledgements}

\bibliographystyle{aa}
\bibliography{biblio.bib}

\begin{thebibliography}{113}
\expandafter\ifx\csname natexlab\endcsname\relax\def\natexlab#1{#1}\fi

\bibitem[{{Adam} {et~al.}(2021){Adam}, {Goksu}, {Brown}, {Rudnick}, \&
  {Ferrari}}]{adam+21}
{Adam}, R., {Goksu}, H., {Brown}, S., {Rudnick}, L., \& {Ferrari}, C. 2021,
  \aap, 648, A60

\bibitem[{{Akamatsu} {et~al.}(2015){Akamatsu}, {van Weeren}, {Ogrean},
  {Kawahara}, {Stroe}, {Sobral}, {Hoeft}, {R{\"o}ttgering}, {Br{\"u}ggen}, \&
  {Kaastra}}]{akamatsu+15}
{Akamatsu}, H., {van Weeren}, R.~J., {Ogrean}, G.~A., {et~al.} 2015, \aap, 582,
  A87

\bibitem[{{Akritas} \& {Bershady}(1996)}]{arkitas+96}
{Akritas}, M.~G. \& {Bershady}, M.~A. 1996, \apj, 470, 706

\bibitem[{{Amodeo} {et~al.}(2018){Amodeo}, {Mei}, {Stanford}, {Lawrence},
  {Bartlett}, {Stern}, {Chary}, {Shim}, {Marleau}, {Melin}, \&
  {Rodr{\'\i}guez-Gonz{\'a}lvez}}]{amodeo+18}
{Amodeo}, S., {Mei}, S., {Stanford}, S.~A., {et~al.} 2018, \apj, 853, 36

\bibitem[{{Artis} {et~al.}(2022){Artis}, {Adam}, {Ade}, {Ajeddig}, {Andr{\'e}},
  {Arnaud}, {Aussel}, {Bartalucci}, {Beelen}, {Beno{\^\i}t}, {Berta}, {Bing},
  {Bourrion}, {Calvo}, {Catalano}, {De Petris}, {D{\'e}sert}, {Doyle},
  {Driessen}, {Ferragamo}, {Gomez}, {Goupy}, {K{\'e}ruzor{\'e}}, {Kramer},
  {Ladjelate}, {Lagache}, {Leclercq}, {Lestrade}, {Mac{\'\i}as-P{\'e}rez},
  {Maury}, {Mauskopf}, {Mayet}, {Monfardini}, {Mu{\~n}oz-Echeverr{\'\i}a},
  {Paliwal}, {Perotto}, {Pisano}, {Pointecouteau}, {Ponthieu}, {Pratt},
  {Rev{\'e}ret}, {Rigby}, {Ritacco}, {Romero}, {Roussel}, {Ruppin}, {Schuster},
  {Shu}, {Sievers}, {Tucker}, \& {Yepes}}]{artis+22}
{Artis}, E., {Adam}, R., {Ade}, P., {et~al.} 2022, in European Physical Journal
  Web of Conferences, Vol. 257, European Physical Journal Web of Conferences,
  00003

\bibitem[{{Asplund} {et~al.}(2009){Asplund}, {Grevesse}, {Sauval}, \&
  {Scott}}]{asplund+09}
{Asplund}, M., {Grevesse}, N., {Sauval}, A.~J., \& {Scott}, P. 2009, \araa, 47,
  481

\bibitem[{{Beresnyak} \& {Miniati}(2016)}]{beresnyak+miniati16}
{Beresnyak}, A. \& {Miniati}, F. 2016, \apj, 817, 127

\bibitem[{{Blandford} \& {Eichler}(1987)}]{blandford+eichler87}
{Blandford}, R. \& {Eichler}, D. 1987, \physrep, 154, 1

\bibitem[{{Bonafede} {et~al.}(2012){Bonafede}, {Br{\"u}ggen}, {van Weeren},
  {Vazza}, {Giovannini}, {Ebeling}, {Edge}, {Hoeft}, \& {Klein}}]{bonafede+12}
{Bonafede}, A., {Br{\"u}ggen}, M., {van Weeren}, R., {et~al.} 2012, \mnras,
  426, 40

\bibitem[{{Bonafede} {et~al.}(2022){Bonafede}, {Brunetti}, {Rudnick}, {Vazza},
  {Bourdin}, {Giovannini}, {Shimwell}, {Zhang}, {Mazzotta}, {Simionescu},
  {Biava}, {Bonnassieux}, {Brienza}, {Br{\"u}ggen}, {Rajpurohit}, {Riseley},
  {Stuardi}, {Feretti}, {Tasse}, {Botteon}, {Carretti}, {Cassano}, {Cuciti},
  {Gasperin}, {Gastaldello}, {Rossetti}, {Rottgering}, {Venturi}, \&
  {Weeren}}]{bonafede+22}
{Bonafede}, A., {Brunetti}, G., {Rudnick}, L., {et~al.} 2022, \apj, 933, 218

\bibitem[{{Bonafede} {et~al.}(2013){Bonafede}, {Vazza}, {Br{\"u}ggen},
  {Murgia}, {Govoni}, {Feretti}, {Giovannini}, \& {Ogrean}}]{bonafede+13}
{Bonafede}, A., {Vazza}, F., {Br{\"u}ggen}, M., {et~al.} 2013, \mnras, 433,
  3208

\bibitem[{{Botteon} {et~al.}(2020){Botteon}, {Brunetti}, {Ryu}, \&
  {Roh}}]{botteon+20b}
{Botteon}, A., {Brunetti}, G., {Ryu}, D., \& {Roh}, S. 2020, \aap, 634, A64

\bibitem[{{Brentjens} \& {de Bruyn}(2005)}]{brentjens+debruyn05}
{Brentjens}, M.~A. \& {de Bruyn}, A.~G. 2005, \aap, 441, 1217

\bibitem[{{Brunetti} \& {Blasi}(2005)}]{brunetti+blasi05}
{Brunetti}, G. \& {Blasi}, P. 2005, \mnras, 363, 1173

\bibitem[{{Brunetti} {et~al.}(2008){Brunetti}, {Giacintucci}, {Cassano},
  {Lane}, {Dallacasa}, {Venturi}, {Kassim}, {Setti}, {Cotton}, \&
  {Markevitch}}]{brunetti+08}
{Brunetti}, G., {Giacintucci}, S., {Cassano}, R., {et~al.} 2008, \nat, 455, 944

\bibitem[{{Brunetti} \& {Jones}(2014)}]{brunetti+jones14}
{Brunetti}, G. \& {Jones}, T.~W. 2014, International Journal of Modern Physics
  D, 23, 1430007

\bibitem[{{Brunetti} \& {Lazarian}(2007)}]{brunetti+lazarian07}
{Brunetti}, G. \& {Lazarian}, A. 2007, \mnras, 378, 245

\bibitem[{{Brunetti} \& {Lazarian}(2011)}]{brunetti+lazarian11}
{Brunetti}, G. \& {Lazarian}, A. 2011, \mnras, 410, 127

\bibitem[{{Brunetti} \& {Lazarian}(2016)}]{brunetti+lazarian16}
{Brunetti}, G. \& {Lazarian}, A. 2016, \mnras, 458, 2584

\bibitem[{{Brunetti} {et~al.}(2001){Brunetti}, {Setti}, {Feretti}, \&
  {Giovannini}}]{brunetti+01}
{Brunetti}, G., {Setti}, G., {Feretti}, L., \& {Giovannini}, G. 2001, \mnras,
  320, 365

\bibitem[{{Brunetti} {et~al.}(2017){Brunetti}, {Zimmer}, \&
  {Zandanel}}]{brunetti+17}
{Brunetti}, G., {Zimmer}, S., \& {Zandanel}, F. 2017, \mnras, 472, 1506

\bibitem[{{Bruno} {et~al.}(2021){Bruno}, {Rajpurohit}, {Brunetti},
  {Gastaldello}, {Botteon}, {Ignesti}, {Bonafede}, {Dallacasa}, {Cassano}, {van
  Weeren}, {Cuciti}, {Di Gennaro}, {Shimwell}, \& {Br{\"u}ggen}}]{bruno+21}
{Bruno}, L., {Rajpurohit}, K., {Brunetti}, G., {et~al.} 2021, \aap, 650, A44

\bibitem[{{Burn}(1966)}]{burn66}
{Burn}, B.~J. 1966, \mnras, 133, 67

\bibitem[{{Cash}(1979)}]{cash79}
{Cash}, W. 1979, \apj, 228, 939

\bibitem[{{Cassano} {et~al.}(2019){Cassano}, {Botteon}, {Di Gennaro},
  {Brunetti}, {Sereno}, {Shimwell}, {van Weeren}, {Br{\"u}ggen}, {Gastaldello},
  {Izzo}, {B{\^\i}rzan}, {Bonafede}, {Cuciti}, {de Gasperin}, {R{\"o}ttgering},
  {Hardcastle}, {Mechev}, \& {Tasse}}]{cassano+19}
{Cassano}, R., {Botteon}, A., {Di Gennaro}, G., {et~al.} 2019, \apjl, 881, L18

\bibitem[{{Cassano} {et~al.}(2012){Cassano}, {Brunetti}, {Norris},
  {R{\"o}ttgering}, {Johnston-Hollitt}, \& {Trasatti}}]{cassano+12}
{Cassano}, R., {Brunetti}, G., {Norris}, R.~P., {et~al.} 2012, \aap, 548, A100

\bibitem[{{Cassano} {et~al.}(2010){Cassano}, {Brunetti}, {R{\"o}ttgering}, \&
  {Br{\"u}ggen}}]{cassano+10}
{Cassano}, R., {Brunetti}, G., {R{\"o}ttgering}, H.~J.~A., \& {Br{\"u}ggen}, M.
  2010, \aap, 509, A68

\bibitem[{{Cassano} {et~al.}(2006){Cassano}, {Brunetti}, \&
  {Setti}}]{cassano+06}
{Cassano}, R., {Brunetti}, G., \& {Setti}, G. 2006, \mnras, 369, 1577

\bibitem[{{Cassano} {et~al.}(2023){Cassano}, {Cuciti}, {Brunetti}, {Botteon},
  {Rossetti}, {Bruno}, {Simionescu}, {Gastaldello}, {van Weeren}, {Brueggen},
  {Dallacasa}, {Zhang}, {Akamatsu}, {Bonafede}, {Di Gennaro}, {Shimwell}, {de
  Gasperin}, {Roettgering}, \& {Jones}}]{cassano+23}
{Cassano}, R., {Cuciti}, V., {Brunetti}, G., {et~al.} 2023, arXiv e-prints,
  arXiv:2301.08052

\bibitem[{{Cassano} {et~al.}(2013){Cassano}, {Ettori}, {Brunetti},
  {Giacintucci}, {Pratt}, {Venturi}, {Kale}, {Dolag}, \&
  {Markevitch}}]{cassano+13}
{Cassano}, R., {Ettori}, S., {Brunetti}, G., {et~al.} 2013, \apj, 777, 141

\bibitem[{{Chandra} {et~al.}(2004){Chandra}, {Ray}, \&
  {Bhatnagar}}]{chandra+04}
{Chandra}, P., {Ray}, A., \& {Bhatnagar}, S. 2004, \apj, 612, 974

\bibitem[{{Cuciti} {et~al.}(2021){Cuciti}, {Cassano}, {Brunetti}, {Dallacasa},
  {de Gasperin}, {Ettori}, {Giacintucci}, {Kale}, {Pratt}, {van Weeren}, \&
  {Venturi}}]{cuciti+21b}
{Cuciti}, V., {Cassano}, R., {Brunetti}, G., {et~al.} 2021, \aap, 647, A51

\bibitem[{{Dallacasa} {et~al.}(2009){Dallacasa}, {Brunetti}, {Giacintucci},
  {Cassano}, {Venturi}, {Macario}, {Kassim}, {Lane}, \& {Setti}}]{dallacasa+09}
{Dallacasa}, D., {Brunetti}, G., {Giacintucci}, S., {et~al.} 2009, \apj, 699,
  1288

\bibitem[{{de Gasperin} {et~al.}(2022){de Gasperin}, {Rudnick}, {Finoguenov},
  {Wittor}, {Akamatsu}, {Br{\"u}ggen}, {Chibueze}, {Clarke}, {Cotton},
  {Cuciti}, {Dom{\'\i}nguez-Fern{\'a}ndez}, {Knowles}, {O'Sullivan}, \&
  {Sebokolodi}}]{degasperin+22}
{de Gasperin}, F., {Rudnick}, L., {Finoguenov}, A., {et~al.} 2022, \aap, 659,
  A146

\bibitem[{{Di Gennaro} {et~al.}(2019){Di Gennaro}, {van Weeren},
  {Andrade-Santos}, {Akamatsu}, {Randall}, {Forman}, {Kraft}, {Brunetti},
  {Dawson}, {Golovich}, \& {Jones}}]{digennaro+19}
{Di Gennaro}, G., {van Weeren}, R.~J., {Andrade-Santos}, F., {et~al.} 2019,
  \apj, 873, 64

\bibitem[{{Di Gennaro} {et~al.}(2021{\natexlab{a}}){Di Gennaro}, {van Weeren},
  {Brunetti}, {Cassano}, {Br{\"u}ggen}, {Hoeft}, {Shimwell}, {R{\"o}ttgering},
  {Bonafede}, {Botteon}, {Cuciti}, {Dallacasa}, {de Gasperin},
  {Dom{\'\i}nguez-Fern{\'a}ndez}, {En{\ss}lin}, {Gastaldello}, {Mandal},
  {Rossetti}, \& {Simionescu}}]{digennaro+21a}
{Di Gennaro}, G., {van Weeren}, R.~J., {Brunetti}, G., {et~al.}
  2021{\natexlab{a}}, Nature Astronomy, 5, 268

\bibitem[{{Di Gennaro} {et~al.}(2021{\natexlab{b}}){Di Gennaro}, {van Weeren},
  {Cassano}, {Brunetti}, {Br{\"u}ggen}, {Hoeft}, {Osinga}, {Botteon}, {Cuciti},
  {de Gasperin}, {R{\"o}ttgering}, \& {Tasse}}]{digennaro+21c}
{Di Gennaro}, G., {van Weeren}, R.~J., {Cassano}, R., {et~al.}
  2021{\natexlab{b}}, \aap, 654, A166

\bibitem[{{Di Gennaro} {et~al.}(2018){Di Gennaro}, {van Weeren}, {Hoeft},
  {Kang}, {Ryu}, {Rudnick}, {Forman}, {R{\"o}ttgering}, {Br{\"u}ggen},
  {Dawson}, {Golovich}, {Hoang}, {Intema}, {Jones}, {Kraft}, {Shimwell}, \&
  {Stroe}}]{digennaro+18}
{Di Gennaro}, G., {van Weeren}, R.~J., {Hoeft}, M., {et~al.} 2018, \apj, 865,
  24

\bibitem[{{Di Gennaro} {et~al.}(2021{\natexlab{c}}){Di Gennaro}, {van Weeren},
  {Rudnick}, {Hoeft}, {Br{\"u}ggen}, {Ryu}, {R{\"o}ttgering}, {Forman},
  {Stroe}, {Shimwell}, {Kraft}, {Jones}, \& {Hoang}}]{digennaro+21b}
{Di Gennaro}, G., {van Weeren}, R.~J., {Rudnick}, L., {et~al.}
  2021{\natexlab{c}}, \apj, 911, 3

\bibitem[{{Dom{\'\i}nguez-Fern{\'a}ndez}
  {et~al.}(2021){Dom{\'\i}nguez-Fern{\'a}ndez}, {Br{\"u}ggen}, {Vazza},
  {Hoeft}, {Banda-Barrag{\'a}n}, {Rajpurohit}, {Wittor}, {Mignone},
  {Mukherjee}, \& {Vaidya}}]{dominguez-fernandez+21}
{Dom{\'\i}nguez-Fern{\'a}ndez}, P., {Br{\"u}ggen}, M., {Vazza}, F., {et~al.}
  2021, \mnras, 507, 2714

\bibitem[{{Donnert} {et~al.}(2013){Donnert}, {Dolag}, {Brunetti}, \&
  {Cassano}}]{donnert+13}
{Donnert}, J., {Dolag}, K., {Brunetti}, G., \& {Cassano}, R. 2013, \mnras, 429,
  3564

\bibitem[{{Donnert} {et~al.}(2018){Donnert}, {Vazza}, {Br{\"u}ggen}, \&
  {ZuHone}}]{donnert+18}
{Donnert}, J., {Vazza}, F., {Br{\"u}ggen}, M., \& {ZuHone}, J. 2018, \ssr, 214,
  122

\bibitem[{{Duchesne} {et~al.}(2021){Duchesne}, {Johnston-Hollitt}, \&
  {Wilber}}]{duchesne+21}
{Duchesne}, S.~W., {Johnston-Hollitt}, M., \& {Wilber}, A.~G. 2021, \pasa, 38,
  e031

\bibitem[{{Eckert}(2016)}]{eckert16}
{Eckert}, D. 2016, {PROFFIT: Analysis of X-ray surface-brightness profiles},
  Astrophysics Source Code Library, record ascl:1608.011

\bibitem[{{Ensslin} {et~al.}(1998){Ensslin}, {Biermann}, {Klein}, \&
  {Kohle}}]{ensslin+98}
{Ensslin}, T.~A., {Biermann}, P.~L., {Klein}, U., \& {Kohle}, S. 1998, \aap,
  332, 395

\bibitem[{{Finoguenov} {et~al.}(2010){Finoguenov}, {Sarazin}, {Nakazawa},
  {Wik}, \& {Clarke}}]{finoguenov+10}
{Finoguenov}, A., {Sarazin}, C.~L., {Nakazawa}, K., {Wik}, D.~R., \& {Clarke},
  T.~E. 2010, \apj, 715, 1143

\bibitem[{{Foreman-Mackey} {et~al.}(2013){Foreman-Mackey}, {Hogg}, {Lang}, \&
  {Goodman}}]{foreman-mackey+13}
{Foreman-Mackey}, D., {Hogg}, D.~W., {Lang}, D., \& {Goodman}, J. 2013, \pasp,
  125, 306

\bibitem[{{Fruscione} {et~al.}(2006){Fruscione}, {McDowell}, {Allen},
  {Brickhouse}, {Burke}, {Davis}, {Durham}, {Elvis}, {Galle}, {Harris},
  {Huenemoerder}, {Houck}, {Ishibashi}, {Karovska}, {Nicastro}, {Noble},
  {Nowak}, {Primini}, {Siemiginowska}, {Smith}, \& {Wise}}]{fruscione+06}
{Fruscione}, A., {McDowell}, J.~C., {Allen}, G.~E., {et~al.} 2006, in
  \procspie, Vol. 6270, Society of Photo-Optical Instrumentation Engineers
  (SPIE) Conference Series, 62701V

\bibitem[{{Fryxell} {et~al.}(2000){Fryxell}, {Olson}, {Ricker}, {Timmes},
  {Zingale}, {Lamb}, {MacNeice}, {Rosner}, {Truran}, \& {Tufo}}]{fryxell+00}
{Fryxell}, B., {Olson}, K., {Ricker}, P., {et~al.} 2000, \apjs, 131, 273

\bibitem[{{George} {et~al.}(2012){George}, {Stil}, \& {Keller}}]{george+12}
{George}, S.~J., {Stil}, J.~M., \& {Keller}, B.~W. 2012, \pasa, 29, 214

\bibitem[{{Govoni} {et~al.}(2001){Govoni}, {En{\ss}lin}, {Feretti}, \&
  {Giovannini}}]{govoni+01}
{Govoni}, F., {En{\ss}lin}, T.~A., {Feretti}, L., \& {Giovannini}, G. 2001,
  \aap, 369, 441

\bibitem[{{Govoni} \& {Feretti}(2004)}]{govoni+04}
{Govoni}, F. \& {Feretti}, L. 2004, International Journal of Modern Physics D,
  13, 1549

\bibitem[{{Gu} {et~al.}(2019){Gu}, {Akamatsu}, {Shimwell}, {Intema}, {van
  Weeren}, {de Gasperin}, {Mernier}, {Mao}, {Urdampilleta}, {de Plaa},
  {Parekh}, {R{\"o}ttgering}, \& {Kaastra}}]{gu+19}
{Gu}, L., {Akamatsu}, H., {Shimwell}, T.~W., {et~al.} 2019, Nature Astronomy,
  3, 838

\bibitem[{{Ha} {et~al.}(2018){Ha}, {Ryu}, \& {Kang}}]{ha+18}
{Ha}, J.-H., {Ryu}, D., \& {Kang}, H. 2018, \apj, 857, 26

\bibitem[{{Hoang} {et~al.}(2017){Hoang}, {Shimwell}, {Stroe}, {Akamatsu},
  {Brunetti}, {Donnert}, {Intema}, {Mulcahy}, {R{\"o}ttgering}, {van Weeren},
  {Bonafede}, {Br{\"u}ggen}, {Cassano}, {Chy{\.z}y}, {En{\ss}lin}, {Ferrari},
  {de Gasperin}, {Gu}, {Hoeft}, {Miley}, {Orr{\'u}}, {Pizzo}, \&
  {White}}]{hoang+17}
{Hoang}, D.~N., {Shimwell}, T.~W., {Stroe}, A., {et~al.} 2017, \mnras, 471,
  1107

\bibitem[{{Hoang} {et~al.}(2019){Hoang}, {Shimwell}, {van Weeren}, {Brunetti},
  {R{\"o}ttgering}, {Andrade-Santos}, {Botteon}, {Br{\"u}ggen}, {Cassano},
  {Drabent}, {de Gasperin}, {Hoeft}, {Intema}, {Rafferty}, {Shweta}, \&
  {Stroe}}]{hoang+19}
{Hoang}, D.~N., {Shimwell}, T.~W., {van Weeren}, R.~J., {et~al.} 2019, \aap,
  622, A20

\bibitem[{{Hoang} {et~al.}(2021){Hoang}, {Zhang}, {Stuardi}, {Shimwell},
  {Bonafede}, {Br{\"u}ggen}, {Brunetti}, {Botteon}, {Cassano}, {de Gasperin},
  {Di Gennaro}, {Hoeft}, {Intema}, {Rajpurohit}, {R{\"o}ttgering},
  {Simionescu}, \& {van Weeren}}]{hoang+21}
{Hoang}, D.~N., {Zhang}, X., {Stuardi}, C., {et~al.} 2021, \aap, 656, A154

\bibitem[{{Hoeft} {et~al.}(2022){Hoeft}, {Rajpurohit}, {Wittor}, {di Gennaro},
  \& {Dom{\'\i}nguez-Fern{\'a}ndez}}]{hoeft+22}
{Hoeft}, M., {Rajpurohit}, K., {Wittor}, D., {di Gennaro}, G., \&
  {Dom{\'\i}nguez-Fern{\'a}ndez}, P. 2022, Galaxies, 10, 10

\bibitem[{{Hutschenreuter} {et~al.}(2022){Hutschenreuter}, {Anderson}, {Betti},
  {Bower}, {Brown}, {Br{\"u}ggen}, {Carretti}, {Clarke}, {Clegg}, {Costa},
  {Croft}, {Van Eck}, {Gaensler}, {de Gasperin}, {Haverkorn}, {Heald}, {Hull},
  {Inoue}, {Johnston-Hollitt}, {Kaczmarek}, {Law}, {Ma}, {MacMahon}, {Mao},
  {Riseley}, {Roy}, {Shanahan}, {Shimwell}, {Stil}, {Sobey}, {O'Sullivan},
  {Tasse}, {Vacca}, {Vernstrom}, {Williams}, {Wright}, \&
  {En{\ss}lin}}]{hutschenreuter+22}
{Hutschenreuter}, S., {Anderson}, C.~S., {Betti}, S., {et~al.} 2022, \aap, 657,
  A43

\bibitem[{{Iapichino} \& {Br{\"u}ggen}(2012)}]{iapichino+bruggen12}
{Iapichino}, L. \& {Br{\"u}ggen}, M. 2012, \mnras, 423, 2781

\bibitem[{{Ignesti}(2022)}]{ignesti22}
{Ignesti}, A. 2022, \na, 92, 101732

\bibitem[{{Jaffe} \& {Perola}(1973)}]{jaffe+perola73}
{Jaffe}, W.~J. \& {Perola}, G.~C. 1973, \aap, 26, 423

\bibitem[{{Jones} {et~al.}(2023){Jones}, {de Gasperin}, {Cuciti}, {Botteon},
  {Zhang}, {Gastaldello}, {Shimwell}, {Simionescu}, {Rossetti}, {Cassano},
  {Akamatsu}, {Bonafede}, {Br{\"u}ggen}, {Brunetti}, {Camillini}, {Di Gennaro},
  {Drabent}, {Hoang}, {Rajpurohit}, {Natale}, {Tasse}, \& {van
  Weeren}}]{jones+23}
{Jones}, A., {de Gasperin}, F., {Cuciti}, V., {et~al.} 2023, arXiv e-prints,
  arXiv:2301.07814

\bibitem[{{Kang} {et~al.}(2017){Kang}, {Ryu}, \& {Jones}}]{kang+17}
{Kang}, H., {Ryu}, D., \& {Jones}, T. 2017, in International Cosmic Ray
  Conference, Vol. 301, 35th International Cosmic Ray Conference (ICRC2017),
  283

\bibitem[{{Kardashev}(1962)}]{kardashev62}
{Kardashev}, N.~S. 1962, \sovast, 6, 317

\bibitem[{{Katz-Stone} {et~al.}(1993){Katz-Stone}, {Rudnick}, \&
  {Anderson}}]{katz-stone+93}
{Katz-Stone}, D.~M., {Rudnick}, L., \& {Anderson}, M.~C. 1993, \apj, 407, 549

\bibitem[{{Kierdorf} {et~al.}(2017){Kierdorf}, {Beck}, {Hoeft}, {Klein}, {van
  Weeren}, {Forman}, \& {Jones}}]{kierdorf+17}
{Kierdorf}, M., {Beck}, R., {Hoeft}, M., {et~al.} 2017, \aap, 600, A18

\bibitem[{{Komissarov} \& {Gubanov}(1994)}]{komissarov+gubanov94}
{Komissarov}, S.~S. \& {Gubanov}, A.~G. 1994, \aap, 285, 27

\bibitem[{{Lindner} {et~al.}(2014){Lindner}, {Baker}, {Hughes}, {Battaglia},
  {Gupta}, {Knowles}, {Marriage}, {Menanteau}, {Moodley}, {Reese}, \&
  {Srianand}}]{lindner+14}
{Lindner}, R.~R., {Baker}, A.~J., {Hughes}, J.~P., {et~al.} 2014, \apj, 786, 49

\bibitem[{{Markevitch} {et~al.}(2005){Markevitch}, {Govoni}, {Brunetti}, \&
  {Jerius}}]{markevitch+05}
{Markevitch}, M., {Govoni}, F., {Brunetti}, G., \& {Jerius}, D. 2005, \apj,
  627, 733

\bibitem[{{Markevitch} \& {Vikhlinin}(2007)}]{markevitch+vikhlinin07}
{Markevitch}, M. \& {Vikhlinin}, A. 2007, \physrep, 443, 1

\bibitem[{{McMullin} {et~al.}(2007){McMullin}, {Waters}, {Schiebel}, {Young},
  \& {Golap}}]{mcmullin+07}
{McMullin}, J.~P., {Waters}, B., {Schiebel}, D., {Young}, W., \& {Golap}, K.
  2007, in Astronomical Society of the Pacific Conference Series, Vol. 376,
  Astronomical Data Analysis Software and Systems XVI, ed. R.~A. {Shaw},
  F.~{Hill}, \& D.~J. {Bell}, 127

\bibitem[{{Offringa} {et~al.}(2010){Offringa}, {de Bruyn}, {Biehl}, {Zaroubi},
  {Bernardi}, \& {Pandey}}]{offringa+10}
{Offringa}, A.~R., {de Bruyn}, A.~G., {Biehl}, M., {et~al.} 2010, \mnras, 405,
  155

\bibitem[{{Offringa} {et~al.}(2014){Offringa}, {McKinley}, {Hurley-Walker},
  {Briggs}, {Wayth}, {Kaplan}, {Bell}, {Feng}, {Neben}, {Hughes}, {Rhee},
  {Murphy}, {Bhat}, {Bernardi}, {Bowman}, {Cappallo}, {Corey}, {Deshpande},
  {Emrich}, {Ewall-Wice}, {Gaensler}, {Goeke}, {Greenhill}, {Hazelton},
  {Hindson}, {Johnston-Hollitt}, {Jacobs}, {Kasper}, {Kratzenberg}, {Lenc},
  {Lonsdale}, {Lynch}, {McWhirter}, {Mitchell}, {Morales}, {Morgan},
  {Kudryavtseva}, {Oberoi}, {Ord}, {Pindor}, {Procopio}, {Prabu}, {Riding},
  {Roshi}, {Shankar}, {Srivani}, {Subrahmanyan}, {Tingay}, {Waterson},
  {Webster}, {Whitney}, {Williams}, \& {Williams}}]{offringa+14}
{Offringa}, A.~R., {McKinley}, B., {Hurley-Walker}, N., {et~al.} 2014, \mnras,
  444, 606

\bibitem[{{Offringa} \& {Smirnov}(2017)}]{offringa+17}
{Offringa}, A.~R. \& {Smirnov}, O. 2017, \mnras, 471, 301

\bibitem[{{Osinga} {et~al.}(2022){Osinga}, {van Weeren}, {Andrade-Santos},
  {Rudnick}, {Bonafede}, {Clarke}, {Duncan}, {Giacintucci}, {Mroczkowski}, \&
  {R{\"o}ttgering}}]{osinga+22}
{Osinga}, E., {van Weeren}, R.~J., {Andrade-Santos}, F., {et~al.} 2022, \aap,
  665, A71

\bibitem[{{Owen} {et~al.}(2014){Owen}, {Rudnick}, {Eilek}, {Rau}, {Bhatnagar},
  \& {Kogan}}]{owen+14}
{Owen}, F.~N., {Rudnick}, L., {Eilek}, J., {et~al.} 2014, \apj, 794, 24

\bibitem[{{Pearce} {et~al.}(2017){Pearce}, {van Weeren}, {Andrade-Santos},
  {Jones}, {Forman}, {Br{\"u}ggen}, {Bulbul}, {Clarke}, {Kraft}, {Medezinski},
  {Mroczkowski}, {Nonino}, {Nulsen}, {Rand all}, \& {Umetsu}}]{pearce+17}
{Pearce}, C.~J.~J., {van Weeren}, R.~J., {Andrade-Santos}, F., {et~al.} 2017,
  \apj, 845, 81

\bibitem[{{Perley} \& {Butler}(2013)}]{perley+butler13}
{Perley}, R.~A. \& {Butler}, B.~J. 2013, \apjs, 206, 16

\bibitem[{{Petrosian}(2001)}]{petrosian01}
{Petrosian}, V. 2001, \apj, 557, 560

\bibitem[{{Pinzke} {et~al.}(2017){Pinzke}, {Oh}, \& {Pfrommer}}]{pinzke+17}
{Pinzke}, A., {Oh}, S.~P., \& {Pfrommer}, C. 2017, \mnras, 465, 4800

\bibitem[{{Planck Collaboration} {et~al.}(2016){Planck Collaboration}, {Ade},
  {Aghanim}, {Arnaud}, {Ashdown}, {Aumont}, {Baccigalupi}, {Banday},
  {Barreiro}, {Barrena}, {Bartlett}, {Bartolo}, {Battaner}, {Battye},
  {Benabed}, {Beno{\^\i}t}, {Benoit-L{\'e}vy}, {Bernard}, {Bersanelli},
  {Bielewicz}, {Bikmaev}, {B{\"o}hringer}, {Bonaldi}, {Bonavera}, {Bond},
  {Borrill}, {Bouchet}, {Bucher}, {Burenin}, {Burigana}, {Butler}, {Calabrese},
  {Cardoso}, {Carvalho}, {Catalano}, {Challinor}, {Chamballu}, {Chary},
  {Chiang}, {Chon}, {Christensen}, {Clements}, {Colombi}, {Colombo}, {Combet},
  {Comis}, {Couchot}, {Coulais}, {Crill}, {Curto}, {Cuttaia}, {Dahle},
  {Danese}, {Davies}, {Davis}, {de Bernardis}, {de Rosa}, {de Zotti},
  {Delabrouille}, {D{\'e}sert}, {Dickinson}, {Diego}, {Dolag}, {Dole},
  {Donzelli}, {Dor{\'e}}, {Douspis}, {Ducout}, {Dupac}, {Efstathiou},
  {Eisenhardt}, {Elsner}, {En{\ss}lin}, {Eriksen}, {Falgarone}, {Fergusson},
  {Feroz}, {Ferragamo}, {Finelli}, {Forni}, {Frailis}, {Fraisse}, {Franceschi},
  {Frejsel}, {Galeotta}, {Galli}, {Ganga}, {G{\'e}nova-Santos}, {Giard},
  {Giraud-H{\'e}raud}, {Gjerl{\o}w}, {Gonz{\'a}lez-Nuevo}, {G{\'o}rski},
  {Grainge}, {Gratton}, {Gregorio}, {Gruppuso}, {Gudmundsson}, {Hansen},
  {Hanson}, {Harrison}, {Hempel}, {Henrot-Versill{\'e}},
  {Hern{\'a}ndez-Monteagudo}, {Herranz}, {Hildebrandt}, {Hivon}, {Hobson},
  {Holmes}, {Hornstrup}, {Hovest}, {Huffenberger}, {Hurier}, {Jaffe}, {Jaffe},
  {Jin}, {Jones}, {Juvela}, {Keih{\"a}nen}, {Keskitalo}, {Khamitov}, {Kisner},
  {Kneissl}, {Knoche}, {Kunz}, {Kurki-Suonio}, {Lagache}, {Lamarre}, {Lasenby},
  {Lattanzi}, {Lawrence}, {Leonardi}, {Lesgourgues}, {Levrier}, {Liguori},
  {Lilje}, {Linden-V{\o}rnle}, {L{\'o}pez-Caniego}, {Lubin},
  {Mac{\'\i}as-P{\'e}rez}, {Maggio}, {Maino}, {Mak}, {Mandolesi}, {Mangilli},
  {Martin}, {Mart{\'\i}nez-Gonz{\'a}lez}, {Masi}, {Matarrese}, {Mazzotta},
  {McGehee}, {Mei}, {Melchiorri}, {Melin}, {Mendes}, {Mennella}, {Migliaccio},
  {Mitra}, {Miville-Desch{\^e}nes}, {Moneti}, {Montier}, {Morgante},
  {Mortlock}, {Moss}, {Munshi}, {Murphy}, {Naselsky}, {Nastasi}, {Nati},
  {Natoli}, {Netterfield}, {N{\o}rgaard-Nielsen}, {Noviello}, {Novikov},
  {Novikov}, {Olamaie}, {Oxborrow}, {Paci}, {Pagano}, {Pajot}, {Paoletti},
  {Pasian}, {Patanchon}, {Pearson}, {Perdereau}, {Perotto}, {Perrott},
  {Perrotta}, {Pettorino}, {Piacentini}, {Piat}, {Pierpaoli}, {Pietrobon},
  {Plaszczynski}, {Pointecouteau}, {Polenta}, {Pratt}, {Pr{\'e}zeau}, {Prunet},
  {Puget}, {Rachen}, {Reach}, {Rebolo}, {Reinecke}, {Remazeilles}, {Renault},
  {Renzi}, {Ristorcelli}, {Rocha}, {Rosset}, {Rossetti}, {Roudier}, {Rozo},
  {Rubi{\~n}o-Mart{\'\i}n}, {Rumsey}, {Rusholme}, {Rykoff}, {Sandri}, {Santos},
  {Saunders}, {Savelainen}, {Savini}, {Schammel}, {Scott}, {Seiffert},
  {Shellard}, {Shimwell}, {Spencer}, {Stanford}, {Stern}, {Stolyarov},
  {Stompor}, {Streblyanska}, {Sudiwala}, {Sunyaev}, {Sutton}, {Suur-Uski},
  {Sygnet}, {Tauber}, {Terenzi}, {Toffolatti}, {Tomasi}, {Tramonte},
  {Tristram}, {Tucci}, {Tuovinen}, {Umana}, {Valenziano}, {Valiviita}, {Van
  Tent}, {Vielva}, {Villa}, {Wade}, {Wandelt}, {Wehus}, {White}, {Wright},
  {Yvon}, {Zacchei}, \& {Zonca}}]{planckcoll16}
{Planck Collaboration}, {Ade}, P.~A.~R., {Aghanim}, N., {et~al.} 2016, \aap,
  594, A27

\bibitem[{{Press} \& {Schechter}(1974)}]{press+schecter74}
{Press}, W.~H. \& {Schechter}, P. 1974, \apj, 187, 425

\bibitem[{{Rajpurohit} {et~al.}(2021{\natexlab{a}}){Rajpurohit}, {Brunetti},
  {Bonafede}, {van Weeren}, {Botteon}, {Vazza}, {Hoeft}, {Riseley},
  {Bonnassieux}, {Brienza}, {Forman}, {R{\"o}ttgering}, {Rajpurohit},
  {Locatelli}, {Shimwell}, {Cassano}, {Di Gennaro}, {Br{\"u}ggen}, {Wittor},
  {Drabent}, \& {Ignesti}}]{rajpurohit+21b}
{Rajpurohit}, K., {Brunetti}, G., {Bonafede}, A., {et~al.} 2021{\natexlab{a}},
  \aap, 646, A135

\bibitem[{{Rajpurohit} {et~al.}(2018){Rajpurohit}, {Hoeft}, {van Weeren},
  {Rudnick}, {R{\"o}ttgering}, {Forman}, {Br{\"u}ggen}, {Croston},
  {Andrade-Santos}, {Dawson}, {Intema}, {Kraft}, {Jones}, \&
  {Jee}}]{rajpurohit+18}
{Rajpurohit}, K., {Hoeft}, M., {van Weeren}, R.~J., {et~al.} 2018, \apj, 852,
  65

\bibitem[{{Rajpurohit} {et~al.}(2022{\natexlab{a}}){Rajpurohit}, {Hoeft},
  {Wittor}, {van Weeren}, {Vazza}, {Rudnick}, {Rajpurohit}, {Forman},
  {Riseley}, {Brienza}, {Bonafede}, {Rajpurohit},
  {Dom{\'\i}nguez-Fern{\'a}ndez}, {Eilek}, {Bonnassieux}, {Br{\"u}ggen}, {Loi},
  {R{\"o}ttgering}, {Drabent}, {Locatelli}, {Botteon}, {Brunetti}, \&
  {Clarke}}]{rajpurohit+22b}
{Rajpurohit}, K., {Hoeft}, M., {Wittor}, D., {et~al.} 2022{\natexlab{a}}, \aap,
  657, A2

\bibitem[{{Rajpurohit} {et~al.}(2022{\natexlab{b}}){Rajpurohit}, {Osinga},
  {Brienza}, {Botteon}, {Brunetti}, {Forman}, {Riseley}, {Vazza}, {Bonafede},
  {van Weeren}, {Br{\"u}ggen}, {Rajpurohit}, {Drabent}, {Dallacasa},
  {Rossetti}, {Rajpurohit}, {Hoeft}, {Bonnassieux}, {Cassano}, \&
  {Miley}}]{rajpurohit+22c}
{Rajpurohit}, K., {Osinga}, E., {Brienza}, M., {et~al.} 2022{\natexlab{b}},
  arXiv e-prints, arXiv:2209.03288

\bibitem[{{Rajpurohit} {et~al.}(2022{\natexlab{c}}){Rajpurohit}, {van Weeren},
  {Hoeft}, {Vazza}, {Brienza}, {Forman}, {Wittor},
  {Dom{\'\i}nguez-Fern{\'a}ndez}, {Rajpurohit}, {Riseley}, {Botteon}, {Osinga},
  {Brunetti}, {Bonnassieux}, {Bonafede}, {Rajpurohit}, {Stuardi}, {Drabent},
  {Br{\"u}ggen}, {Dallacasa}, {Shimwell}, {R{\"o}ttgering}, {Gasperin},
  {Miley}, \& {Rossetti}}]{rajpurohit+22a}
{Rajpurohit}, K., {van Weeren}, R.~J., {Hoeft}, M., {et~al.}
  2022{\natexlab{c}}, \apj, 927, 80

\bibitem[{{Rajpurohit} {et~al.}(2020){Rajpurohit}, {Vazza}, {Hoeft}, {Loi},
  {Beck}, {Vacca}, {Kierdorf}, {van Weeren}, {Wittor}, {Govoni}, {Murgia},
  {Riseley}, {Locatelli}, {Drabent}, \& {Bonnassieux}}]{rajpurohit+20}
{Rajpurohit}, K., {Vazza}, F., {Hoeft}, M., {et~al.} 2020, \aap, 642, L13

\bibitem[{{Rajpurohit} {et~al.}(2021{\natexlab{b}}){Rajpurohit}, {Vazza}, {van
  Weeren}, {Hoeft}, {Brienza}, {Bonnassieux}, {Riseley}, {Brunetti},
  {Bonafede}, {Br{\"u}ggen}, {Formann}, {Rajpurohit}, {R{\"o}ttgering},
  {Drabent}, {Dom{\'\i}nguez-Fern{\'a}ndez}, {Wittor}, \&
  {Andrade-Santos}}]{rajpurohit+21a}
{Rajpurohit}, K., {Vazza}, F., {van Weeren}, R.~J., {et~al.}
  2021{\natexlab{b}}, \aap, 654, A41

\bibitem[{{Rajpurohit} {et~al.}(2021{\natexlab{c}}){Rajpurohit}, {Wittor}, {van
  Weeren}, {Vazza}, {Hoeft}, {Rudnick}, {Locatelli}, {Eilek}, {Forman},
  {Bonafede}, {Bonnassieux}, {Riseley}, {Brienza}, {Brunetti}, {Br{\"u}ggen},
  {Loi}, {Rajpurohit}, {R{\"o}ttgering}, {Botteon}, {Clarke}, {Drabent},
  {Dom{\'\i}nguez-Fern{\'a}ndez}, {Di Gennaro}, \&
  {Gastaldello}}]{rajpurohit+21}
{Rajpurohit}, K., {Wittor}, D., {van Weeren}, R.~J., {et~al.}
  2021{\natexlab{c}}, \aap, 646, A56

\bibitem[{{Rice}(1945)}]{rice45}
{Rice}, S.~O. 1945, Bell System Technical Journal, 24, 46

\bibitem[{{Robitaille} \& {Bressert}(2012)}]{aplpy}
{Robitaille}, T. \& {Bressert}, E. 2012, {APLpy: Astronomical Plotting Library
  in Python}, Astrophysics Source Code Library

\bibitem[{{Roettiger} {et~al.}(1999){Roettiger}, {Burns}, \&
  {Stone}}]{roettiger+99}
{Roettiger}, K., {Burns}, J.~O., \& {Stone}, J.~M. 1999, \apj, 518, 603

\bibitem[{{Shimwell} {et~al.}(2022){Shimwell}, {Hardcastle}, {Tasse}, {Best},
  {R{\"o}ttgering}, {Williams}, {Botteon}, {Drabent}, {Mechev}, {Shulevski},
  {van Weeren}, {Bester}, {Br{\"u}ggen}, {Brunetti}, {Callingham}, {Chy{\.z}y},
  {Conway}, {Dijkema}, {Duncan}, {de Gasperin}, {Hale}, {Haverkorn}, {Hugo},
  {Jackson}, {Mevius}, {Miley}, {Morabito}, {Morganti}, {Offringa}, {Oonk},
  {Rafferty}, {Sabater}, {Smith}, {Schwarz}, {Smirnov}, {O'Sullivan},
  {Vedantham}, {White}, {Albert}, {Alegre}, {Asabere}, {Bacon}, {Bonafede},
  {Bonnassieux}, {Brienza}, {Bilicki}, {Bonato}, {Calistro Rivera}, {Cassano},
  {Cochrane}, {Croston}, {Cuciti}, {Dallacasa}, {Danezi}, {Dettmar}, {Di
  Gennaro}, {Edler}, {En{\ss}lin}, {Emig}, {Franzen}, {Garc{\'\i}a-Vergara},
  {Grange}, {G{\"u}rkan}, {Hajduk}, {Heald}, {Heesen}, {Hoang}, {Hoeft},
  {Horellou}, {Iacobelli}, {Jamrozy}, {Jeli{\'c}}, {Kondapally}, {Kukreti},
  {Kunert-Bajraszewska}, {Magliocchetti}, {Mahatma}, {Ma{\l}ek}, {Mandal},
  {Massaro}, {Meyer-Zhao}, {Mingo}, {Mostert}, {Nair}, {Nakoneczny},
  {Nikiel-Wroczy{\'n}ski}, {Orr{\'u}}, {Pajdosz-{\'S}mierciak}, {Pasini},
  {Prandoni}, {van Piggelen}, {Rajpurohit}, {Retana-Montenegro}, {Riseley},
  {Rowlinson}, {Saxena}, {Schrijvers}, {Sweijen}, {Siewert}, {Timmerman},
  {Vaccari}, {Vink}, {West}, {Wo{\l}owska}, {Zhang}, \& {Zheng}}]{shimwell+22}
{Shimwell}, T.~W., {Hardcastle}, M.~J., {Tasse}, C., {et~al.} 2022, \aap, 659,
  A1

\bibitem[{{Sokoloff} {et~al.}(1998){Sokoloff}, {Bykov}, {Shukurov},
  {Berkhuijsen}, {Beck}, \& {Poezd}}]{sokoloff+98}
{Sokoloff}, D.~D., {Bykov}, A.~A., {Shukurov}, A., {et~al.} 1998, \mnras, 299,
  189

\bibitem[{{Stuardi} {et~al.}(2021){Stuardi}, {Bonafede}, {Lovisari},
  {Dom{\'\i}nguez-Fern{\'a}ndez}, {Vazza}, {Br{\"u}ggen}, {van Weeren}, \& {de
  Gasperin}}]{stuardi+21}
{Stuardi}, C., {Bonafede}, A., {Lovisari}, L., {et~al.} 2021, \mnras, 502, 2518

\bibitem[{{Stuardi} {et~al.}(2019){Stuardi}, {Bonafede}, {Wittor}, {Vazza},
  {Botteon}, {Locatelli}, {Dallacasa}, {Golovich}, {Hoeft}, {van Weeren},
  {Br{\"u}ggen}, \& {de Gasperin}}]{stuardi+19}
{Stuardi}, C., {Bonafede}, A., {Wittor}, D., {et~al.} 2019, \mnras, 2080

\bibitem[{{Tribble}(1991)}]{tribble91}
{Tribble}, P.~C. 1991, \mnras, 250, 726

\bibitem[{{Urdampilleta} {et~al.}(2018){Urdampilleta}, {Akamatsu}, {Mernier},
  {Kaastra}, {de Plaa}, {Ohashi}, {Ishisaki}, \& {Kawahara}}]{urdampilleta+18}
{Urdampilleta}, I., {Akamatsu}, H., {Mernier}, F., {et~al.} 2018, \aap, 618,
  A74

\bibitem[{{van Haarlem} {et~al.}(2013){van Haarlem}, {Wise}, {Gunst}, {Heald},
  {McKean}, {Hessels}, {de Bruyn}, {Nijboer}, {Swinbank}, {Fallows},
  {Brentjens}, {Nelles}, {Beck}, {Falcke}, {Fender}, {H{\"o}randel},
  {Koopmans}, {Mann}, {Miley}, {R{\"o}ttgering}, {Stappers}, {Wijers},
  {Zaroubi}, {van den Akker}, {Alexov}, {Anderson}, {Anderson}, {van Ardenne},
  {Arts}, {Asgekar}, {Avruch}, {Batejat}, {B{\"a}hren}, {Bell}, {Bell}, {van
  Bemmel}, {Bennema}, {Bentum}, {Bernardi}, {Best}, {B{\^i}rzan}, {Bonafede},
  {Boonstra}, {Braun}, {Bregman}, {Breitling}, {van de Brink}, {Broderick},
  {Broekema}, {Brouw}, {Br{\"u}ggen}, {Butcher}, {van Cappellen}, {Ciardi},
  {Coenen}, {Conway}, {Coolen}, {Corstanje}, {Damstra}, {Davies}, {Deller},
  {Dettmar}, {van Diepen}, {Dijkstra}, {Donker}, {Doorduin}, {Dromer}, {Drost},
  {van Duin}, {Eisl{\"o}ffel}, {van Enst}, {Ferrari}, {Frieswijk}, {Gankema},
  {Garrett}, {de Gasperin}, {Gerbers}, {de Geus}, {Grie{\ss}meier}, {Grit},
  {Gruppen}, {Hamaker}, {Hassall}, {Hoeft}, {Holties}, {Horneffer}, {van der
  Horst}, {van Houwelingen}, {Huijgen}, {Iacobelli}, {Intema}, {Jackson},
  {Jelic}, {de Jong}, {Juette}, {Kant}, {Karastergiou}, {Koers}, {Kollen},
  {Kondratiev}, {Kooistra}, {Koopman}, {Koster}, {Kuniyoshi}, {Kramer},
  {Kuper}, {Lambropoulos}, {Law}, {van Leeuwen}, {Lemaitre}, {Loose}, {Maat},
  {Macario}, {Markoff}, {Masters}, {McFadden}, {McKay-Bukowski}, {Meijering},
  {Meulman}, {Mevius}, {Middelberg}, {Millenaar}, {Miller-Jones}, {Mohan},
  {Mol}, {Morawietz}, {Morganti}, {Mulcahy}, {Mulder}, {Munk}, {Nieuwenhuis},
  {van Nieuwpoort}, {Noordam}, {Norden}, {Noutsos}, {Offringa}, {Olofsson},
  {Omar}, {Orr{\'u}}, {Overeem}, {Paas}, {Pandey-Pommier}, {Pandey}, {Pizzo},
  {Polatidis}, {Rafferty}, {Rawlings}, {Reich}, {de Reijer}, {Reitsma},
  {Renting}, {Riemers}, {Rol}, {Romein}, {Roosjen}, {Ruiter}, {Scaife}, {van
  der Schaaf}, {Scheers}, {Schellart}, {Schoenmakers}, {Schoonderbeek},
  {Serylak}, {Shulevski}, {Sluman}, {Smirnov}, {Sobey}, {Spreeuw}, {Steinmetz},
  {Sterks}, {Stiepel}, {Stuurwold}, {Tagger}, {Tang}, {Tasse}, {Thomas},
  {Thoudam}, {Toribio}, {van der Tol}, {Usov}, {van Veelen}, {van der Veen},
  {ter Veen}, {Verbiest}, {Vermeulen}, {Vermaas}, {Vocks}, {Vogt}, {de Vos},
  {van der Wal}, {van Weeren}, {Weggemans}, {Weltevrede}, {White}, {Wijnholds},
  {Wilhelmsson}, {Wucknitz}, {Yatawatta}, {Zarka}, {Zensus}, \& {van
  Zwieten}}]{vanhaarlem+13}
{van Haarlem}, M.~P., {Wise}, M.~W., {Gunst}, A.~W., {et~al.} 2013, \aap, 556,
  A2

\bibitem[{{van Weeren} {et~al.}(2017){van Weeren}, {Andrade-Santos}, {Dawson},
  {Golovich}, {Lal}, {Kang}, {Ryu}, {Br{\`i}ggen}, {Ogrean}, {Forman}, {Jones},
  {Placco}, {Santucci}, {Wittman}, {Jee}, {Kraft}, {Sobral}, {Stroe}, \&
  {Fogarty}}]{vanweeren+17a}
{van Weeren}, R.~J., {Andrade-Santos}, F., {Dawson}, W.~A., {et~al.} 2017,
  Nature Astronomy, 1, 0005

\bibitem[{{van Weeren} {et~al.}(2011){van Weeren}, {Br{\"u}ggen},
  {R{\"o}ttgering}, \& {Hoeft}}]{vanweeren+11b}
{van Weeren}, R.~J., {Br{\"u}ggen}, M., {R{\"o}ttgering}, H.~J.~A., \& {Hoeft},
  M. 2011, \mnras, 418, 230

\bibitem[{{van Weeren} {et~al.}(2019){van Weeren}, {de Gasperin}, {Akamatsu},
  {Br{\"u}ggen}, {Feretti}, {Kang}, {Stroe}, \& {Zandanel}}]{vanweeren+19}
{van Weeren}, R.~J., {de Gasperin}, F., {Akamatsu}, H., {et~al.} 2019, \ssr,
  215, 16

\bibitem[{{van Weeren} {et~al.}(2010){van Weeren}, {R{\"o}ttgering},
  {Br{\"u}ggen}, \& {Hoeft}}]{vanweeren+10}
{van Weeren}, R.~J., {R{\"o}ttgering}, H.~J.~A., {Br{\"u}ggen}, M., \& {Hoeft},
  M. 2010, Science, 330, 347

\bibitem[{{Vazza} {et~al.}(2018){Vazza}, {Brunetti}, {Br{\"u}ggen}, \&
  {Bonafede}}]{vazza+18}
{Vazza}, F., {Brunetti}, G., {Br{\"u}ggen}, M., \& {Bonafede}, A. 2018, \mnras,
  474, 1672

\bibitem[{{Wilber} {et~al.}(2018){Wilber}, {Br{\"u}ggen}, {Bonafede}, {Savini},
  {Shimwell}, {van Weeren}, {Rafferty}, {Mechev}, {Intema}, {Andrade-Santos},
  {Clarke}, {Mahony}, {Morganti}, {Prand oni}, {Brunetti}, {R{\"o}ttgering},
  {Mandal}, {de Gasperin}, \& {Hoeft}}]{wilber+18}
{Wilber}, A., {Br{\"u}ggen}, M., {Bonafede}, A., {et~al.} 2018, \mnras, 473,
  3536

\bibitem[{{Willingale} {et~al.}(2013){Willingale}, {Starling}, {Beardmore},
  {Tanvir}, \& {O'Brien}}]{willingale+13}
{Willingale}, R., {Starling}, R.~L.~C., {Beardmore}, A.~P., {Tanvir}, N.~R., \&
  {O'Brien}, P.~T. 2013, \mnras, 431, 394

\bibitem[{Wittor {et~al.}(2021)Wittor, Ettori, Vazza, Rajpurohit, Hoeft, \&
  Domínguez-Fernández}]{wittor+21}
Wittor, D., Ettori, S., Vazza, F., {et~al.} 2021

\bibitem[{{Wittor} {et~al.}(2019){Wittor}, {Hoeft}, {Vazza}, {Br{\"u}ggen}, \&
  {Dom{\'\i}nguez-Fern{\'a}ndez}}]{wittor+19}
{Wittor}, D., {Hoeft}, M., {Vazza}, F., {Br{\"u}ggen}, M., \&
  {Dom{\'\i}nguez-Fern{\'a}ndez}, P. 2019, \mnras, 490, 3987

\bibitem[{{Xie} {et~al.}(2020){Xie}, {van Weeren}, {Lovisari},
  {Andrade-Santos}, {Botteon}, {Br{\"u}ggen}, {Bulbul}, {Churazov}, {Clarke},
  {Forman}, {Intema}, {Jones}, {Kraft}, {Lal}, {Mroczkowski}, \&
  {Zitrin}}]{xie+20}
{Xie}, C., {van Weeren}, R.~J., {Lovisari}, L., {et~al.} 2020, \aap, 636, A3

\bibitem[{{Zhuravleva} {et~al.}(2019){Zhuravleva}, {Churazov}, {Schekochihin},
  {Allen}, {Vikhlinin}, \& {Werner}}]{zhuravleva+19}
{Zhuravleva}, I., {Churazov}, E., {Schekochihin}, A.~A., {et~al.} 2019, Nature
  Astronomy, 3, 832

\bibitem[{ZuHone(2010)}]{zuhone10}
ZuHone, J. 2010

\end{thebibliography}

\clearpage
\onecolumn
\begin{appendix}
\clearpage
\section{Peeling on L-band data}\label{apx:Lband_extracal}
\begin{figure*}[h!]
\centering
\includegraphics[width=0.3\textwidth]{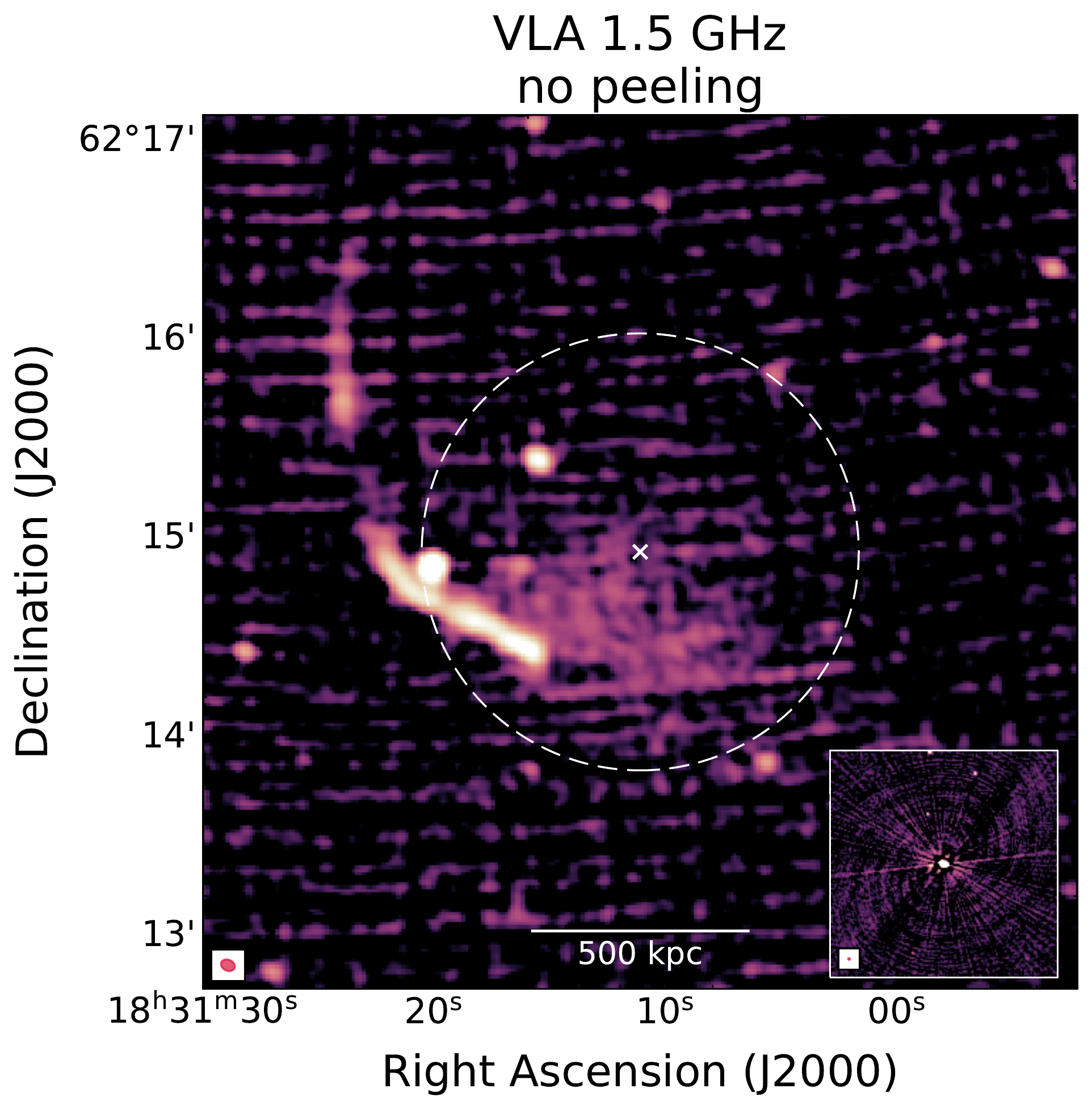}
\includegraphics[width=0.3\textwidth]{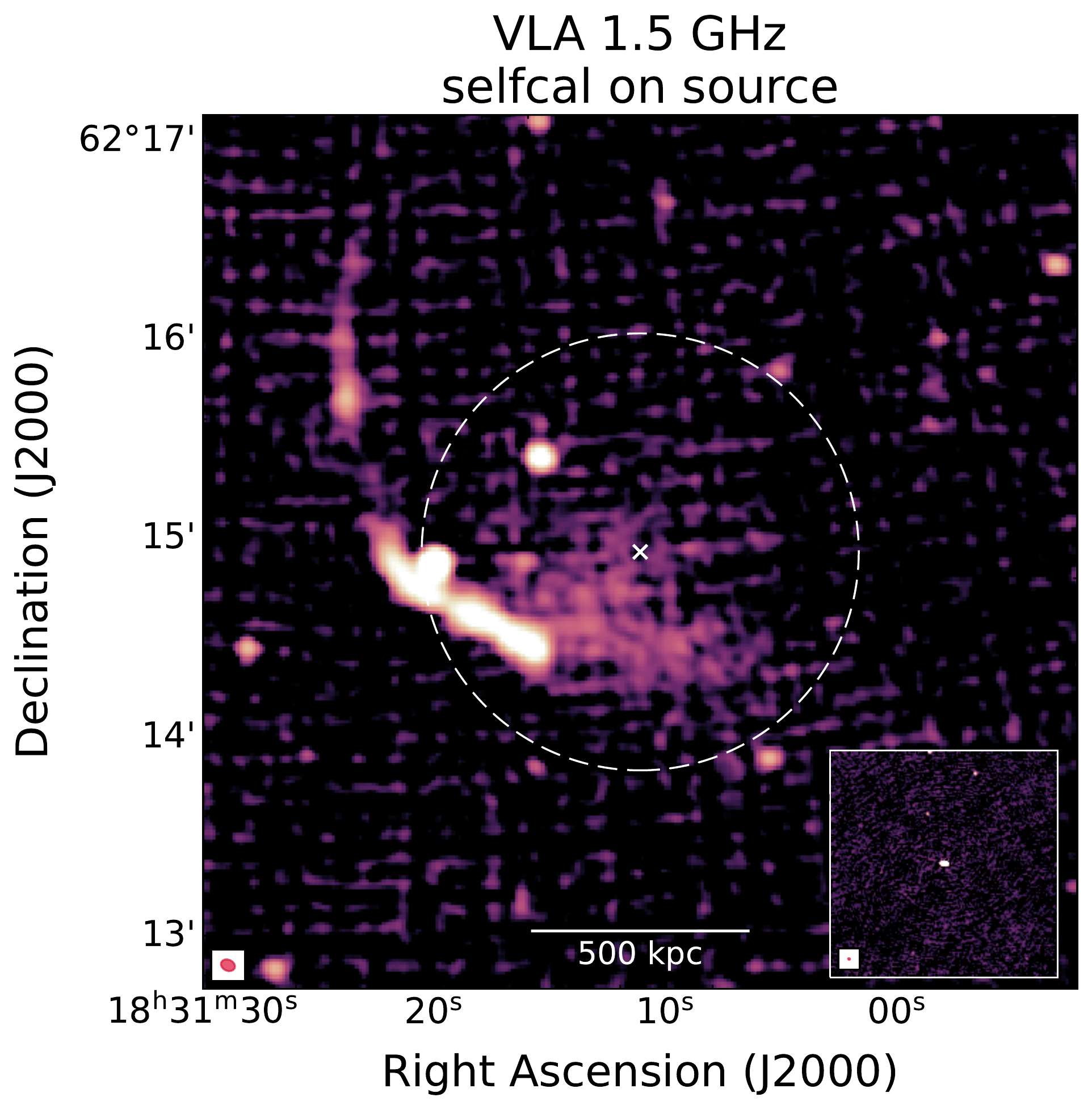}
\includegraphics[width=0.3\textwidth]{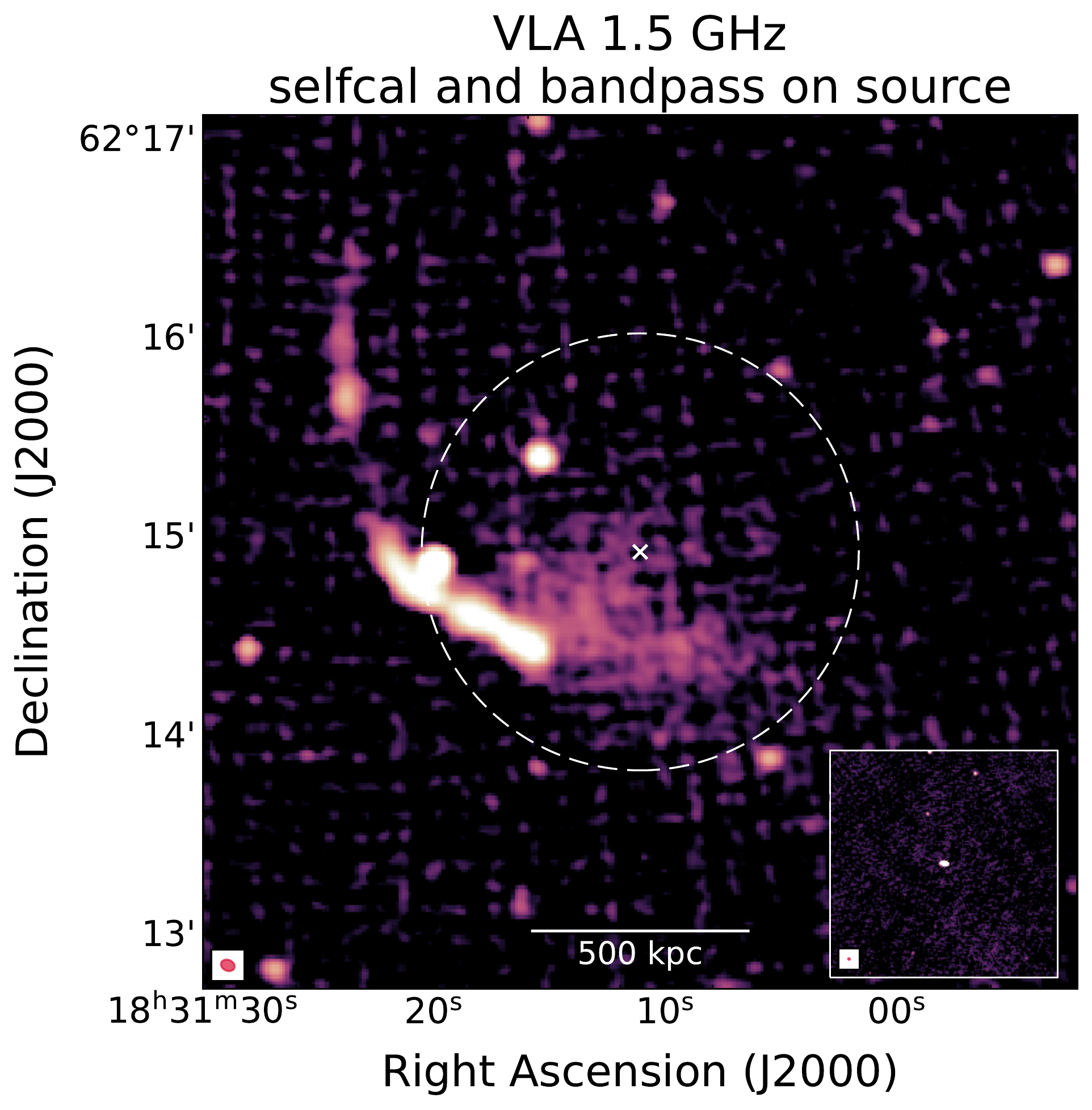}
\caption{Peeling results on the 1--2 GHz observations of \cluster. Left panel: cluster image after standard selfcal. Middle panel: cluster image after standard peeling (i.e. only rounds of selfcal on the troubling source). Right panel: cluster image after additional rounds of bandpass calibration on the troubling source. In the inset of each panel we show the improvement on the calibration of the troubling source, with the limits on the colourmap set to the last step.}
\label{fig:apx_lband_cal}
\end{figure*}

In this section, we present the additional calibration performed on the point source at the location $\rm RA=18^h31^m24.6^s~DEC=+62^\circ30^\prime34.32''$, whose side-lobes were affecting the visibilities on the observing target. In Fig.~\ref{fig:apx_lband_cal}, we present the images of the target and peeled source after the ``standard'' self-cal calibration and after the on-source self-cal and bandpass calibrations. After the additional calibration, the source is subtracted from the visibilities.

\section{Additional frequency observations}\label{apx:other_freqs}
\begin{figure*}[h!]
\centering
\includegraphics[width=\textwidth]{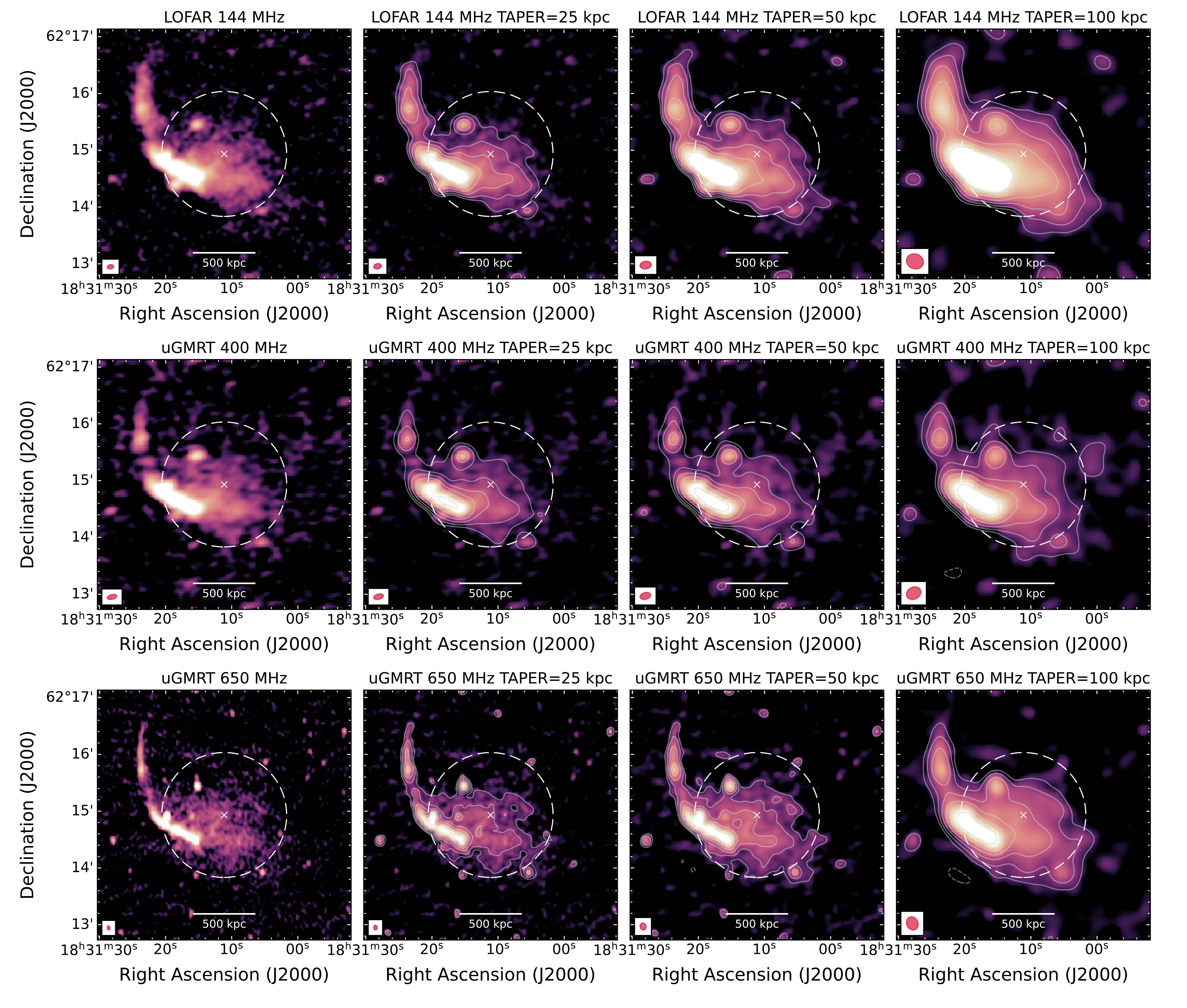}
\caption{As Fig. \ref{fig:vla_full_low_res}, but for the LOFAR 144 MHz and uGMRT 400 MHz and 650 MHz observations (top to bottom rows).}
\label{fig:otherfreqs_full_low_res}
\end{figure*}

\begin{figure*}
\centering
\includegraphics[width=\textwidth]{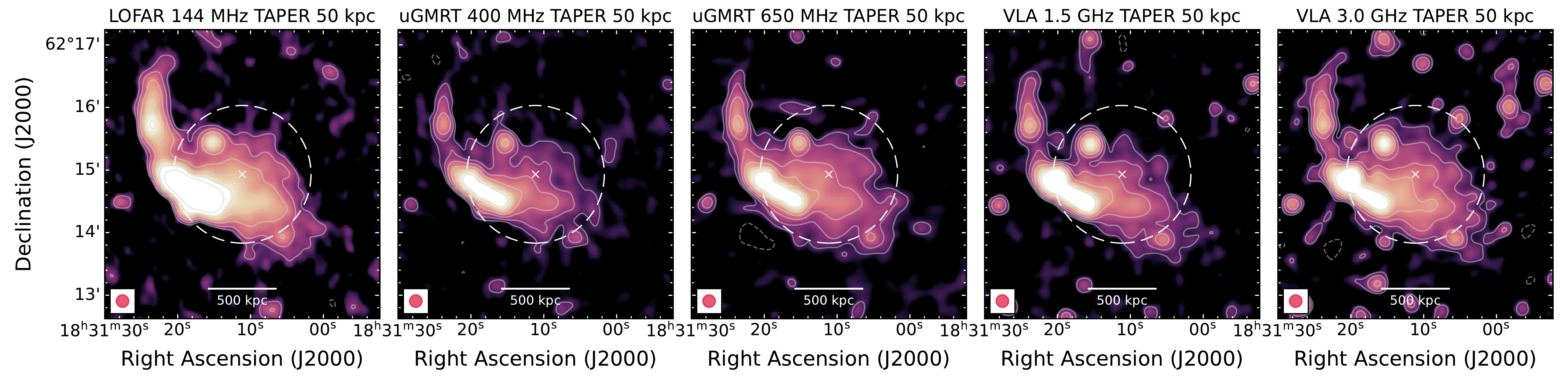}
\caption{Comparison of the $12''$ observations at each frequencies. Radio contours start from $2.5\sigma_{\rm rms,\nu}$, where $\rm\sigma_{rms,144MHz}=137.1~\mu Jy~beam^{-1}$, $\rm\sigma_{rms,400MHz}=98.7~\mu Jy~beam^{-1}$, $\rm\sigma_{rms,650MHz}=41.6~\mu Jy~beam^{-1}$, $\rm\sigma_{rms,1.5GHz}=17.4~\mu Jy~beam^{-1}$ and $\rm\sigma_{rms,3.0GHz}=6.2~\mu Jy~beam^{-1}$.}\label{fig:allfreqs}
\end{figure*}

In this section, we present the images for the additional frequency observations used for the analysis (Fig. \ref{fig:otherfreqs_full_low_res}). Details on these images are reported in Table \ref{tab:images_otherfreqs}. A comparison between all the observations used for this study at $12''$ is shown in Fig. \ref{fig:allfreqs}.

\begin{table}[h!]
\caption{Radio imaging details for the LOFAR 144 MHz and uGMRT 400 MHz and 650 MHz observations.}
\vspace{-5mm}
\begin{center}
\resizebox{0.5\textwidth}{!}{
\begin{tabular}{cccccccccc}
\hline
\hline
Central frequency & Resolution & $uv$-taper & Map noise \\
$\nu$ [MHz] & $\Theta$ [$''\times'',^\circ$] & [kpc] & $\rm \sigma_{rms}~[\mu Jy~beam^{-1}]$ \\
\hline
144 & $4.4\times3.3$, 93 &  None$^\dagger$ & 137.5 \\
    & $6.8\times4.3$, 102 & None & 91.6 \\
    & $8.3\times5.4$, 101 & 25 & 93.9 \\
    & $11.9\times8.3$, 98 & 50 & 114.5 \\
    & $18.4\times15.5$, 69 & 100 & 164.8 \\
400  & $8.6\times4.5$, 103 & None$^\dagger$ & 70.0\\
    & $10.1\times5.0$, 104 & None & 50.0 \\
    & $10.4\times5.4$, 104 & 25 & 50.0 \\
    & $11.5\times7.2$, 106 & 50 & 58.3 \\
    & $15.8\times12.6$, 117 & 100 & 101.2 \\
650 & $4.0\times2.4$, 14 & None$^\dagger$ & 15.4 \\
    & $4.3\times2.9$, 10 & None & 11.3 \\
    & $5.0\times3.6$, 3 & 25 & 11.3 \\
    & $7.6\times6.1$, 14 & 50 & 19.3 \\
    & $14.4\times11.9$, 25 & 100 & 44.5 \\    
\hline
\end{tabular}
}
\end{center}
\vspace{-5mm}
\tablefoot{$^\dagger$These images are obtained with \texttt{robust=-1.25}.}
\label{tab:images_otherfreqs}
\end{table}

\clearpage

\section{Polarisation $QU$-fitting}\label{apx:qufitting_plots}
In this section we briefly summarise the {\it QU}-fitting procedure to retrieve the polarisation parameters. We also present additional fitting plots for single high- and low-SNR pixels, and for the full relic.

The $Q(\lambda^2)$ and $U(\lambda^2)$ emission is simultaneously fitted with cosine and sine models, while the $I(\lambda^2)$ emission is fitted with a log-parabolic model\footnote{We follow the notation in Sect. \ref{sec:spixmaps}, with $A$ the curvature parameter and $\alpha=2 A\log\nu_{\rm ref}+B$ the spectral index.} to take into account the curvature in the spectrum.
The fitting procedure uses then a Markov Chain Monte Carlo \citep[MCMC;][]{foreman-mackey+13} approach to evaluate the posterior probability of the polarisation parameters (i.e. $p_0$, $\chi_0$, RM and, eventually, either $\sigma_{\rm RM}$ or $\varsigma_{\rm RM}$), using the following initial priors:
\begin{equation}
\begin{cases}
I_0 \in [0, +\infty] \\
A \in [-\infty, +\infty] \\
B \in [-\infty, +\infty] \\
p_0 \in [0,1] \\
\chi_0 \in [-\infty,+\infty] \\
{\rm RM} \in [-400,+400] \\
\sigma_{\rm RM}^2 \in [0,+\infty] ~{\rm or}~ \varsigma_{\rm RM}^2 \in [0,+\infty] \, .
\end{cases}
\end{equation}
We note that, with this prior on $\chi_0$, we need to subsequently fold the polarisation angle to $\pi$ to take into account for multiple angle rotation. Angle values of 0 or $\pi$ and $\pm\pi/2$ reflect the north/south and east/west magnetic field directions, respectively.

\begin{figure*}
\centering
\includegraphics[width=\textwidth]{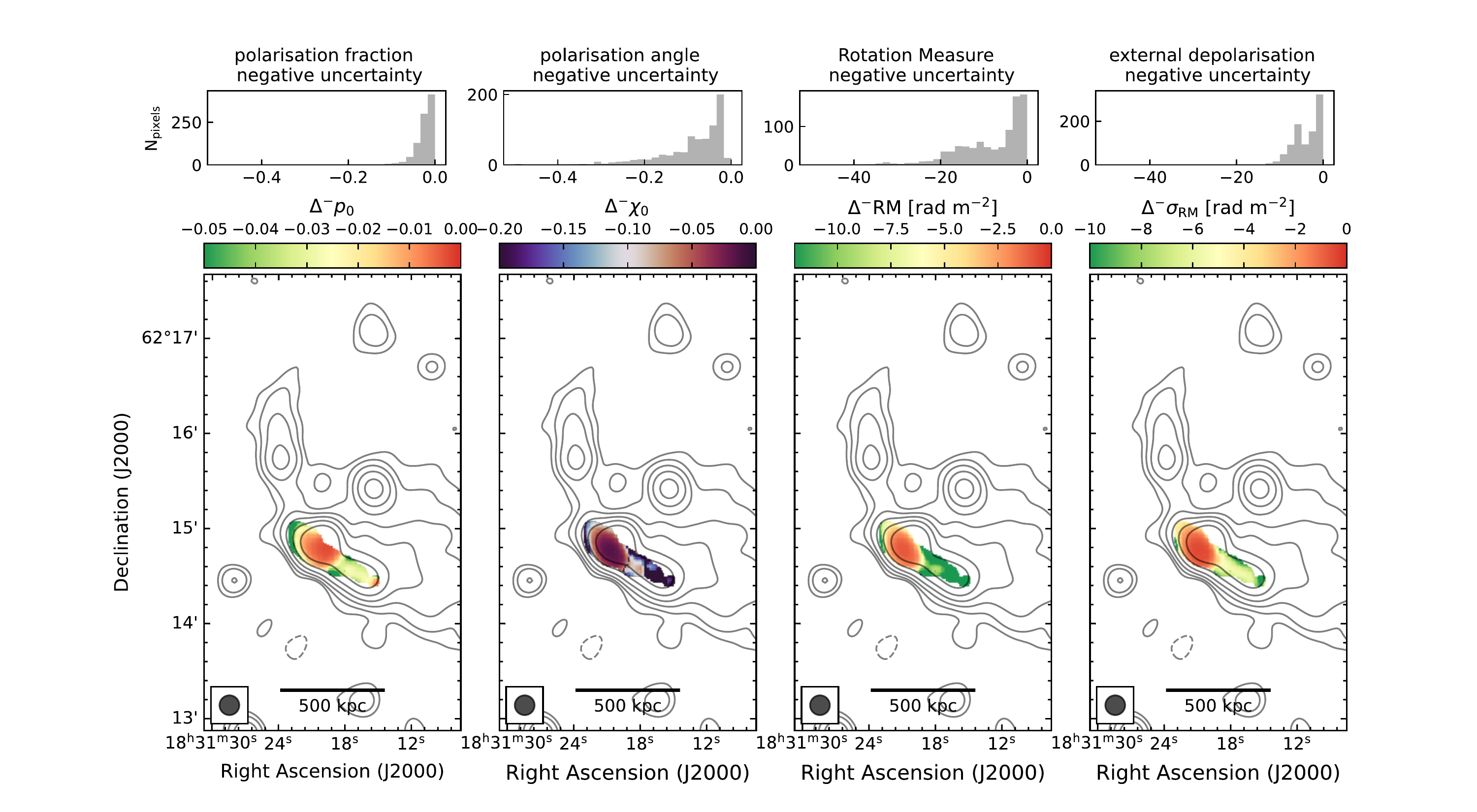}\\
\includegraphics[width=\textwidth]{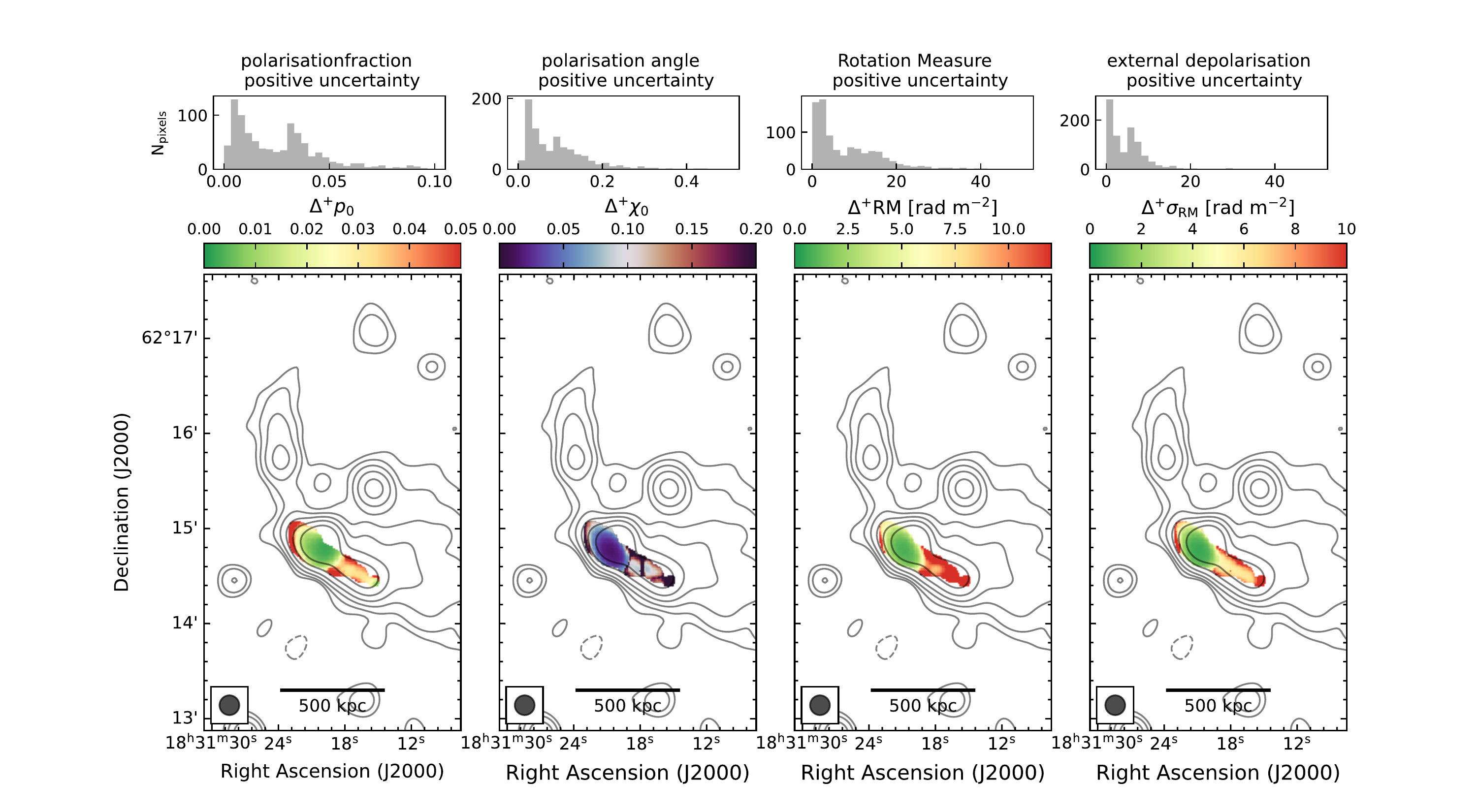}
\caption{Corresponding negative (top) and positive (bottom) uncertainties maps of Fig.~\ref{fig:polparam}}
\label{fig:polparamerrs}
\end{figure*}

\begin{figure*}
\centering
\includegraphics[height=0.4\textwidth]{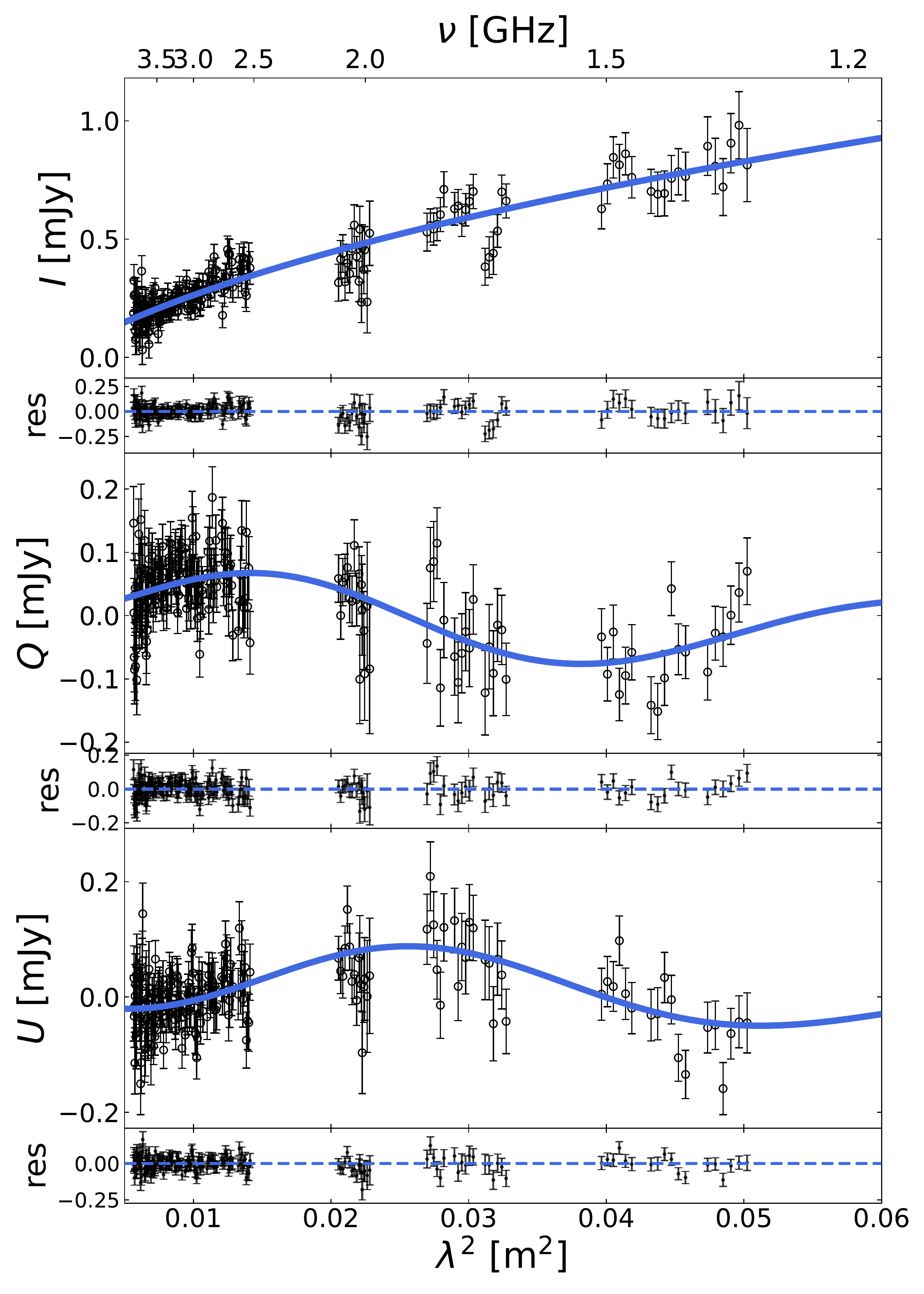}
\includegraphics[height=0.4\textwidth]{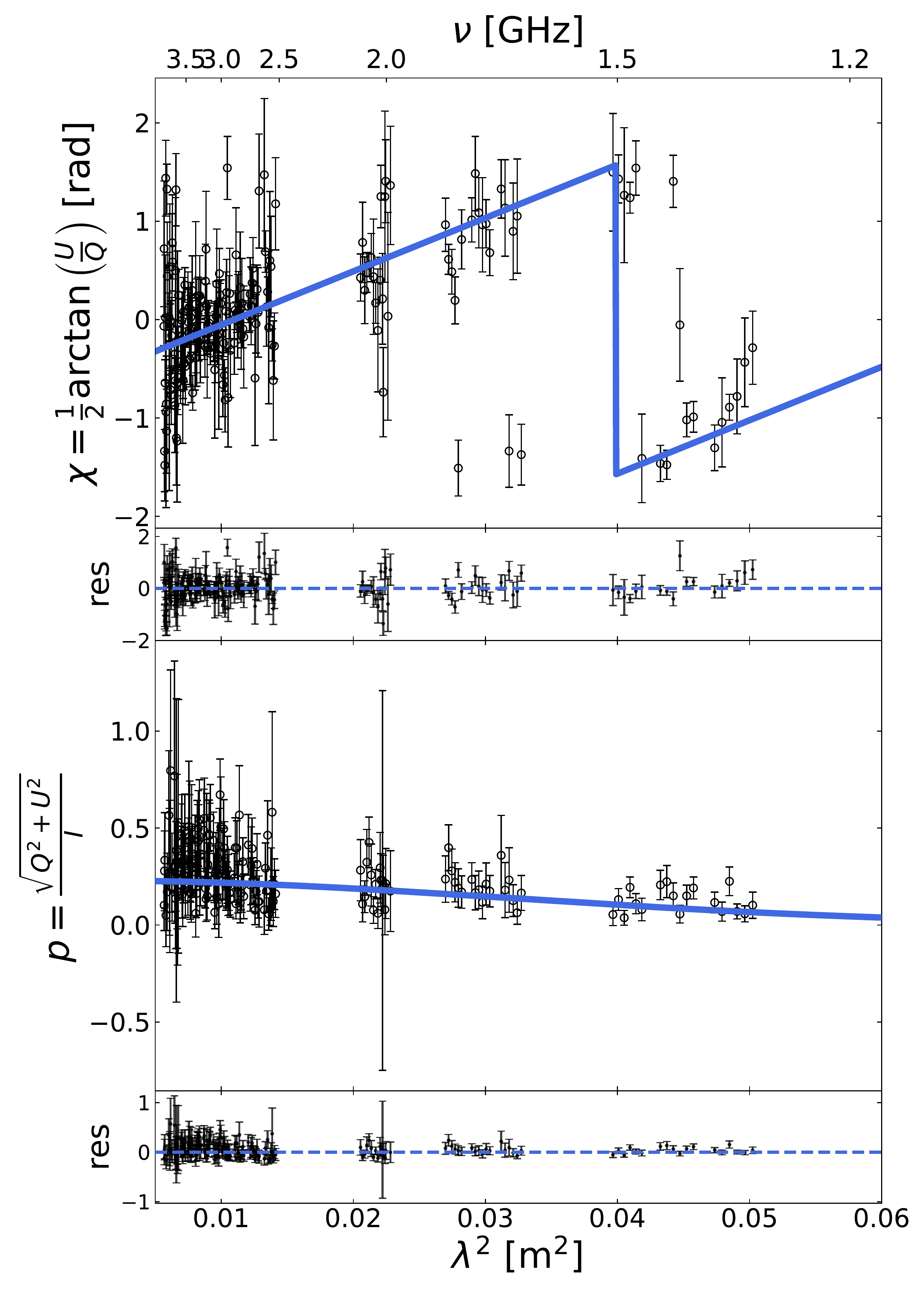}
\includegraphics[height=0.4\textwidth]{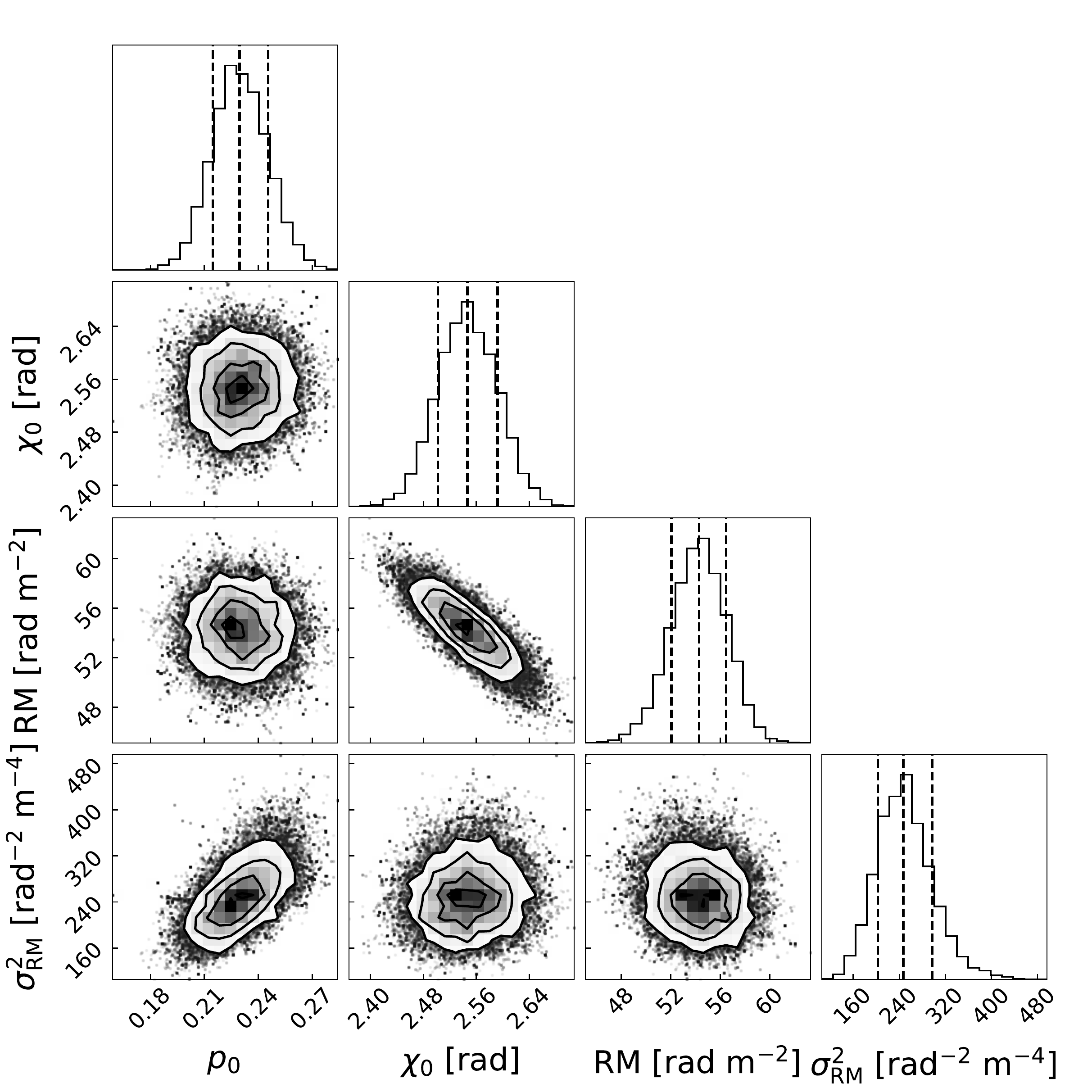}\\
\includegraphics[height=0.4\textwidth]{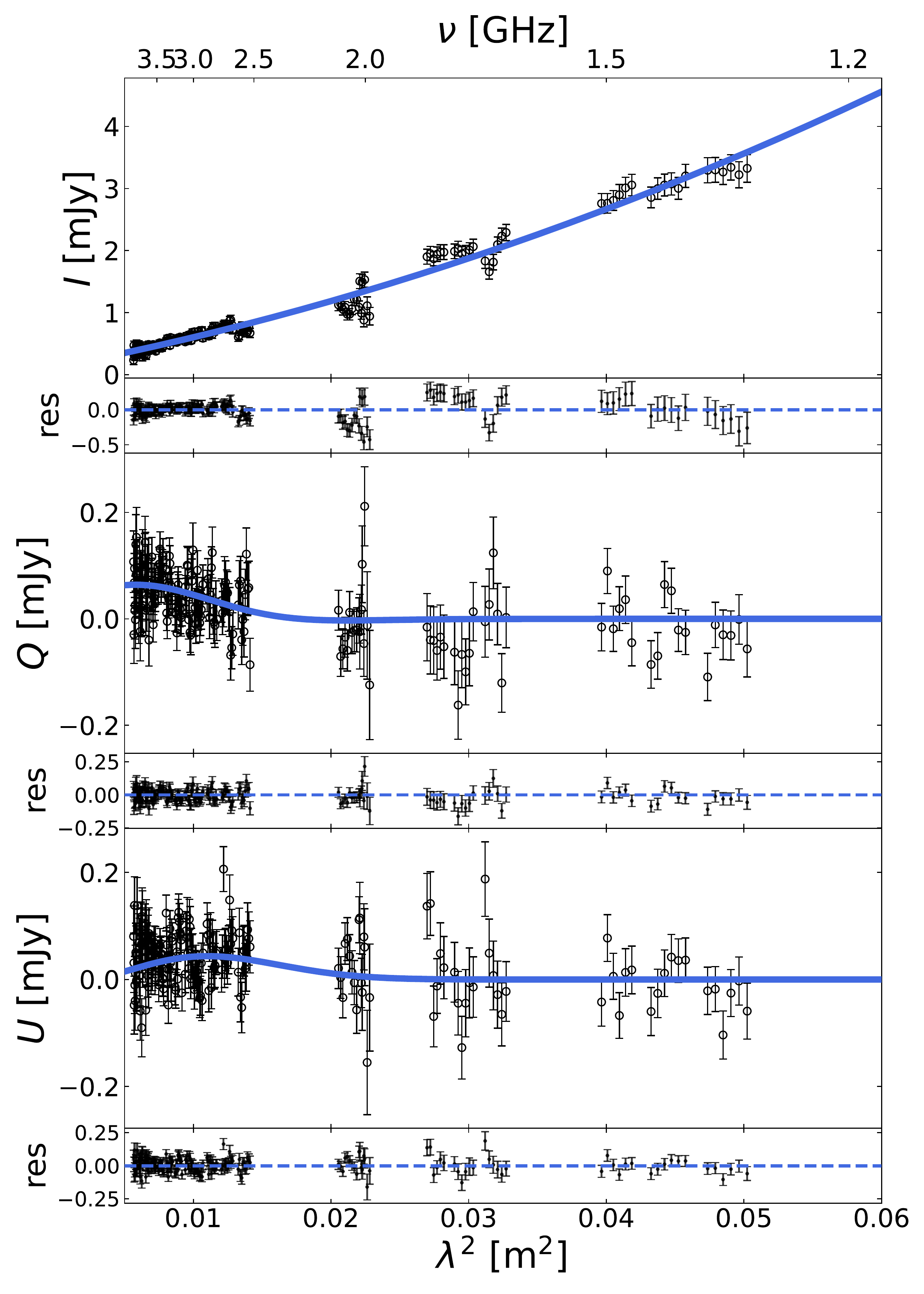}
\includegraphics[height=0.4\textwidth]{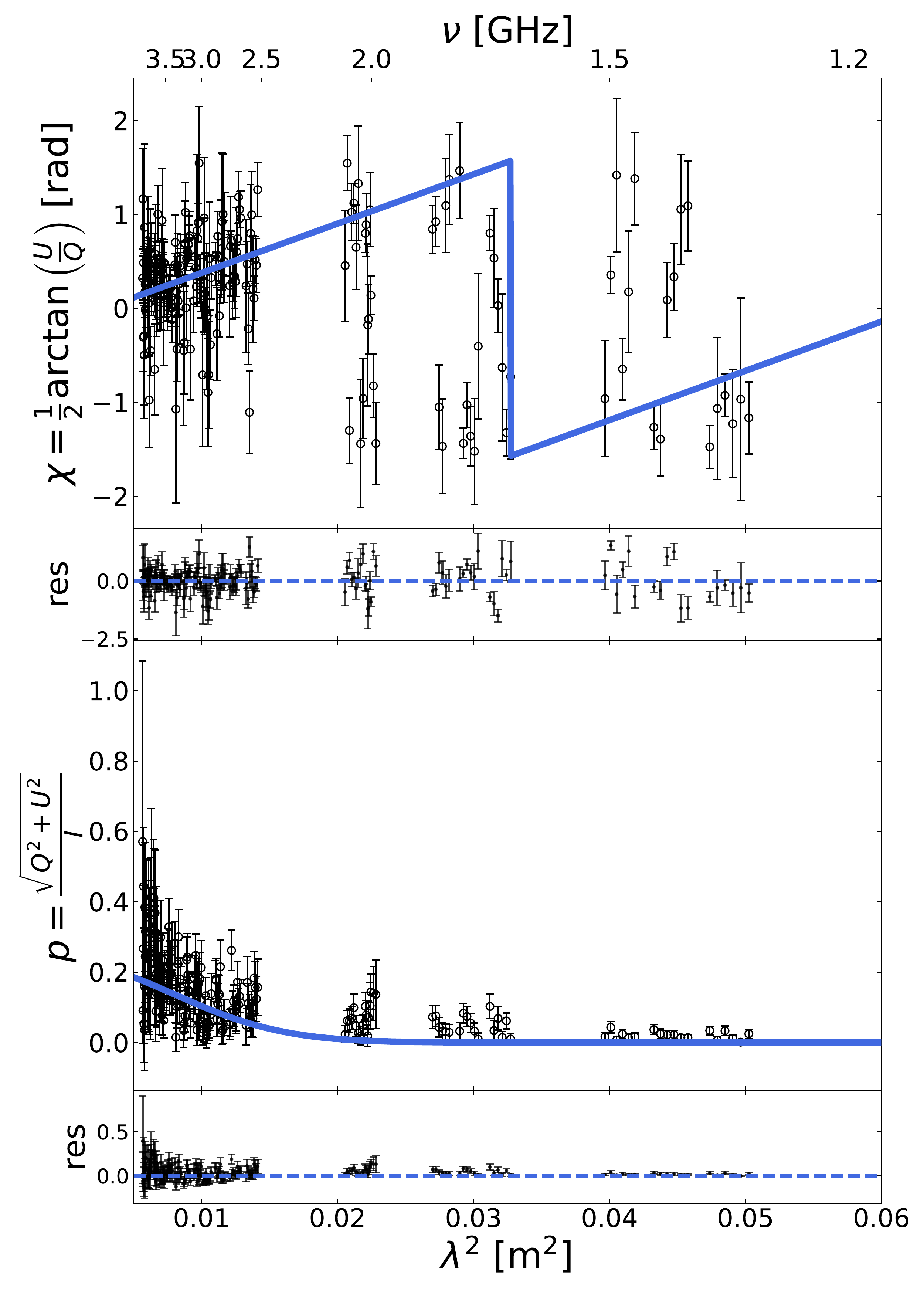}
\includegraphics[height=0.4\textwidth]{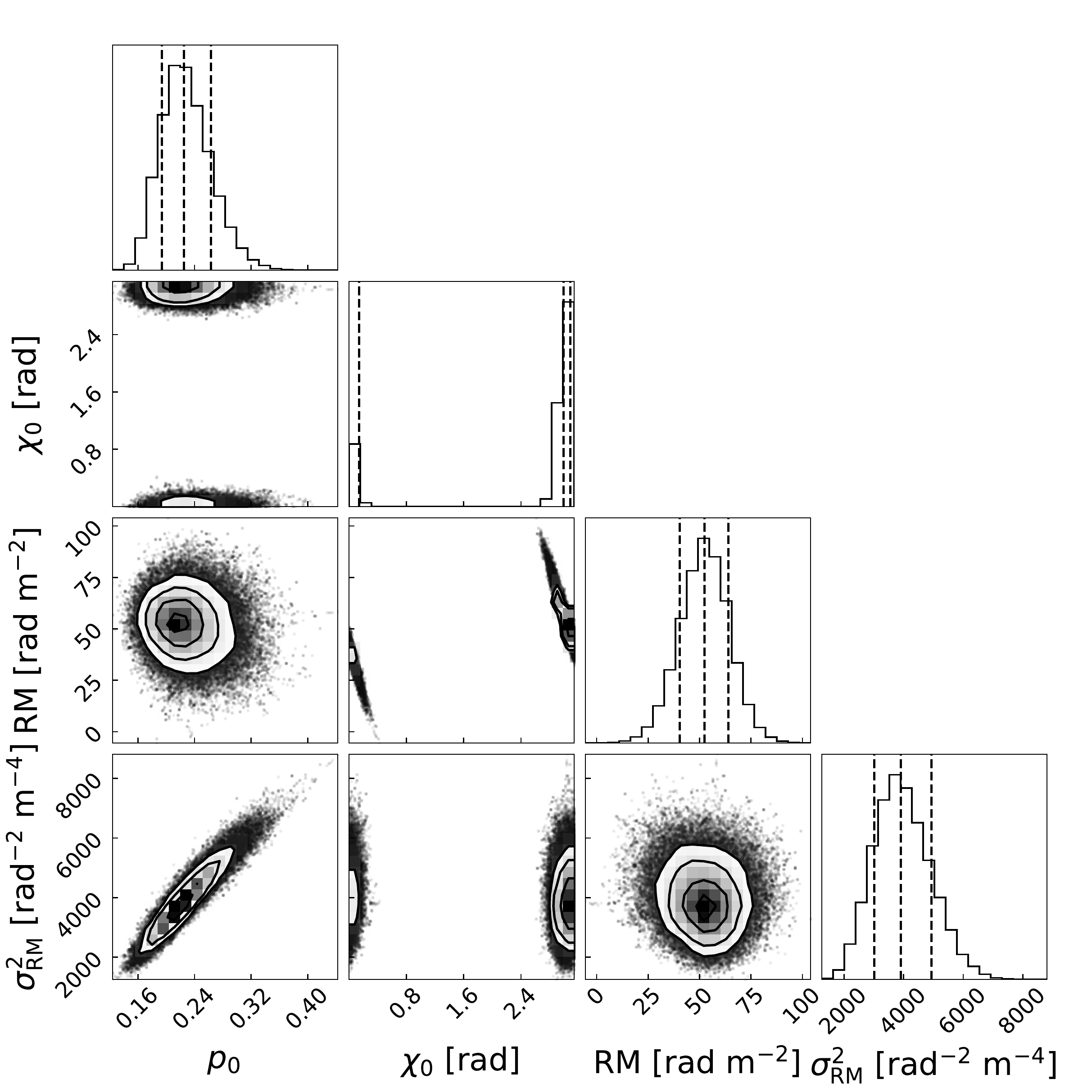}
\caption{{\it QU}-fitting results for a single pixel in ${\rm R2_N}$ (top) and for a single pixel in ${\rm R2_S}$ (bottom), assuming the external depolarisation model (EDF, Eq. \ref{eq:polmodels}). Left panel: Fits on Stokes $I$, $Q$ and $U$ fluxes. Middle panel: Resulting fractional polarisation, $p(\lambda^2)$, and polarisation angle, $\chi(\lambda^2)$. Right panel: Corner plot for the distribution of the uncertainties in the fitted polarisation parameters (i.e. $p_0$, $\chi_0$, RM and $\sigma_{\rm RM}^2$); contour levels are drawn at $[0.5,1.0,1.5,2.0]\sigma$, with $\sigma$ the 68\% statistical uncertainty (see dashed lines in the 1D histogram).}
\label{fig:polsinglepixel}
\end{figure*}

\begin{figure*}
\centering
\includegraphics[height=0.4\textwidth]{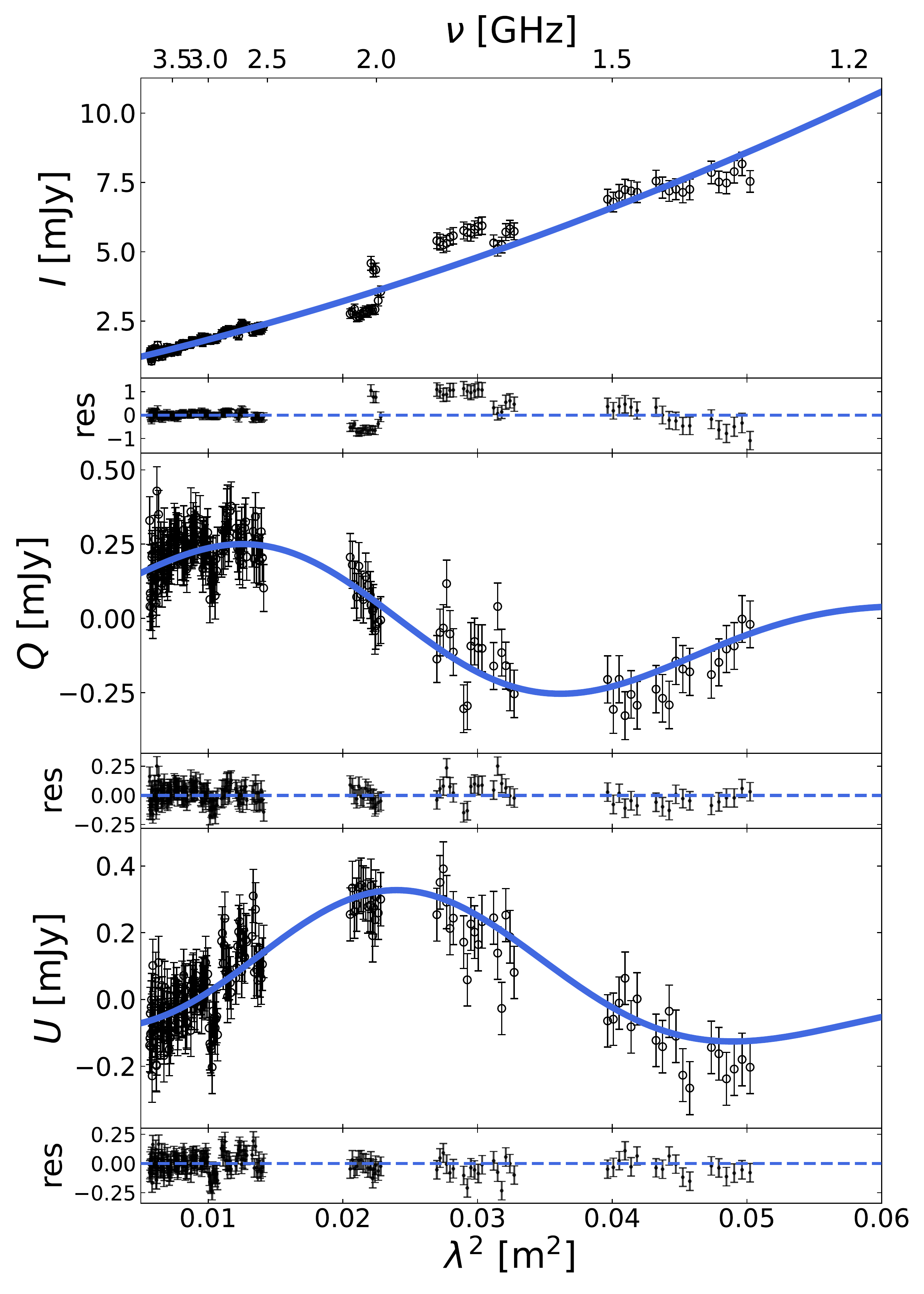}
\includegraphics[height=0.4\textwidth]{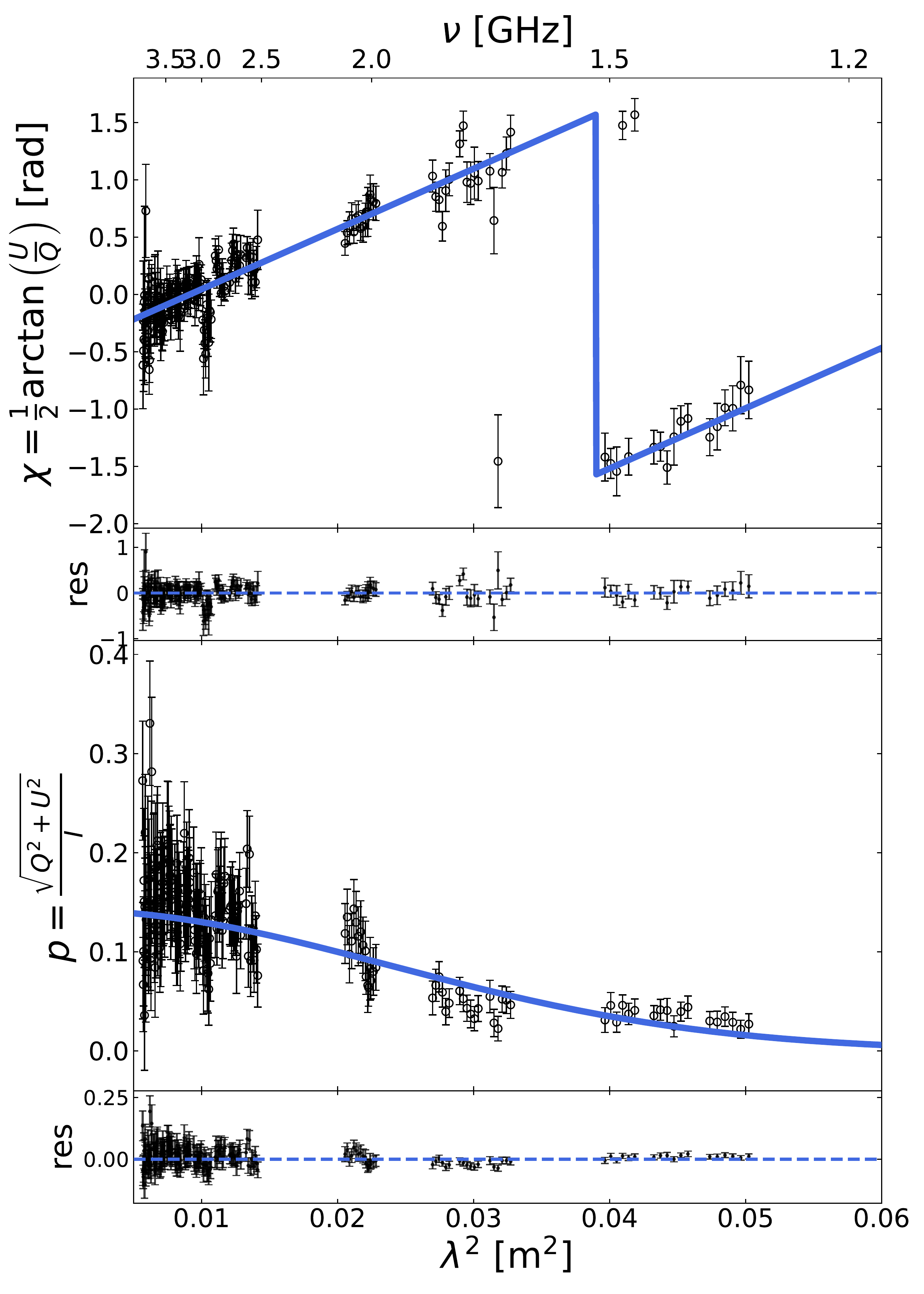}
\includegraphics[height=0.4\textwidth]{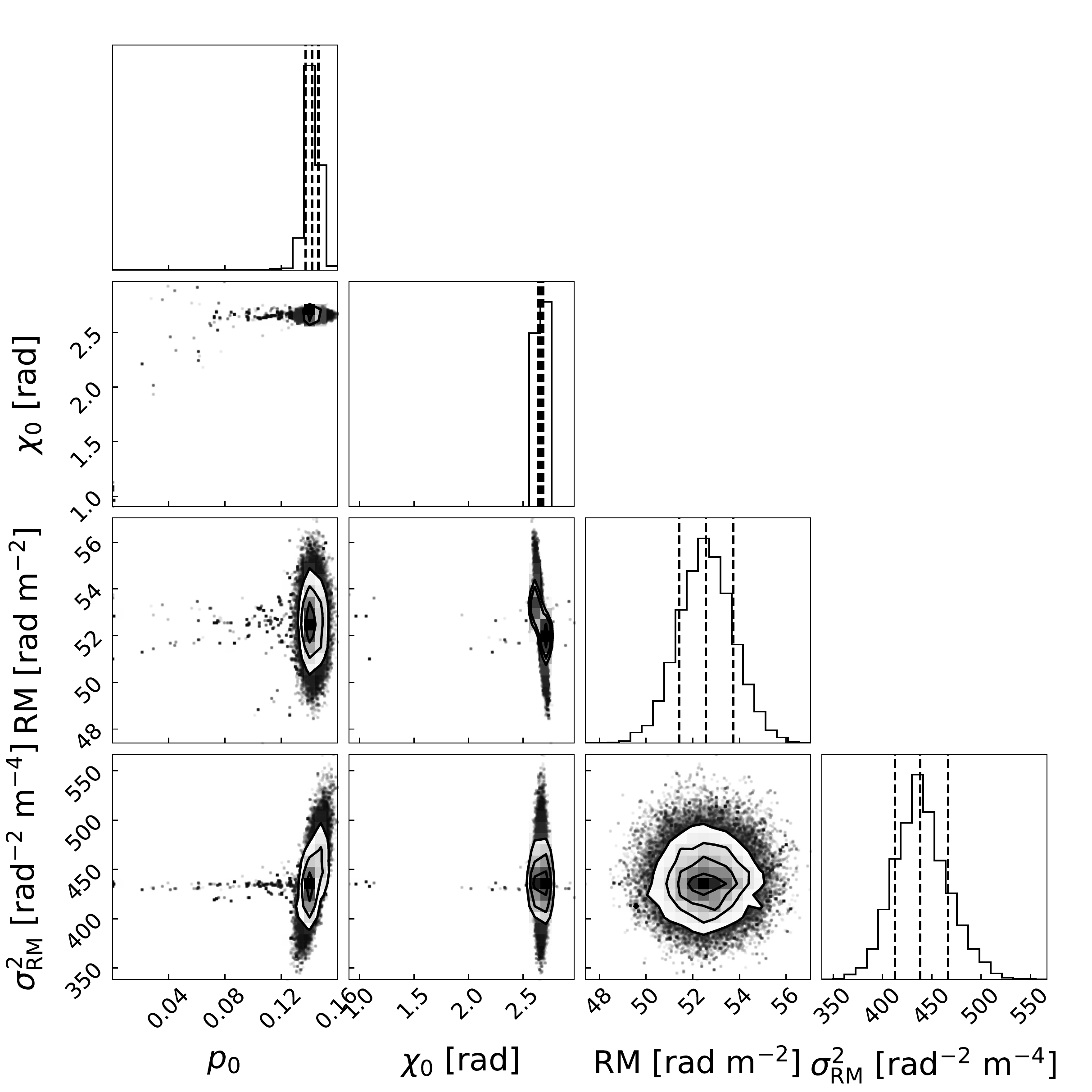}\\
\includegraphics[height=0.4\textwidth]{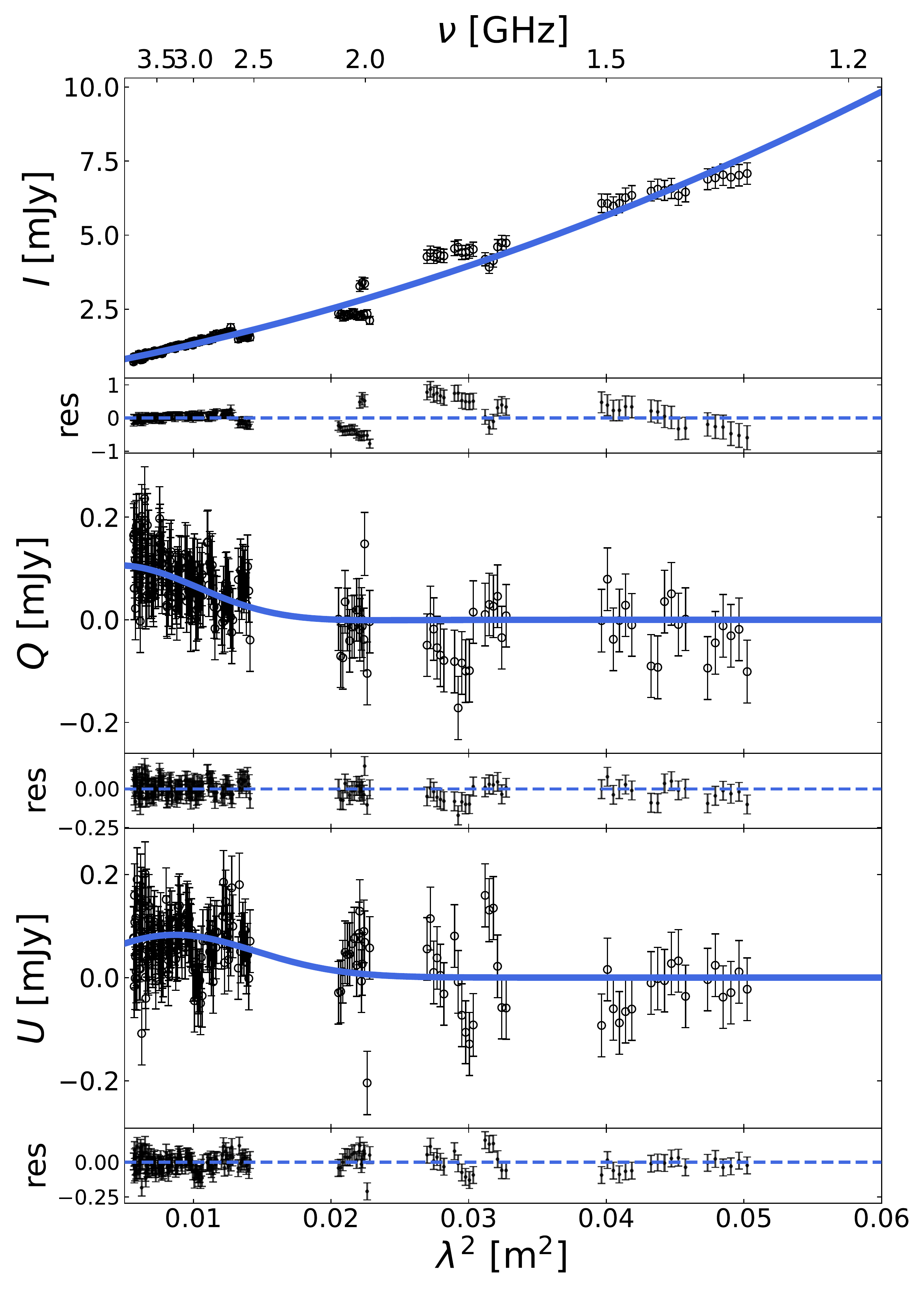}
\includegraphics[height=0.4\textwidth]{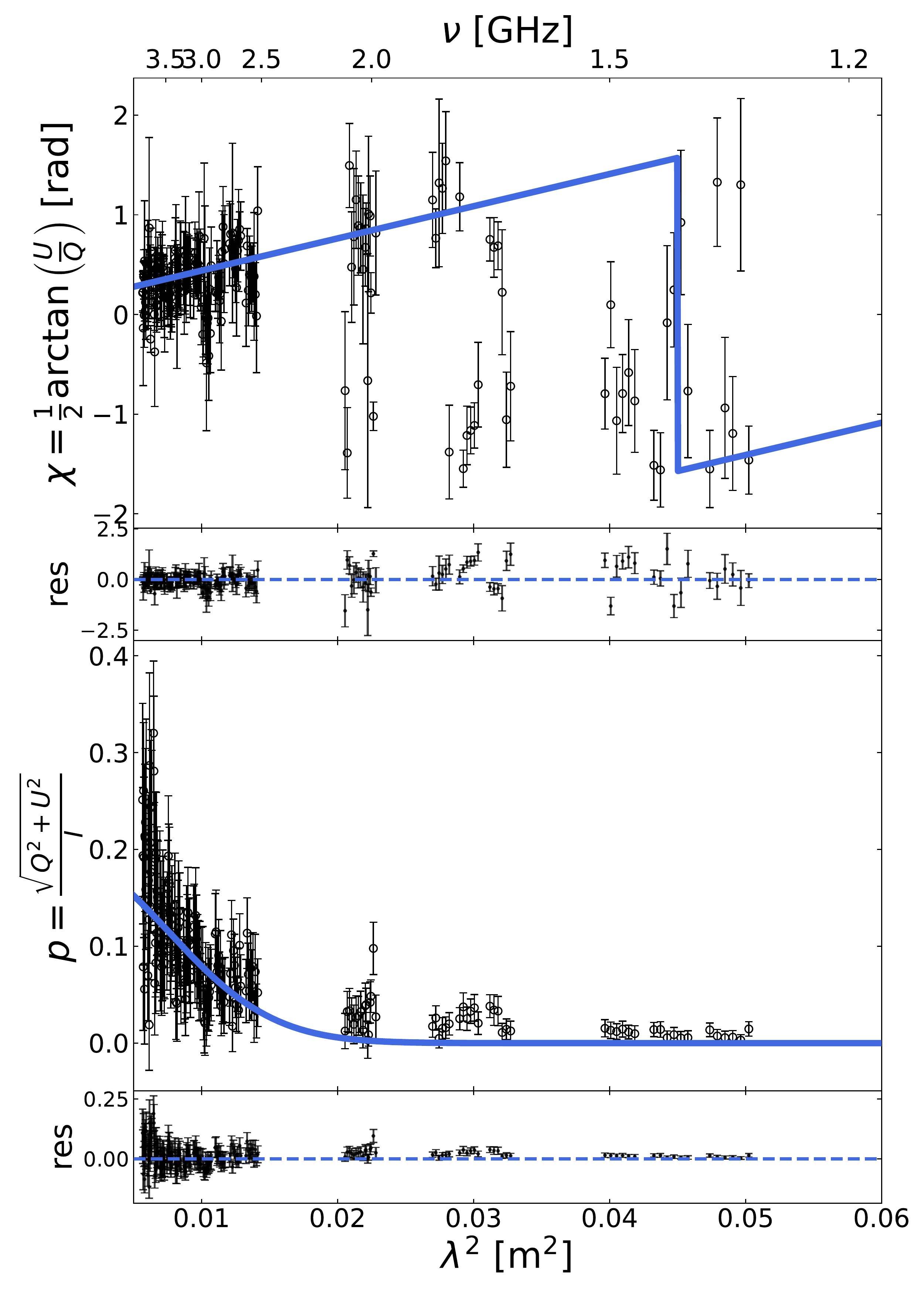}
\includegraphics[height=0.4\textwidth]{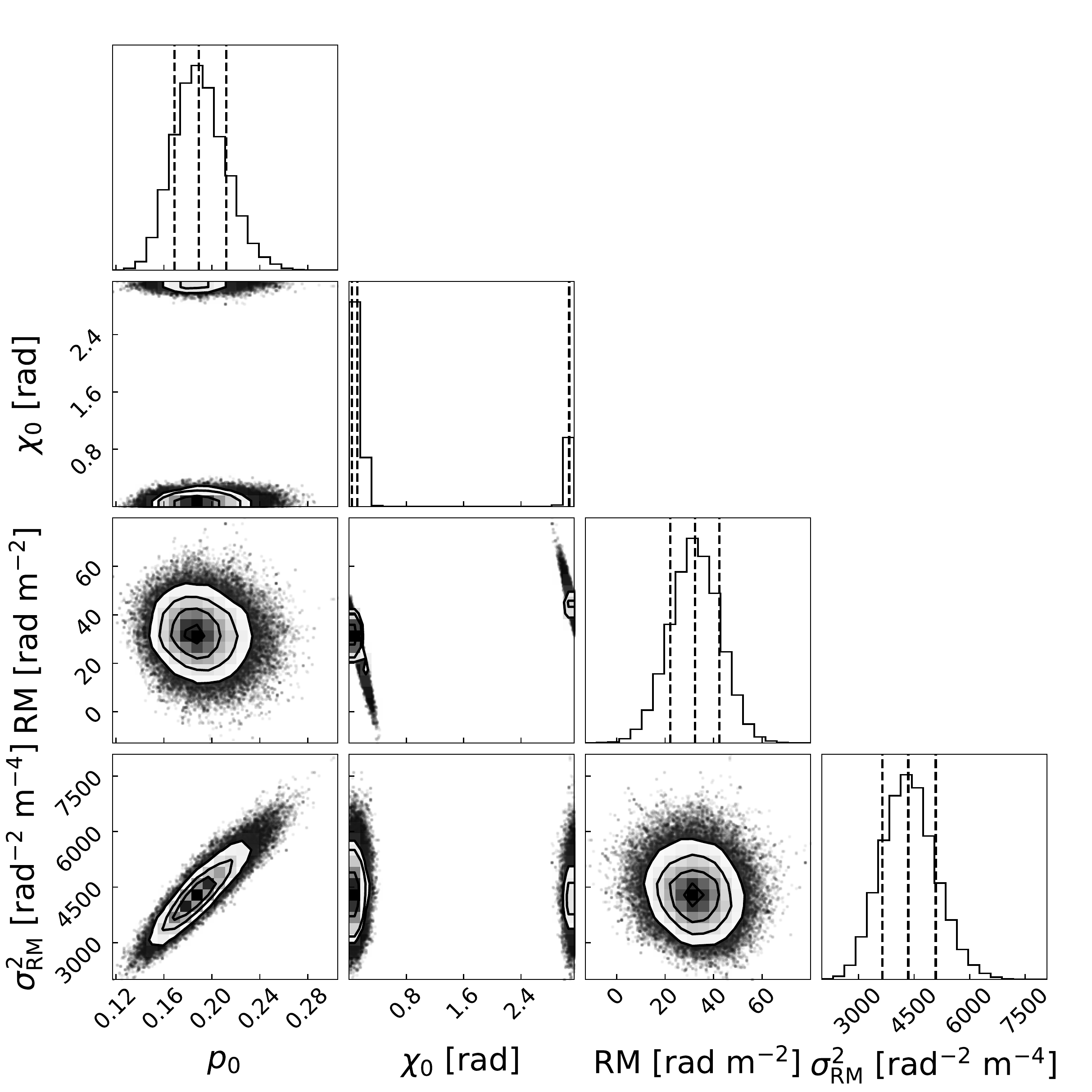}\\
\caption{As Fig. \ref{fig:polsinglepixel} but for ${\rm R2_N}$ (top) and ${\rm R2_S}$ integrated (bottom).}
\label{fig:polintegr}
\end{figure*}

In Fig. \ref{fig:polparamerrs} we show the corresponding negative (top) and positive (bottom) uncertainties maps of Fig.~\ref{fig:polparam}. In Fig. \ref{fig:polsinglepixel} we show examples of fitting results on a single pixel on ${\rm R2_N}$ and ${\rm R2_S}$, while in Fig. \ref{fig:polintegr} we show the integrated fitting results.

\end{appendix}

\end{document}